\documentclass{article}

\usepackage{graphicx} 
\usepackage[utf8]{inputenc}
\usepackage{jheppub}
\usepackage{amsthm}
\usepackage{amsmath}
\usepackage[usenames, dvipsnames]{xcolor}
\usepackage{bbm}
\usepackage{tabularx}
\usepackage{multirow}
\usepackage{hyperref}
\usepackage{comment}
\usepackage{amssymb,amsfonts}
\usepackage{mathtools}
\usepackage{mathrsfs}
\usepackage{subcaption}
\usepackage{graphicx}
\usepackage{tikz}
\usepackage{tikz-cd}

\newcommand{\vdm}{\Delta}

\def\half{{\scriptstyle \frac 12}}

\newcommand{\mc}{\mathcal}

\newcommand{\exg}[1]{\llangle #1 \rrangle}

\newcommand{\exlc}[1]{\llangle #1 \rrangle_{\!\!\phantom{\vert}_{\rm c}}}

\newcommand{\SL}{{\rm SL}}

\renewcommand{\(}{\left(}
\renewcommand{\)}{\right)}
\renewcommand{\[}{\left[}
\renewcommand{\]}{\right]}

\makeatletter
\DeclareFontFamily{OMX}{MnSymbolE}{}
\DeclareSymbolFont{MnLargeSymbols}{OMX}{MnSymbolE}{m}{n}
\SetSymbolFont{MnLargeSymbols}{bold}{OMX}{MnSymbolE}{b}{n}
\DeclareFontShape{OMX}{MnSymbolE}{m}{n}{
    <-6>  MnSymbolE5
   <6-7>  MnSymbolE6
   <7-8>  MnSymbolE7
   <8-9>  MnSymbolE8
   <9-10> MnSymbolE9
  <10-12> MnSymbolE10
  <12->   MnSymbolE12
}{}
\DeclareFontShape{OMX}{MnSymbolE}{b}{n}{
    <-6>  MnSymbolE-Bold5
   <6-7>  MnSymbolE-Bold6
   <7-8>  MnSymbolE-Bold7
   <8-9>  MnSymbolE-Bold8
   <9-10> MnSymbolE-Bold9
  <10-12> MnSymbolE-Bold10
  <12->   MnSymbolE-Bold12
}{}

\let\llangle\@undefined
\let\rrangle\@undefined
\DeclareMathDelimiter{\llangle}{\mathopen}%
                     {MnLargeSymbols}{'164}{MnLargeSymbols}{'164}
\DeclareMathDelimiter{\rrangle}{\mathclose}%
                     {MnLargeSymbols}{'171}{MnLargeSymbols}{'171}

\makeatother

\newcommand{\bW}{\mathbb{W}}
\newcommand{\be}{\begin{equation}}
\newcommand{\ee}{\end{equation}}
\newcommand{\ba}{\begin{aligned}}
\newcommand{\ea}{\end{aligned}}

\def\ie{\begin{equation}\begin{aligned}}
\def\fe{\end{aligned}\end{equation}}

\title{Electromagnetic Duality for Line Defect Correlators\\
in $\mathcal{N}=4$ Super Yang--Mills Theory}

\author[1,2]{Daniele Dorigoni}
\author[3]{$\!\!$, Zhihao Duan}
\author[4]{$\!\!$, Daniele  R. Pavarini}
\author[3]{$\!\!$, Congkao Wen}
\author[5]{$\!\!$, Haitian Xie}

\affiliation[1]{\small \vspace{0.2cm} Centre for Particle Theory \& Department of Mathematical Sciences Durham University, Lower Mountjoy, Stockton Road, Durham, DH1 3LE, UK}
\affiliation[2]{\small Max-Planck-Institut f\"ur Gravitationsphysik (Albert-Einstein-Institut), 
am M\"uhlenberg 1, Potsdam, 14476, Germany}
\affiliation[3]{\small School of Physics and Astronomy, Queen Mary University of London, Mile End Road, London, E1 4NS, UK}
\affiliation[4]{\small Scuola Normale Superiore, Piazza dei Cavalieri 7, Pisa, 56126,  Italy}
\affiliation[5]{\small Jefferson Physical Laboratory, Harvard University, Cambridge, MA 02138, USA}

\emailAdd{daniele.dorigoni@durham.ac.uk}
\emailAdd{z.duan@qmul.ac.uk}
\emailAdd{daniele.pavarini@sns.it}
\emailAdd{c.wen@qmul.ac.uk}
\emailAdd{hxie@fas.harvard.edu}

\abstract{We study particular integrated correlation functions of two superconformal primary operators of the stress tensor multiplet in the presence of a half-BPS line defect labelled by electromagnetic charges $(p,q)$ in $\mathcal{N}=4$ supersymmetric Yang--Mills theory (SYM) with gauge group $SU(N)$. An important consequence of ${\rm SL}(2,\mathbb{Z})$ electromagnetic duality in $\mathcal{N}=4$ SYM is that correlators of line defect operators with different charges $(p,q)$ must be related in a non-trivial manner when the complex coupling $\tau = \theta/(2\pi) + 4\pi i /g_{_{\rm YM}}^2$ is transformed appropriately. In this work we introduce a novel class of real-analytic functions whose automorphic properties with respect to ${\rm SL}(2,\mathbb{Z})$ match the expected transformations of line defect operators in $\mathcal{N}=4$ SYM under electromagnetic duality. At large $N$ and fixed $\tau$, the correlation functions we consider are related to scattering amplitudes of two gravitons from extended $(p,q)$-strings in the holographic dual type IIB superstring theory.
We show that the large-$N$ expansion coefficients of the integrated two-point line defect correlators are given by finite linear combinations with rational coefficients of elements belonging to this class of automorphic functions. On the other hand, for any fixed value of $N$  we conjecture that the line defect integrated correlators can be expressed as formal infinite series over such automorphic functions. The resummation of this series produces a simple lattice sum representation for the integrated line defect correlator that manifests its automorphic properties. We explicitly demonstrate this construction for the cases with gauge group $SU(2)$ and $SU(3)$. Our results give direct access to non-perturbative integrated correlators in the presence of an 't Hooft-line defect, observables otherwise very difficult to compute by other means.
}

\begin{document}

\begin{flushright}
{\small QMUL-PH-24-20}
\end{flushright}

\maketitle
\section{Introduction}

Four-dimensional $\mathcal{N}=4$ supersymmetric Yang--Mills theory (SYM) \cite{Brink:1976bc} is famously conjectured to display a generalisation of electromagnetic duality usually called $S$-duality \cite{Montonen:1977sn, Witten:1978mh, Osborn:1979tq, Goddard:1976qe}. $S$-duality is non-perturbative in nature, connecting in a non-trivial way the theory at weak coupling with the strong coupling regime. As a consequence of this non-perturbative aspect, it is challenging to investigate directly the implications of $S$-duality on the non-trivial physical observables of the theory, such as correlation functions of local operators as well as of extended defect operators. Observables in $\mathcal{N}=4$ SYM are in general, functions of the space-time coordinates as well as functions of the complexified Yang--Mills coupling,
\be 
\tau= \tau_1 + i \tau_2 \coloneqq \frac{\theta}{2\pi} +\frac{4\pi i}{g^2_{_{\rm YM}}} \, ,
\ee 
with $\theta$ the topological theta angle and $g_{_{\rm YM}}$ the Yang--Mills coupling constant.
 A striking consequence of $S$-duality is that different correlators might be be related to one another upon transforming the coupling $\tau$  via the usual ${\rm SL}(2, \mathbb{Z})$ action
 \begin{align}
 \tau \rightarrow \tau' &\label{eq:GNOtau}= \gamma\cdot \tau \coloneqq \frac{a\tau+b}{c\tau+d}\,,  \qquad {\rm with} \quad   \gamma=\left(\begin{matrix} a& b \\ c & d \end{matrix}\right) \in {\rm SL}(2, \mathbb{Z})\,.
 \end{align}
It appears manifest that to explicitly verify or exploit the predictions coming from $S$-duality, it is necessary to compute non-trivial correlation functions at finite $\tau$, which is of course an extremely difficult task to achieve even for a highly symmetric theory like $\mathcal{N}=4$ SYM.  

In recent years it has been shown that with the aid of $S$-duality predictions, it is possible to provide exact expressions as functions of the coupling constant $\tau$ for a class of observables closely related to correlation functions. These observables, often referred to as \textit{integrated correlators}, are correlation functions of certain half-BPS local operators whose spacetime dependence is integrated out over some specific integration measures~\cite{Binder:2019jwn,Chester:2020dja}. Remarkably, such integrated correlators  can be computed using supersymmetric localisation~\cite{Pestun:2007rz} and are expressible~\cite{Binder:2019jwn,Chester:2020dja} as $(N-1)$-dimensional matrix model integrals, where $(N-1)$ is the rank of the gauge group, which for the rest of this paper we will take to be $SU(N)$. For a large class of cases, the matrix model integral can be evaluated explicitly and the corresponding integrated correlators are expressed in terms of modular forms with a non-holomorphic dependence from the coupling constant $\tau$, thus providing manifest realisations of $S$-duality on physical observables in $\mathcal{N}=4$ SYM \cite{Chester:2019jas, Chester:2020vyz, Dorigoni:2021guq,Dorigoni:2021bvj,Dorigoni:2021rdo, Dorigoni:2022cua, Collier:2022emf, Dorigoni:2022zcr, Alday:2023pet, Dorigoni:2023ezg, Paul:2022piq, Paul:2023rka, Brown:2023cpz}, see also  \cite{Dorigoni:2022iem} for a review and further references therein.
Similar analysis have then been carried out for four-point functions in $4$-dimensional theories with fewer supersymmetries \cite{Chester:2022sqb, Behan:2023fqq, Billo:2023kak, Billo:2024ftq, Pini:2024uia}, as well as for $3$-dimensional theories such as the ABJM model \cite{Chester:2018aca, Binder:2018yvd, Agmon:2019imm,Alday:2021ymb}.  
While integrated correlators provide crucial exact results which bring us one step closer to understanding the non-integrated correlators, we stress that they are also extremely  important from a practical perspective since they furnish additional constraints which can be effectively included in the numerical bootstrap programme, aimed for example at obtaining coupling-dependent bounds on the anomalous dimensions and OPE coefficients  of unprotected operators \cite{Chester:2021aun,Chester:2022sqb,Chester:2023ehi, Behan:2024vwg}.

While so far most of these successful studies have focused on the analysis of integrated correlation functions for local operators, the implications of $S$-duality extend in a very non-trivial way to the case of extended objects such as defect operators. A key example of an extended operator is the Wilson-line defect, which can be thought of as a measure of the coupling between a heavy electric particle probe and the gauge field of the theory. Under $\mathcal{N}=4$ SYM electromagnetic duality we have that upon the inversion $\tau\to -1/\tau$, a Wilson-line defect must be related to an 't Hooft-line defect \cite{Kapustin:2005py}, which instead describes the insertion in the path-integral of a magnetic monopole. 
A natural question then arises: what is the class of ${\rm SL}(2,\mathbb{Z})$ automorphic forms relevant for discussing correlation functions of local operators in the presence of line operators such as Wilson and 't Hooft defects? 
The aim of this work is precisely to  address this question for a family of integrated correlation functions of local operators in the presence of particular line defect operators which we now introduce.

While Wilson-line defect operators can be described in the path-integral via a path-ordered exponential of local fields along the line supporting the defect, an ’t Hooft operator is an example of a disorder operator and its path-integral definition \cite{tHooft:1977nqb} involves specifying a certain singular gauge transformation around a path that links non-trivially the line supporting the 't Hooft defect, thus creating a magnetic flux tube along the loop effectively measured by $\pi_1(G)$.
As a consequence of electromagnetic duality in $\mathcal{N}=4$ SYM, Kapustin \cite{Kapustin:2005py} has shown that while Wilson-line defects are labelled by a representation $R$ of the gauge group $G$, 't Hooft defects must be labelled by a representation $\!\,^LR$ of the Langlands (or GNO \cite{Goddard:1976qe}) dual magnetic group $\!\,^LG$. A Wilson-line defect can then be thought of as having inserted in the path-integral the world-line of a point-like electric particle transforming in the representation $R$ of the group $G$, and analogously  an 't Hooft-line defect inserts a magnetic monopole transforming in the representation $\!\,^LR$ of the dual group $\!\,^L G$.
Furthermore, given that $\pi_1(G)$ is isomorphic to the centre $Z(^LG)$ of the dual group $\!\,^LG$, we have that the magnetic flux  created by an 't Hooft-line defect operator in the representation $\!\,^LR$ is given
by the charge $q\in Z(\!\,^LG) \subset \!\,^LG$ of the representation $\!\,^LR$ of the magnetic group $\!\,^LG$.

In this paper we consider half-BPS Wilson-line defect operators in the fundamental representation of the gauge group which we will take to be $SU(N)$.
Thanks to the holographic principle, $\mathcal{N}=4$ SYM has a dual description in terms of type IIB superstring theory on $AdS_5\times S^5$ where the Wilson-line defect operators is identified with a fundamental string \cite{Maldacena:1998im,Rey:1998ik}.
More in general  we consider a dyonic half-BPS line defect operator \cite{Kapustin:2005py} labelled by electromagnetic charges $(p,q)$, with $(p,q)$ two co-prime integers. The half-BPS 't Hooft-line defect operator correspond to the case $(p,q)=(0,1)$, which is holographically dual to a $D1$-brane, while generic $(p,q)$ Wilson-'t Hooft-line defects are dual to $(p,q)$-strings in the bulk \cite{Schwarz:1995dk}.

It is important to emphasise that the fundamental ’t Hooft-line defect should really be labelled by a weight vector corresponding to the fundamental representation of the Langland magnetic dual group. To properly understand defect operators in $\mathcal{N}=4$ SYM, one must keep track of how the global form of the gauge group as well as the associated discrete theta angles transform under the electromagnetic ${\rm SL}(2,\mathbb{Z})$ action \cite{Aharony:2013hda}. In particular, under $S$-duality the $SU(N)$ Wilson loop is mapped to the 't Hooft loop of the Langland dual gauge group $(PSU(N))_0$, where the subscript denotes the value of the discrete $\mathbb{Z}_N$ theta angle for the $PSU(N)$ gauge theory.
In the present work we consider only the case where the $4$-dimensional space-time is either $\mathbb{R}^4$ or $S^4$ for which these subtleties are not important. We will then study the relations imposed by electromagnetic duality on correlation functions of line defects
operators defined in the same $SU(N)$  gauge theory.

The physical observables we here consider are correlation functions of local operators in the presence of a half-BPS line defect in the fundamental representation of $SU(N)$ with electromagnetic charges $(p,q)$ with $p$ and $q$ coprime, schematically denoted by $\mathbb{L}_{(p,q)}$. To be precise, the main character of our story is the integrated line defect correlator in $\mathcal{N}=4$ SYM with $SU(N)$ gauge group first introduced in \cite{Pufu:2023vwo},\footnote{We also consider simpler observables such as $ \langle\,   \mathbb{L}_{(p,q)}   \,  \rangle$, i.e. the expectation value of the line defect without any insertions of local operators, as well as the one and two-point functions  $ \langle\,     \mathcal{O}_2(x_1) \,\mathbb{L}_{(p,q)} \rangle $ and $ \langle\,    \mathcal{O}_2(x_1) \mathcal{O}_2(x_2) \, \mathbb{L}_{(p,q)} \rangle$ with special choices for the space-time insertion points and for the R-charge polarisation vectors of the local operators and the line defect. These simpler line defect correlation functions can all be computed from $ \langle\,   \mathbb{L}_{(p,q)}   \,  \rangle$ by acting with certain covariant derivatives with respect to $\tau$. These observables can all be expressed in terms of a special simpler subclass of the more general automorphic forms we introduce in this work. } 
\begin{equation}
\begin{aligned} \label{eq:DD}
 \mathcal{I}_{\mathbb{L}, N} (p,q; \tau) &=   \int   \langle\,     \mathcal{O}_2(x_1) \mathcal{O}_2(x_2) \,\mathbb{L}_{(p,q)}  \rangle\, {\rm d}\mu(x_i) \, ,
\end{aligned}
\end{equation}
where $\mathcal{O}_2$ is the dimension-two half-BPS superconformal primary operator in the stress tensor multiplet. 
The Wilson-line defect, with charges $(1,0)$, will be denoted as $\mathbb{W} = \mathbb{L}_{(1,0)}$  while $\mathbb{T} = \mathbb{L}_{(0,1)}$ refers to the 't Hooft-line defect, with charges $(0,1)$.

Similar to the integrated correlators of four local operators $\mathcal{O}_2$ introduced in \cite{Binder:2019jwn, Chester:2020dja}, here  as well in \eqref{eq:DD} we integrate out the non-trivial spacetime dependence of the insertion points of the two local half-BPS operators $\mathcal{O}_2$. 
The explicit form of the integration measure ${\rm d}\mu(x_i)$ in \eqref{eq:DD} as well as the precise form for the correlator can be found in \cite{Billo:2024kri, Dempsey:2024vkf}, importantly we stress that this measure is dictated entirely by supersymmetry.
More precisely, the half-BPS Wilson-line defect integrated correlator is derived from the well-known matrix model formulation for the partition function of $\mathcal{N}=2^*$ SYM, which is a mass deformation of the superconformal $\mathcal{N}= 4$ SYM theory with mass parameter $m$. 
The expectation value of the half-BPS fundamental Wilson loop in $SU(N)$  $\mathcal{N}=2^*$ SYM on $S^4$, denoted by $\langle \mathbb{W} \rangle_{\mathcal{N}=2^*}$,    
was determined by Pestun using supersymmetric localisation \cite{Pestun:2007rz}.
As shown in \cite{Pufu:2023vwo}, the Wilson-line defect integrated correlator
$\mathcal{I}_{\mathbb{W}, N} ( \tau)$ introduced in \eqref{eq:DD} is then related to the  expectation value of the Wilson-line defect in $\mathcal{N}=2^*$ SYM as follows,
\begin{equation}\label{eq:IntroW}
\begin{aligned}
\mathcal{I}_{\mathbb{W}, N} ( \tau)= \Big[\partial_m^2 \log   \langle\,   \mathbb{W}   \,  \rangle_{\mathcal{N}=2^*}(m,\tau)\Big]_{m=0}\, . 
\end{aligned}
\end{equation}
It is important to stress that even though we have not made it manifest, the quantity $\mathcal{I}_{\mathbb{W}, N} ( \tau)$ is a non-holomorphic function of the coupling constant $\tau$. The main goal of this paper is exploiting supersymmetric localisation combined with the electromagnetic duality properties of $\mathcal{N}=4$ SYM to understand which class of ${\rm SL}(2,\mathbb{Z})$ automorphic forms can possibly describe the line-defect integrated correlators  $\mathcal{I}_{\mathbb{L}, N} (p,q; \tau)$.

The transformation properties of $\mathcal{I}_{\mathbb{L}, N} (p,q; \tau)$ under $\mathcal{N}=4$ SYM $S$-duality are determined by the modular properties of the corresponding line defect $\mathbb{L}_{(p,q)}$ and of the local operator $\mathcal{O}_2$. While the local operator $\mathcal{O}_2$ is invariant under  ${\rm SL}(2, \mathbb{Z})$, the line defect  $\mathbb{L}_{(p,q)}$ transforms non-trivially. In particular the integers $(p,q)$, which denote the electromagnetic charges of the defect, do transform under the action of the $\SL(2,\mathbb{Z})$ electromagnetic duality group, as we have commented. For the theory with coupling constant $\tau'=\gamma \cdot \tau$ given in \eqref{eq:GNOtau}, the line defect $\mathbb{L}_{(p,q)}$ is mapped into a defect with charges $(p',q')$ given by
 \begin{align}
\mathbb{L}_{(p,q)} \rightarrow  \mathbb{L}_{(p',q')} \, , \quad {\rm with} \quad  (p',q') = (p,q) \left(\begin{matrix} a& -c \\ -b & d \end{matrix}\right) \, . \label{eq:GNOcharges}
 \end{align}
 This implies that correlation functions in the presence of a line defect operator must obey the following transformation properties, 
 \begin{equation}
\begin{aligned} \label{eq:autoformIntro}
 \mathcal{I}_{\mathbb{L}, N} (p,q;  \tau) =  \mathcal{I}_{\mathbb{L}, N} (p',q';  \tau')\, ,
\end{aligned}
 \end{equation}
 valid for all $\gamma\in {\rm SL}(2,\mathbb{Z})$ where the coupling constant $\tau' = \gamma\cdot \tau$ and the charges $(p',q')$ have been transformed accordingly to \eqref{eq:GNOtau} and \eqref{eq:GNOcharges}.

The key result of our analysis is that by exploiting the above relation \eqref{eq:autoformIntro} and the explicit matrix model computation from supersymmetric localisation, we conjecture the lattice sum integral representation of $ \mathcal{I}_{\mathbb{L},N}(p,q;\tau)$ valid for any $N$ and any defect-charges  $(p,q)$,
\begin{align}
&\label{eq:LatticeSumRes2} \mathcal{I}_{\mathbb{L},N}(p,q;\tau)= \\
 &\notag \frac{N}{L^1_{N-1}( -\frac{\pi | q\tau+p|^2}{\tau_2})}   \sum_{(n,m)\in \mathbb{Z}^2}  \int_0^\infty e^{-t_1 \frac{\tau_2}{\pi |q\tau+p|^2} }  e^{- t_2 \pi \frac{|n\tau+m|^2}{\tau_2}}  e^{-t_3 \pi\frac{ \tau_2 }{|q\tau+p|^2}   (np-mq)^2} \mathcal{B}_N(t_1, t_2,t_3) \,{\rm d}^3t\,,
\end{align}
where $\mathcal{B}_N(t_1, t_2,t_3)$ is a function of the three real variables $t_1, t_2, t_3$ and of the number of colours $N$. The overall factor in \eqref{eq:LatticeSumRes2} given in terms of a Laguerre polynomial, $L^1_{N-1}(x)$, arises from the normalisation for the integrated correlator~\eqref{eq:DD} where we divide by the vacuum expectation value of the line defect operator.

We will show that the lattice sum integral representation \eqref{eq:LatticeSumRes2} for $ \mathcal{I}_{\mathbb{L},N}(p,q;\tau) $ satisfies precisely the correct automorphic property \eqref{eq:autoformIntro} under ${\rm SL}(2,\mathbb{Z})$ electromagnetic duality. The function $\mathcal{B}_N(t_1, t_2,t_3)$,  whose general properties will be discussed in section \ref{sec:FiniteN}, holds all the dynamic information of the integrated correlators in the presence of the line defect. 
In support of the conjectural expression \eqref{eq:LatticeSumRes2}, we first outline the general construction for $\mathcal{B}_N(t_1, t_2,t_3)$ with generic $N$, and then present a complete analysis for the cases of $SU(2)$ and $SU(3)$ gauge groups, for which we derive the corresponding functions $\mathcal{B}_N(t_1, t_2,t_3)$ in \eqref{eq:B2} and \eqref{eq:B3}. 
Furthermore, we verify the consistency of our proposal with the large-$N$ fixed-$\tau$ expansion of the line defect integrated correlator which was the subject of \cite{Pufu:2023vwo}.

\subsection*{Outline}

The rest of the paper is organised as follows. 
In section \ref{sec:EasyCorr}, we introduce various integrated and non-integrated correlation functions of local operators in the presence of a line defect and discuss how these physical quantities can be computed from a matrix model derived from supersymmetric localisation and their modularity properties under electromagnetic duality. 

Section \ref{sec:LargeN} is devoted to the analysis of the large-$N$ fixed-$\tau$ expansion of the integrated Wilson-line defect correlator \eqref{eq:IntroW}. We show that each order in the $1/N$ expansion receives contributions from an infinite number of instantons. Combining these non-perturbative results with the expected transformation properties of line defect correlators under ${\rm SL}(2, \mathbb{Z})$, we introduce a novel class of real-analytic automorphic forms and show that the expansion coefficients of the large-$N$ expansion at fixed $\tau$ can be expressed as finite rational linear combinations of such automorphic functions. 

In section \ref{sec:FiniteN}, we propose an extremely simple lattice sum integral representation for the integrated line defect correlators \eqref{eq:DD} for generic values of $N$ and arbitrary $\tau$. Interestingly, we can rewrite this lattice sum integral representation as a formal series over the same class of automorphic functions appearing in the large-$N$ fixed-$\tau$ expansion. We show the validity of our conjecture by explicitly determining the lattice sum integral representation for the integrated line defect correlators \eqref{eq:DD} with gauge group $SU(2)$ and $SU(3)$ starting from their corresponding matrix model formulations. 

We conclude in section \ref{sec:conclusion} where we also comment on future research directions. The paper also includes three appendices. In appendix \ref{sec:AppMM} we provide details on the perturbative and non-perturbative contributions to correlators involving a Wilson-line defect from their matrix model formulations. The novel class of automorphic functions relevant for describing integrated line defect correlators is introduced in appendix \ref{sec:AppF}, where we also discuss their key properties such as modular transformations, Fourier modes decomposition and asymptotic expansions. In appendix \ref{sec:AppendixDecomp}, we provide more technical details for the cases of the integrated line defect correlators with gauge group $SU(2)$ and $SU(3)$ expressed as formal infinite series over the new class of automorphic functions.

\section{Line Defect Correlators from Matrix Model Integrals}
\label{sec:EasyCorr}

In this section we briefly review how certain correlation functions of half-BPS local operators in the presence of line defects can be computed in $\mathcal{N}=4$ SYM with gauge group $SU(N)$ starting from an $(N-1)$-dimensional matrix model. We will furthermore highlight how such correlation functions transform under electromagnetic duality.

\subsection{Matrix model setup}
\label{sec:MM}

Our analysis of $\mathcal{N}=4$ SYM line defect correlators, in particular half-BPS Wilson-line defect correlators, is based on the well-known matrix model formulation for the partition function of a different supersymmetric theory known as $\mathcal{N}=2^*$ SYM, which is a mass deformation of the superconformal $\mathcal{N}= 4$ SYM theory with mass parameter $m$. 
The partition function of $SU(N)$  $\mathcal{N}=2^*$ SYM on $S^4$, denoted by $Z_{N}(m, \tau)$,    
was determined by Pestun using supersymmetric localisation \cite{Pestun:2007rz}, where it was shown to have the form 
\begin{equation}
\label{Zdef}
  Z_{N}(m, \tau) =  \int   e^{- 2\pi \tau_2 \sum_i a_i^2} \,\vdm(a_i)\, {\hat{Z}}_{N}^{{\rm  pert}}(m, a_{ij})\,  \vert {\hat{Z}}_{N}^{{\rm inst}}(m, \tau, a_{ij})\vert^2\,{\rm d}^{N-1} a \,.
 \end{equation}
The integration is over $N$ real variables $a_i$, with $i = 1, \ldots, N$ parametrising the Cartan subalgebra of the gauge group.  For $SU(N)$ the $a_i$ are  subject to the constraint $\sum_i a_i = 0$, whereas in the case of $U(N)$ the $a_i$ are free variables without this constraint, furthermore we define $a_{ij}\coloneqq a_i - a_j$. 

We denote the square of the Vandermonde determinant as 
\begin{equation}\label{eq:vdm}
\vdm(a_i) \coloneqq  \prod_{i < j}|a_i-a_j|^2 \,,
\end{equation}
while the perturbative factor in \eqref{Zdef} is given by 
  \begin{equation}
 \label{Zpertdef}
  {\hat{Z}}_{N}^{{\rm pert}}(m, a_{ij}) \coloneqq H(m)\prod_{i,j} \frac{   H(a_{ij})}{ H(a_{ij}+ m)} \,,
  \end{equation}
 where the function $H(z)$ is given by $H(z)\coloneqq e^{-(1+\gamma)z^2}\, G(1+iz)\, G(1-iz)$ with $G(z)$ is a Barnes G-function, and $\gamma$ is the Euler-Mascheroni constant. 
 The factors of  $|{\hat Z}_{N}^{\rm inst}|^2 = {\hat Z}_{N}^{{\rm inst}} \,\overline{{{\hat Z}}_{N}^{{\rm inst}}}$ are the Nekrasov partition function \cite{Nekrasov:2002qd} describing the contributions from instantons and anti-instantons localised at the poles of $S^4$; we will discuss the explicit form of Nekrasov partition function later, especially in the small-$m$ expansion which is relevant for the integrated correlators. 
 We notice for future reference that when $m=0$ which is exactly the case in which the deformed $\mathcal{N}=2^*$ theory reduces  back to $\mathcal{N}=4$ SYM, we have:
 \begin{equation}
 {\hat{Z}}_{N}^{{\rm pert}}(m{=}0, a_{ij}) = {\hat{Z}}_{N}^{{\rm inst}}(m{=}0, \tau, a_{ij}) = 1\,.\label{eq:Zmzero}
 \end{equation}

 In what follows, we will denote expectation values in the hermitian matrix model using the double parenthesis notation $\llangle \mathcal{O}(a_k)\rrangle$ defined as 
 \begin{equation}\label{eq:MM}
 \llangle \mathcal{O}(a_k)\rrangle \coloneqq \frac{1}{  Z_{N}( \tau_2)} \int   e^{- 2\pi \tau_2 \sum_i a_i^2} \,\vdm(a_i)\, \mathcal{O}(a_k)\,{\rm d}^{N-1} a \,.
 \end{equation}
 The normalisation factor $Z_N(\tau_2) $ is simply given by
\begin{equation}
Z_N(\tau_2) \coloneqq Z_N(m{=}0,\tau)  =  \int   e^{- 2\pi \tau_2 \sum_i a_i^2} \,\vdm(a_i)\,{\rm d}^{N-1} a  \,,\label{eq:Norm}
\end{equation}
and it is chosen so that $\llangle 1 \rrangle = 1$.
The $\mathcal{N}=2^*$ SYM partition function on $S^4$ presented in \eqref{Zdef} can then be written as
 \begin{equation}
 \label{eq:ZN2s}
 Z_N(m,\tau) =Z_N(\tau_2)\, \llangle{\hat{Z}}_{N}^{{\rm  pert}}(m, a_{ij})\,  \vert {\hat{Z}}_{N}^{{\rm inst}}(m, \tau, a_{ij})\vert^2 \rrangle\,.
 \end{equation}

More details on the one-loop determinant factor $\hat{Z}^{\rm pert}_N$, the instanton partition function $\hat{Z}^{\rm inst}_N$ and computations of such matrix model integrals are presented in appendix \ref{sec:AppMM}, while in the next section we focus on what type of defect correlators can be computed starting from this matrix model formulation.

\subsection{Expectation value of line defect operators}

In \cite{Pestun:2007rz} it was shown that it is possible to compute the expectation value of a half-BPS supersymmetric Wilson-line defect wrapping the equator of the space-time $S^4$ directly from the hermitian matrix model \eqref{Zdef}.
In particular, the expectation value of a fundamental half-BPS Wilson-line defect in $\mathcal{N}=2^*$ SYM can be computed via the matrix model \eqref{eq:MM} and it is given by 
\begin{align}
\langle \mathbb{W} \rangle_{\mathcal{N}=2^*}(m,\tau) & = \frac{\llangle {\hat{Z}}_{N}^{{\rm  pert}}(m, a_{ij})\,  \vert {\hat{Z}}_{N}^{{\rm inst}}(m, \tau, a_{ij})\vert^2 \Big(\frac{1}{N}\sum_{i=1}^N e^{2\pi a_i} \Big) \rrangle}{\llangle {\hat{Z}}_{N}^{{\rm  pert}}(m, a_{ij})\,  \vert {\hat{Z}}_{N}^{{\rm inst}}(m, \tau, a_{ij})\vert^2  \rrangle}\,.\label{eq:WilsonVEVN2}
\end{align}
Setting $m=0$ and using \eqref{eq:Zmzero} we derive the expectation value of a fundamental half-BPS Wilson-line defect in $\mathcal{N}=4$ SYM,
\begin{align}
\langle \mathbb{W} \rangle(\tau) & =\langle \mathbb{W} \rangle_{\mathcal{N}=2^*}(m{=}0,\tau) = \llangle \frac{1}{N}\Big(\sum_{i=1}^N e^{2\pi a_i} \Big) \rrangle\,.\label{eq:WilsonVEV}
\end{align}

From now on we focus our attention solely to the case where the gauge group is $SU(N)$, although a similar analysis can be carried out for $U(N)$, furthermore since we are setting the deformation mass parameter $m\to 0$, the matrix model simplifies dramatically due to \eqref{eq:Zmzero}. More concretely, using \eqref{eq:MM}, the fundamental Wilson-line defect expectation value can be written in the following form
\begin{align}
\langle \mathbb{W} \rangle(\tau) 
&\label{eq:W1} =\frac{1}{Z_N(\tau_2)} \int  \vdm(a) \exp\big[-2\pi \tau_2 \,\mbox{Tr}(a^2) -2\pi\, \mbox{Tr}(B\,a)\big]\,\delta(\mbox{Tr} \,a)\,{\rm d}^N a\,,
\end{align}
where we use the matrix notation $a\coloneqq \mbox{diag}(a_1,...,a_N)$ and introduce  the matrix $B$
\begin{equation}
B \coloneqq \frac{1}{N} \mbox{diag}(1,...,1,1-N) = \frac{1}{N}\mathbbm{1} - \mbox{diag}(0,...,0,-1) \,.\label{eq:MagField}
\end{equation}
Shortly the matrix $B$ will play the role of magnetic field in the case of the fundamental 't Hooft-line defect.
We complete the square at exponent in \eqref{eq:W1} and eventually arrive at the well-known result \cite{Drukker:2000rr},
\begin{align}
\langle \mathbb{W} \rangle(\tau) 
& \label{eq:WilsonInt} =  \exp\Big[ \frac{\pi}{2\tau_2}\frac{N-1}{N} \Big] \frac{1}{Z_N(\tau_2)} \int  \vdm \Big(a - \frac{B}{2\tau_2}\Big) \exp\big[-2\pi \tau_2 \,\mbox{Tr} a^2\big]\, \delta(\mbox{Tr} \,a)\,{\rm d}^N a \,,\\
& = \frac{1}{N}\exp\Big[ \frac{\pi}{2\tau_2}\frac{N-1}{N} \Big]  L^1_{N-1}\left( -\frac{\pi}{\tau_2}\right) \,, \label{eq:WilsonLag}
\end{align}
with $L^1_{N}(x)$ denoting the generalised Laguerre polynomials.
We note that given the form \eqref{eq:MagField} of the matrix $B$, the shifted Vandermonde determinant appearing in \eqref{eq:WilsonInt} simply amounts to
\begin{equation}
\vdm\Big(a - \frac{B}{2\tau_2}\Big)  =  \prod_{i<j} \Big( a_i-a_j - \frac{B_i - B_j}{2\tau_2} \Big)^2  =  \prod_{i<j<N} (a_i-a_j )^2 \, \prod_{i=1}^{N-1} \Big(a_i - a_N -\frac{1}{2\tau_2}\Big)^2\,.\label{eq:vdmW}
\end{equation}


We now turn our attention to the expectation value of a half-BPS fundamental 't Hooft-line defect in $\mathcal{N}=4$ SYM.
As a consequence of modularity \cite{Kapustin:2005py}, we know that under the $S$ duality transformation $\tau\to -1/\tau$, a half-BPS 't Hooft-line defect in the representation $\,\!^LR$ of $\,\!^LG$ is mapped into a half-BPS Wilson-line defect in the representation $\,\!^LR$ of $G$.
Starting from a path integral definition of an ’t Hooft-line defect operator, \cite{Gomis:2009ir}  computed to leading order in perturbation theory the expectation value of a circular half-BPS ’t Hooft defect in $\mathcal{N}=4$ SYM with arbitrary gauge group $G$ and showed that it agrees with the leading order of the $S$-dual Wilson-line defect.

Furthermore using supersymmetric localisation arguments, \cite{Gomis:2011pf} proposed a matrix model formulation for computing the expectation value of a half-BPS 't Hooft-line defect in a general $\mathcal{N}=2$ SYM theory.
For the particular case of $\mathcal{N}=2^*$ SYM with gauge group $SU(2)$, \cite{Gomis:2011pf} checked numerically that this matrix model agrees for different values of coupling $\tau$, mass deformation parameter $m$, and 't Hooft-line defect charge $p$,
with the respective expectation value of the $S$-dual Wilson-line defects.

We now show that using the results of \cite{Gomis:2011pf}  it is possible to prove that the expectation value of the fundamental 't Hooft-line defect in $\mathcal{N}=4$ SYM is indeed equal to the $S$-dual of the fundamental Wilson-line defect \eqref{eq:WilsonLag}.
If we consider a half-BPS 't Hooft-line defect in $\mathcal{N}=2^*$ SYM positioned on the equator of $S^4$ carrying a magnetic charge labeled by a coweight\footnote{Coweights are elements of the Cartan subalgebra $\mathfrak{h}$ of the gauge group $G$ such that $\alpha\cdot B$ is an integer for all roots $\alpha\in \mathfrak{h}^*$.  } $B$ of the gauge group $G$, then its expectation value is given by the matrix model  \cite{Gomis:2011pf},
\begin{equation}
\langle \mathbb{T} \rangle_{\mathcal{N}=2^*} (m,\tau) =\sum_{v} \int \vert \,Z^{\rm north}(\tau, m , a - \frac{i}{2}v)|^2 \,Z^{\rm equator}(m,a,v,B) \,{\rm d}^r a\,,\label{eq:Tvev}
\end{equation}
where as in \eqref{Zdef} the integral runs over the Cartan subalgebra $a\in \mathfrak{h}$ of $G$. 
Since the magnetic charge $B$ is a coweight it can be thought of as the highest weight for a representation $\!\,^LR$ of the GNO magnetic dual group $\!\,^LG$, hence the sum appearing in \eqref{eq:Tvev} is over coweights $v$ of $G$ such that their corresponding weights of $\!\,^LG$ appear in the representation defined by $B$.

The analysis of \cite{Gomis:2011pf} showed that the localised path integral receives two types of contributions. The first is schematically denoted as $Z^{\rm north}(\tau, m , a - \frac{i}{2}v)$ in \eqref{eq:Tvev}, and captures field configurations localised only at the north and south pole (from the complex conjugate term) of $S^4$.
These contributions are directly related to the perturbative one-loop determinant and instanton partition function previously presented in \eqref{Zdef}.
In particular we note that when $m= 0$ the $\mathcal{N}=2^*$ SYM theory reduces back to $\mathcal{N}=4$ for which we have
\begin{equation}
\vert Z^{\rm north}(\tau, m=0 , a - \frac{i}{2}v)|^2 =  \vdm( a - \frac{i}{2}v)\,|Z^{{\rm cl}}(\tau,a - \frac{i}{2}v)|^2\,, \label{eq:ZNorth}
\end{equation}
where the Vandermonde determinant is given in \eqref{eq:vdm} and the classical action $Z^{\rm cl}$ is defined as
\begin{equation}
Z^{\rm {cl}}(\tau,a) = \exp\Big[i\pi\tau\, \mbox{Tr}(a^2) \Big]\,.\label{eq:ZclB}
\end{equation}

The second term appearing in \eqref{eq:Tvev}, denoted with $Z^{\rm equator}$, contains all contributions to the path-integral which have support precisely on the equator of the $S^4$ where the 't Hooft-line defect has been inserted. This equatorial factor can be furthermore split into two different sources.
Firstly, in defining the path-integral with an 't Hooft-line defect insertion one must impose particular boundary conditions along the defect which produce further contributions to the one-loop determinant factor \eqref{Zpertdef}. Secondly, by inserting along the equator an 't Hooft-line defect we source a singular magnetic monopole with charge $B$. The sum over ``smaller'' coweights $v$ in \eqref{eq:Tvev} is due to monopole bubbling effects where we need to include in the path-integral contributions arising from smooth monopoles which screen the magnetic charge $B$ down to $v$. While \cite{Gomis:2011pf} computed exactly the equatorial perturbative contribution to $Z^{\rm equator}$ for a general $\mathcal{N}=2$ theory with arbitrary gauge group, the monopole bubbling effects are more difficult to compute (and beyond the scope of this paper) and in  \cite{Gomis:2011pf} an exact expression for the complete $Z^{\rm equator}$ is provided only in the case of $\mathcal{N}=2^*$ SYM with gauge group $G=SU(2)$.

Importantly, in this case we find that the equatorial contribution reduces to $1$ when $m= 0 $ and the theory becomes $\mathcal{N}=4$ SYM, i.e.
\begin{equation}
Z^{\rm equator}_{SU(2)}(m=0,a,q,p)=1\,,\label{eq:ZeqSU2}
\end{equation}
where in $SU(2)$ the integers $p,q$ denote the coweights $B$ and $v$ respectively, with $p-q$ an even integer.
We claim that this result is enough to show that an $S$-duality transformation of the $\mathcal{N}=4$ SYM fundamental Wilson-line defect given in equation \eqref{eq:W1} reproduces identically the 't Hooft-line defect expectation value for general gauge group $G=SU(N)$ which can be computed from \eqref{eq:Tvev}.

Since we have considered a Wilson-line defect in the fundamental representation, its $S$-dual 't Hooft-line defect must have minimal magnetic field (modulo Weyl reflections)
\begin{equation}
B = \frac{1}{N}\mathbbm{1} - \mbox{diag}(0,0,...,0,1) = \frac{1}{N} \mbox{diag}(1,...,1,1-N) \,.\label{eq:MagField2}
\end{equation}
Given that this magnetic field is effectively selecting an $SU(2)$ sector within the full gauge group $G=SU(N)$, similar to \eqref{eq:ZeqSU2} it is therefore natural to argue that in $SU(N)$ as well
the equatorial contribution for a minimal magnetic field trivialises in $\mathcal{N}=4$ SYM, i.e. we conjecture
\begin{equation}
Z^{\rm equator}_{SU(N)}(m=0,a,B,B)=1\,.\label{eq:ZeqSUN}
\end{equation}
Under this assumption, the expectation value for the minimal fundamental half-BPS 't Hooft-line defect in $\mathcal{N}=4$ SYM is obtained from \eqref{eq:Tvev} by setting the mass deformation parameter $m$ to zero and making use of \eqref{eq:ZNorth}-\eqref{eq:ZeqSUN}, thus taking the simpler form
\begin{equation}
\langle \mathbb{T} \rangle (\tau) =\langle \mathbb{T} \rangle_{\mathcal{N}=2^*} (m{=}0,\tau)  =\frac{1}{Z_N(\tau_2)} \int  \vdm\left(a - \frac{i}{2}B\right) |Z^{\rm {cl}}(\tau,a- \frac{i}{2} B)|^2  \,\delta(\mbox{Tr}\,a)\,{\rm d}^{N} a\,,\label{eq:Tvev2}
\end{equation}
where we introduced back the matrix model normalisation factor $Z_N(0,\tau_2)$ given in \eqref{eq:Norm}.

To show that \eqref{eq:Tvev2} is the $S$-dual of \eqref{eq:WilsonLag} we start by rewriting the classical action as a quadratic form
\begin{equation}
|Z^{\rm cl}(\tau,a- \frac{i}{2} B)|^2 = \exp\big[ 4 \pi  \frac{ \tau \bar{\tau}}{(\tau-\bar{\tau})} \mbox{Tr}(B/2)^2 \big]  \exp\Big[ \pi i (\tau-\bar\tau) \mbox{Tr}\Big( a + \frac{\tau+\bar\tau}{i(\tau-\bar\tau)} \frac{B}{2}\Big)^2 \Big]\,.\label{eq:Zcl}
\end{equation}
We then use \eqref{eq:MagField} to compute the overall $a$-independent term which is given by
\begin{equation}
4 \pi  \frac{ \tau \bar{\tau}}{(\tau-\bar{\tau})} \mbox{Tr}(B/2)^2  = \frac{\pi}{2 \tau_2|_S} \frac{N-1}{N}\,,
\end{equation}
where $\tau_2\vert_S = \mbox{Im}\big(S\cdot \tau\big)= \tau_2/|\tau|^2$ with $S= \left(\begin{smallmatrix} 0 & -1 \\ 1 & 0 \end{smallmatrix}\right)\in {\rm SL}(2,\mathbb{Z})$ is the $S$ transformation of $\tau_2 = \mbox{Im}\tau$.
This factor is exactly the $S$ transformation of the exponential factor present in the Wilson-line defect \eqref{eq:WilsonInt}.

With these observations we simplify \eqref{eq:Tvev2} to
\begin{equation}
\langle \mathbb{T} \rangle (\tau) =  \exp\Big(\frac{\pi}{2 \tau_2|_S} \frac{N-1}{N}\Big) \frac{1}{Z_N(\tau_2)}  \int  \vdm(a -i \frac{B}{2})   \exp\Big[ -2 \pi \tau_2 \mbox{Tr}\Big( a - \frac{\tau_1}{2\tau_2} B\Big)^2 \Big]\, \delta(\mbox{Tr} \,a)\, {\rm d}^N a\,. \label{eq:Thooft1}
\end{equation}
Since the overall exponential factor is already expressed as the $S$-dual of the corresponding factor in the Wilson-line defect, we simply need to check that the remaining matrix model integral produces the $S$-dual transformation of the Laguerre polynomial term \eqref{eq:WilsonLag}.
For simplicity we perform an $S$-duality transformation to \eqref{eq:Thooft1} and consider 
\begin{equation}
\langle \mathbb{T} \rangle \left(-\frac{1}{\tau}\right) =  \exp\Big(\frac{\pi}{2 \tau_2 } \frac{N-1}{N}\Big)\frac{1}{Z_N( \frac{\tau_2}{|\tau|^2})}  \int \vdm(a -i \frac{B}{2})   \exp\Big[ -2 \pi \tau_2 \,\mbox{Tr}\Big( \frac{a}{|\tau|} + \frac{\tau_1}{2\tau_2 |\tau|} B\Big)^2 \Big] \, \delta(\mbox{Tr} \,a) \,{\rm d}^N a\,. \label{eq:Thooft2}
\end{equation}
We then perform the change of variables $a =  \tilde{a} |\tau| - \frac{\tau_1}{2\tau_2 } B$ and notice that after the inversion $\tau_2\to \frac{\tau_2}{|\tau|^2}$ the normalisation factor \eqref{eq:Norm} becomes  $Z_N( \frac{\tau_2}{|\tau|^2}) = |\tau|^{N^2-1} Z_N(\tau_2)$  thus cancelling the overall scale $|\tau|^{N^2-1}$ coming from the change of variables at numerator (and the contribution from the delta function), thus leaving us with
\begin{equation}
\langle \mathbb{T} \rangle\left(-\frac{1}{\tau}\right) =  \exp\Big(\frac{\pi}{2 \tau_2 } \frac{N-1}{N}\Big) \frac{1}{Z_N(\tau_2)} \int \vdm(\tilde{a} - \frac{\tau}{2 \tau_2 |\tau|} B)   \exp\Big[ -2 \pi \tau_2\, \mbox{Tr}\,\tilde{a}^2 \Big] \delta(\mbox{Tr} \,\tilde{a})\,{\rm d}^N \tilde{a} \,. \label{eq:Thooft3}
\end{equation}

We now focus on rewriting the Vandermonde determinant appearing in \eqref{eq:Thooft3}:
\begin{align}
\vdm\big(\tilde{a} - \frac{\tau}{2 \tau_2 |\tau| }B\big) 
 \notag =\!  \prod_{i<j<N} (\tilde{a}_i-\tilde{a}_j)^2 \prod_{i=1}^{N-1}  \Big[(\tilde{a}_i-\tilde{a}_N- \frac{1}{2\tau_2})^2+ \eta \,  (\tilde{a}_i -\tilde{a}_N)\Big]\,,
\end{align}
where for convenience we introduced the combination
\begin{equation}
\eta \coloneqq \frac{|\tau|-\tau_1}{\tau_2|\tau|}\,.
\end{equation}
We find that the right hand side  of \eqref{eq:Thooft3} is independent of $\eta$ due to the following matrix model identity, 
\begin{align}
    \int d^N a\, e^{-\sum_{i=1}^{N} a_i^2} \left( \prod_{i<j< N} a_{ij}^2 \right)   \prod_{i=\ell}^{N-1} a_{iN}^2 \, \prod_{j=1}^k a_{iN}  =0 \, ,
\end{align}
valid for integers $\ell$ and $k$ such that $1\leq k < \ell$. This identity is non-trivial only when $k$ is even and we have verified its validity for all values of $\ell,k$ and $N\leq 7$, although we do not have a proof for it. 

With this observation we can then rewrite the $S$-dual 't Hooft-line defect \eqref{eq:Thooft3} matrix model integral exactly as the Wilson-line defect \eqref{eq:WilsonInt}, i.e.
\begin{align}
\langle \mathbb{T} \rangle\left(-\frac{1}{\tau}\right) &\notag = \exp\Big(\frac{\pi}{2 \tau_2 } \frac{N-1}{N}\Big)   \frac{1}{Z_N(\tau_2)} \int \vdm(\tilde{a} - \frac{1}{2\tau_2}B)   \exp\Big[ -2 \pi \tau_2\, \mbox{Tr}\,\tilde{a}^2 \Big] \, \delta(\mbox{Tr} \,\tilde{a})\,{\rm d}^N \tilde{a} \\
&= \langle \mathbb{W} \rangle(\tau)\,.\label{eq:THooftS}
\end{align}

We now consider correlation functions of local operators in the presence of line defects, for which electromagnetic duality plays a more prominent role.

\subsection{Non-integrated line defect correlators}

An interesting class of correlation functions with a half-BPS Wilson-line defect insertion can be constructed directly from the expectation value of the Wilson-line defect \eqref{eq:WilsonVEV}.
We define two such defect correlators as follows:
\begin{align}
\mathcal{C}_\mathbb{W}(\tau) &\coloneqq \label{eq:Cw}2i\tau_2 \frac{\partial}{\partial \tau } \log \langle \mathbb{W} \rangle (\tau)\,,\\
\mathcal{E}_\mathbb{W}(\tau) &\coloneqq  \label{eq:Ew} \Delta_\tau \log \langle  \mathbb{W} \rangle (\tau) \,,\phantom{\frac{\partial}{\partial\tau}}
\end{align}
with $\Delta_\tau \coloneqq 4\tau_2^2\partial_\tau \partial_{\bar\tau}$ the $\SL(2,\mathbb{R})$-invariant Laplacian. Note that we suppress the explicit dependence of these correlation functions from the number of colours, $N$, since it can be easily retrieved from \eqref{eq:WilsonLag} and it will not play any role for the following discussion.
From a path-integral perspective one would naively think that \eqref{eq:Cw}-\eqref{eq:Ew} naturally yield defect correlation functions with the insertion of respectively one and two $S^4$ integrals over the insertion points of some local operators. However, due to supersymmetry, these particular combinations of derivatives of the $\mathcal{N}=4$ half-BPS Wilson-line defect expectation value, correspond to \textit{non-integrated} one and two-point line defect correlators, with operators inserted in a specific way, corresponding to certain well-known topological correlation functions of twisted local operators and
defects \cite{Giombi:2009ds,Drukker:2007yx,Pestun:2009nn,Giombi:2009ek,Wang:2020seq}.

We begin with \eqref{eq:Cw}. It is know that the one-point function of a superconformal primary operator in the presence of a half-BPS defect determines (modulo a proportionality  constant) all defect one-point functions for the conformal primaries in the same superconformal multiplet \cite{Liendo:2016ymz}.  
In particular, the line defect one-point function of every operator in the stress tensor multiplet is determined by that of the dimension-two superconformal primary operator $\mathcal{O}_2(x,Y)$  which is constructed as a bilinear in the $\mathcal{N}=4$ SYM scalar fields $\Phi_I$, with $I=1,...,6$, belonging to the adjoint representation of the gauge group,
\begin{equation}
\mathcal{O}_2(x,Y) \coloneqq \mbox{Tr} \Big( \Phi_I(x) \Phi_J(x)\Big) Y^I Y^J\,,
\end{equation}
with $Y^I$ a null-polarisation vector for the $SO(6)_R$ $R$-symmetry indices.

If we canonically normalise the operator $\mathcal{O}_2(x,Y)$, it was shown in \cite{Pufu:2023vwo} that
\eqref{eq:Cw} corresponds to the non-integrated one-point function\footnote{We note that line defect one-point functions similar to $\mathcal{C}_\mathbb{W}(\tau)$ have also been studied in perturbation theory for higher scaling-dimension local operators in $\mathcal{N}=2$ SYM theories \cite{Billo:2018oog}. }:
\begin{equation}
\frac{\langle \mathcal{O}_2(x,Y_1)\,\mathbb{W}(Y_2)   \,\rangle}{\langle \mathbb{W}\rangle} = \label{eq:CwCorr}\frac{2\sqrt{2}}{c} \mathcal{C}_{\mathbb{W}}(\tau) \frac{Y_{12}^2}{|x_\perp|^2}\,,
\end{equation}
with $c=(N^2-1)/4$ the central charge of the $SU(N)$ theory, while $Y_i$ denotes the two $R$-symmetry polarisation vectors and $Y_{ij} \coloneqq Y_i \cdot Y_j$.
Furthermore, since a general half-BPS superconformal defect breaks the $SO(6)_R$ $R$-symmetry down to an $SO(5)_R$ subgroup, it must be labelled by an $R$-symmetry polarisation vector of unit norm and real entries, hence we denote this particular defect by $\mathbb{W}(Y)$.

Note that while the Wilson-line defect expectation value \eqref{eq:WilsonLag} does not depend on the particular $R$-symmetry polarisation, the one-point function \eqref{eq:CwCorr} does. However, this dependence is completely fixed by superconformal invariance. 
Furthermore, the four-dimensional Euclidean
coordinate $x$ can be split as $x=(x_\perp,x^4)$. The half-BPS  Wilson-line defect lies along the direction $x^4$ while being localised at $x_\perp=0$, so that the distance $|x_\perp|^2$ appearing in \eqref{eq:CwCorr} denotes the transverse distance of the operator $\mathcal{O}_2(x,Y_1)$ from the Wilson-line defect.

The second quantity \eqref{eq:Ew} yields a particular two-point defect correlator which lies in a special two-dimensional topological sector of the full $\mathcal{N}=4$ SYM theory algebra of local and extended operators \cite{Drukker:2007yx,Pestun:2009nn,Giombi:2009ds,Giombi:2009ek,Wang:2020seq}. 
As reviewed in great detail in \cite{Pufu:2023vwo}, the quantity \eqref{eq:Ew} can be shown to reproduce a second topological and non-integrated two-point defect correlation function, namely
\begin{equation}\label{eq:Ewcorr}
\langle \mathcal{O}_2(\tilde{x}_1,Y) \mathcal{O}_2(\tilde{x}_2,\overline{Y})\,  \mathbb{W}(Y_3)\rangle_{\rm c} = \frac{}{} \frac{8\,\mathcal{E}_\mathbb{W}(\tau)}{c}\,,
\end{equation}
where the subscript ${\rm c}$ indicates that this is the connected part of the correlator, and the insertion points are fixed at $\tilde{x}_1=-\tilde{x}_2 = (1,0,0,0)$ with the Wilson-line defect being located at $x_\perp=0$ and stretching along the $x_4$ direction.
The $R$-symmetry polarisations are fixed to be $Y = (0,0,0,0,-i,1) $ and $Y_3=(0,0,0,0,0,1)$. We refer to \cite{Pufu:2023vwo} for details on the derivation.

Thanks to electromagnetic duality, in $\mathcal{N}=4$ SYM we can relate a defect correlator with a Wilson-line defect insertion to that of a general defect $\mathbb{L}$ with charges $(p,q)$.
Rather than using the vector $(p,q)$ of electromagnetic charges, we find it convenient to label the defect in terms of an equivalence class
\begin{equation}
[ \rho ]=  \left(\begin{matrix}  * &  *  \\ q & p \end{matrix}\right) \in B(\mathbb{Z})\backslash {\rm PSL}(2,\mathbb{Z})\,,
\end{equation}
where ${\rm PSL}(2,\mathbb{Z}) \coloneqq \SL(2,\mathbb{Z})/\{\mathbbm{1},-\mathbbm{1}\} $ while the Borel stabiliser of the cusp is denoted by $B(\mathbb{Z})$ and it is simply given by $B(\mathbb{Z}) \coloneqq  \{ \pm T^n\,\vert\,n\in \mathbb{Z}\}$
with $T \coloneqq \left(\begin{smallmatrix} 1 & 1 \\ 0& 1 \end{smallmatrix}\right)$.
The coset $B(\mathbb{Z})\backslash {\rm PSL}(2,\mathbb{Z})$ results precisely isomorphic to\footnote{Note that throughout this paper we will assume the magnetic charge of the defect to be positive, $q\geq 0$. Defects with negative magnetic charges can be obtained by acting with charge conjugations. However, the integrated correlators here considered are invariant under charge conjugation,~i.e. they will be real analytic functions of the coupling constant~$\tau$.} 
\begin{align}
  B(\mathbb{Z})\backslash {\rm PSL}(2,\mathbb{Z}) \simeq \{(p,q) \in \mathbb{Z}^2 \,\vert\,{\rm gcd}(p,q) =1\,,q\geq 0\} \,.
\end{align}
In the coset notation, a Wilson-line defect corresponds to the coset element $[\rho]= [\mathbbm{1}]$, while an 't Hooft-line defect correspond to $[\rho]=[S]$
where $S = \left(\begin{smallmatrix} 0 & -1 \\ 1 & 0 \end{smallmatrix}\right)$.

Under an electromagnetic duality transformation $\gamma\in {\rm SL}(2,\mathbb{Z})$, the defect $\mathbb{L}_{(p,q)}$ is then mapped into a different defect $\mathbb{L}_{(p',q')}$ whose charges are given in \eqref{eq:GNOcharges} or in coset notation
 \begin{align}
 \label{eq:GNOcoset}
[ \rho ] =\left(\begin{matrix}  * &  *  \\ q & p \end{matrix}\right)\rightarrow  [ \rho' ]=[ \rho\,\gamma^{-1} ]=\left(\begin{matrix}  * &  *  \\ q' & p' \end{matrix}\right) \, .
 \end{align}
We then rewrite the transformation property \eqref{eq:autoformIntro} for an arbitrary correlation functions, 
$\mathcal{I}_\mathbb{L}([\rho];\tau)$, of local modular invariant operators in the presence of a line-defect $\mathbb{L}$ with charges parametrised by $[\rho]$ as
 \begin{equation}
\begin{aligned} \label{eq:autoform}
 \mathcal{I}_{\mathbb{L}} ([ \rho ];  \tau) =  \mathcal{I}_{\mathbb{L}} ([  \rho \,\gamma^{-1} ];  \gamma\cdot\tau) \qquad \Leftrightarrow \qquad \mathcal{I}_{\mathbb{L}} ([ \rho ]; \gamma \cdot \tau) =  \mathcal{I}_{\mathbb{L}} ([  \rho \,\gamma ];  \tau) \, .
\end{aligned}
 \end{equation}
In particular, if we start with a Wilson-line defect insertion and act on it with an $\SL(2,\mathbb{Z})$ transformation $\rho\in \SL(2,\mathbb{Z})$ such that $[\rho] = \left(\begin{smallmatrix} * & * \\ q & p \end{smallmatrix}\right)\in B(\mathbb{Z})\backslash \SL(2,\mathbb{Z})$, we transform the line defect $\mathbb{W}$ to a different line defect $\mathbb{L}$ with charges $(p,q)$, i.e. we derive from \eqref{eq:autoform}
\begin{equation}
\mathcal{I}_\mathbb{W}(\rho\cdot\tau) = \mathcal{I}_{\mathbb{L}} ([ \mathbbm{1} ]; \rho \cdot \tau) =  \mathcal{I}_{\mathbb{L}} ([  \rho  ];  \tau)\,.\label{eq:automGen}
\end{equation}
Note that this equation implies that any correlator of modular invariant local operators, such as $\mathcal{O}_2$, in the presence of a Wilson-line defect must be invariant under the action of $T=\left(\begin{smallmatrix} 1 & 1 \\ 0 & 1 \end{smallmatrix}\right)\in\SL(2,\mathbb{Z})$, i.e. for all $n\in \mathbb{Z}$ we must have
\begin{equation}
\mathcal{I}_\mathbb{W}(\tau+n) = \mathcal{I}_{\mathbb{L}} ([ \mathbbm{1} ]; T^n \cdot \tau) =  \mathcal{I}_{\mathbb{L}} ([  T^n  ];  \tau) = \mathcal{I}_\mathbb{W}(\tau)\,.
\end{equation}

However, we need to be careful on how this $\SL(2,\mathbb{Z})$ action affects the other fields inserted in the correlation function when these are not modular invariant.
For the correlator \eqref{eq:Ew} the consequences of such an $\SL(2,\mathbb{Z})$ action are immediate and we find:
\begin{align}
\mathcal{E}_\mathbb{L}([\rho];\tau)  = \Delta_\tau \log \langle \mathbb{L}\rangle(\tau) = \mathcal{E}_\mathbb{W}(\rho \cdot \tau)\,,
\end{align}
thanks to the fact that the Laplacian $\Delta_\tau$ is invariant under the $\SL(2,\mathbb{Z})$ action, hence exactly of the form \eqref{eq:automGen}.
In particular we must have the  identity between defect correlators
\begin{align}
\langle \mathcal{O}_2(\tilde{x}_1,Y) \mathcal{O}_2(\tilde{x}_2,\overline{Y})\,  \mathbb{L}(Y_3)\rangle_{\rm c} &=\frac{8\,\mathcal{E}_\mathbb{L}([\rho];\tau)}{c}\,.
\end{align}

For the correlator \eqref{eq:Cw} the story is a little bit subtler.
If we choose $\rho \in \SL(2,\mathbb{Z})$ and then apply the corresponding electromagnetic duality transformation directly to \eqref{eq:CwCorr} we deduce:
\begin{align}
 \frac{\langle \mathcal{O}_2(x,Y_1)\,\mathbb{L}(Y_2)   \,\rangle}{\langle \mathbb{L}\rangle}\label{eq:CwLcorr2}= \frac{\langle \mathcal{O}_2(x,Y_1)\,\mathbb{W}(Y_2)   \,\rangle}{\langle \mathbb{W}\rangle}\Big\vert_\rho = \frac{2\sqrt{2}}{c} \mathcal{C}_{\mathbb{W}}(\rho\cdot \tau) \frac{Y_{12}^2}{|x_\perp|^2}\,.
\end{align}
However, we will promptly argue that as a consequence of modularity for a general $[\rho]\neq[\mathbbm{1}]$ we have 
$$ \mathcal{C}_{\mathbb{W}}( \rho \cdot \tau)  \neq  2i \tau_2\partial_\tau \log\langle \mathbb{L}\rangle(\tau)\,.
$$
To better clarify this point, we firstly note from the definition \eqref{eq:Cw} that the action of $\tau_2\partial_\tau$ on the vacuum expectation value of the half-BPS Wilson-line defect $\mathbb{W}$ should really yield an insertion of the integrated chiral Lagrangian operator $\mathcal{O}_\tau$ rather than $\mathcal{O}_2$.

The chiral and anti-chiral Lagrangian operators are super-descendent operators of $\mathcal{O}_2$ given by $\mathcal{O}_\tau = \delta^4 \mathcal{O}_2$ and $\bar{\mathcal{O}}_{\bar \tau} = \bar \delta^4 \mathcal{O}_2$ \cite{Dolan:2001tt},
 where $\delta$ (respectively $\bar\delta$) is a (anti-)chiral supersymmetry transformation.
Just like $\mathcal{O}_2$, these operators belong to the stress tensor supermultiplet but carry non-zero holomorphic/anti-holomorphic modular weights $(1,-1)$ and $(-1,1)$ respectively, so that under $\SL(2,\mathbb{Z})$ they transform via a $U(1)$ rotation usually denoted by $U(1)_Y$.
Here $U(1)_Y$ was termed as the `bonus $U(1)_Y$ symmetry' in \cite{Intriligator:1998ig}, which is the holographic image of the $U(1)$ R-symmetry in type IIB supergravity and breaks to $\mathbb{Z}_4$ when stringy corrections are turned on. 
The $\mathcal{N}=4$ SYM Lagrangian can be expressed as the sum of two complex conjugate parts 
\begin{align}
\mathcal{L}=&-   \frac{i}{2 \tau_2}  \left( \tau \mathcal{O}_\tau-  \bar \tau  \bar{\mathcal{O}}_{\bar \tau}\right)\,, 
\label{loperdef}
\end{align}
where the chiral and anti-chiral Lagrangians are defined by  
\begin{align}
 \label{otaudef} 
 \mathcal{O}_\tau = \frac{ \tau_2}{4\pi}  \,  \mbox{Tr} \left( -{1\over 2}  F_{\alpha\beta} F^{\alpha\beta} +\dots \right)\,,\qquad\qquad
\bar{\mathcal{O}}_{\bar\tau} = \frac{ \tau_2}{4\pi} \,\mbox{Tr} \left( - {1\over 2} \bar F_{{\dot \alpha}{\dot \beta}} \bar  F^{{\dot \alpha}{\dot \beta}} 
+\dots  \right)\, ,
\end{align}
with $F_{\alpha\beta} $, $\bar F_{{\dot \alpha}{\dot \beta}}$ the self-dual and anti self-dual Yang--Mills field strengths and where ``$\dots$" indicates terms involving fermions and scalar fields in the Yang--Mills supermultiplet.

Hence, from a path-integral perspective we expect that \eqref{eq:CwCorr} should be schematically written as\footnote{More precisely, given our definition \eqref{eq:Cw} we have to consider $ \int_{S^4} \mathcal{O}_{\tau}^{{\rm sphere}}(x)$, where the dimension-$4$ operator $\mathcal{O}_{\tau}^{{\rm sphere}}$ is related to the flat-space operator given in \eqref{otaudef} by $ \mathcal{O}_{\tau}^{{\rm sphere}}(x) = \Omega_{\rm sphere}(x)^4 \,\mathcal{O}_\tau(x)$, with the sphere conformal factor given by $\Omega_{\rm sphere}(x) = (1+x^2)/2$. 
However, thanks to conformal invariance we have $\langle  \big(\int_{S^4}\mathcal{O}_{\tau}^{{\rm sphere}}(x)\big)\, \mathbb{W}\rangle_{S^4} =\langle \big(\int_{\mathbb{R}^4} \mathcal{O}_{\tau}(x)\big)\, \mathbb{W}\rangle$. 
For related discussions see
\cite{Dempsey:2024vkf}.}
\begin{equation}
\mathcal{C}_\mathbb{W}(\tau) = \label{eq:Cw2}\frac{\langle \Big( \int_{\mathbb{R}^4} \mathcal{O}_\tau(x) \,{\rm d}x\Big) \mathbb{W}\rangle}{\langle \mathbb{W}\rangle} \,.
\end{equation}
Note that naively such an integrated one-point point function would result in a divergent quantity since the non-integrated one-point function must take the form
\begin{equation}
\langle \mathcal{O}_\tau(x) \,\mathbb{W}\rangle = \frac{a_{\mathcal{O}_\tau} }{|x_\perp|^4}\,,
\end{equation}
with $a_{\mathcal{O}_\tau}$ a coupling dependent space-time constant.
However, supersymmetric localisation provides for a natural regulator of the divergent right hand side of \eqref{eq:Cw2} since no divergences arise by computing $\mathcal{C}_\mathbb{W}(\tau)$ starting from the matrix model result \eqref{eq:WilsonLag}.
We conclude that besides \eqref{eq:CwCorr} we must also have the defect one-point function relation
\begin{equation}
\label{eq:CwOtau}
\frac{\langle \mathcal{O}_\tau(x) \,\mathbb{W} \rangle}{\langle \mathbb{W}\rangle} = \alpha \,\mathcal{C}_\mathbb{W}(\tau) \frac{1}{|x_\perp|^4}\,,
\end{equation}
with $\alpha\in \mathbb{R}$ a constant.
This is not surprising since as mentioned above the defect one-point \eqref{eq:CwCorr} of the superconformal primary operator $\mathcal{O}_2$ fixes all defect one-point functions for conformal primaries in the same multiplet, hence in particular of $\mathcal{O}_\tau$.

It is well-known \cite{Fisher:1982fu} that while the stress tensor sits in the $\mathcal{N}=2$ supercurrent multiplet, the operator $\mathcal{O}_\tau$ sits in the $\mathcal{N}=2$ anomaly multiplet. Although it will not be discussed here, using an argument based on superconformal Ward identities akin to that of \cite{Fiol:2015spa, Bianchi:2018zpb, Bianchi:2019dlw} for the defect one-point function of the stress tensor, it is possible to fix the proportionality constant $\alpha$ in \eqref{eq:CwOtau} for the chiral Lagrangian operator.
Here we are more interested in the modularity properties of the relation \eqref{eq:CwOtau}:
 while it is true that for the case of the Wilson-line defect the insertion of the chiral Lagrangian in \eqref{eq:Cw2} can immediately be related to the insertion of $\mathcal{O}_2$ as in \eqref{eq:CwCorr}, the story is slightly more complicated in the case of a general line defect insertion.

Importantly, as stressed above the superconformal primary operator $\mathcal{O}_2$ is modular invariant but the chiral Lagrangian operator $\mathcal{O}_\tau$ does transform with a $U(1)$ weight corresponding to having holomorphic/anti-holomorphic modular weights $(+1,-1)$.\footnote{See \cite{Green:2020eyj, Dorigoni:2021rdo} for a detailed discussion of the modular properties of correlation functions involving the operator $\mathcal{O}_\tau$.}
Hence if we act on \eqref{eq:Cw}-\eqref{eq:Cw2} with the $\SL(2,\mathbb{Z})$ action given by $\rho$ we find:
\begin{align}
\mathcal{C}_\mathbb{W}(\rho \cdot \tau) & =  \frac{q\tau+p}{q\bar\tau+p} \,\mathcal{C}_\mathbb{L}([\rho];\tau) \,\,,\\
\mathcal{C}_\mathbb{L}([\rho];\tau) &\coloneqq \label{eq:CwLcorr1}2i\tau_2 \frac{\partial}{\partial \tau} \log \langle \mathbb{L}\rangle(\tau)\quad \Rightarrow\quad \frac{\langle \mathcal{O}_\tau(x) \,\mathbb{L} \rangle}{\langle \mathbb{L}\rangle} = \alpha \,\mathcal{C}_\mathbb{L}([\rho];\tau) \frac{1}{|x_\perp|^4}\,,
\end{align}
i.e. the $\rho \in \SL(2,\mathbb{Z})$ action changes the charges of the defect, while simultaneously introducing the $(+1,-1)$ automorphic factor $(q\tau+p)/(q\bar\tau+p)$ due to the $\mathcal{O}_\tau$ insertion, which is of course consistent with how the operator $\tau_2\partial_\tau$ transforms under the $ \SL(2,\mathbb{Z})$ action specified by $\rho$.

By combining \eqref{eq:CwLcorr2} with \eqref{eq:CwLcorr1} we then conclude that while for the Wilson-line defect case the insertion of $\mathcal{O}_2$ and $\mathcal{O}_\tau$ are actually identical, i.e. we can write either \eqref{eq:CwCorr} or \eqref{eq:Cw2}, the same is not true for a generic line defect where instead we have 
\begin{align}
\frac{\langle \mathcal{O}_2(x,Y_1)\,\mathbb{L}(Y_2) \rangle(\tau)}{\langle \mathbb{L}\rangle} = \frac{2\sqrt{2}}{c} \Big(\frac{ q  \tau+p}{q\bar\tau+ p} \Big)\,\mathcal{C}_\mathbb{L}([\rho];\tau)\,\frac{Y_{12}^2}{|x_\perp|^2}   \,.\label{eq:oneptL}
\end{align}
The phase factor $( q  \tau+p)/(q\bar\tau+ p)$ reduces to $1$ for the Wilson-line defect case $(p,q)=(1,0)$, and in general it  compensates the non-trivial ${\rm SL}(2,\mathbb{Z})$ transformation of $\tau_2 \partial_{\tau}$ in the definition of $\mathcal{C}_\mathbb{L}$ as given in \eqref{eq:CwLcorr1}.
The 't Hooft-line defect one-point function has been analysed in the large-$N$ planar limit using integrability methods \cite{Kristjansen:2023ysz}. However, in this regime $\tau_1$ is effectively set to $0$, hence the phase factor in equation \eqref{eq:oneptL} reduces to $-1$ and no difference is practically observed between an $\mathcal{O}_2$ and an $\mathcal{O}_\tau$ one-point function in the presence of an 't Hooft-line defect.

We now move on to discuss a more interesting class of non-topological integrated correlation functions of line defect operators.

\subsection{Integrated line defect correlators}

Lastly, we present the defect integrated correlator firstly introduced in \cite{Pufu:2023vwo} 
and defined starting from the more general matrix model formulation \eqref{eq:WilsonVEVN2} as
\begin{align}
\mathcal{I}_{\mathbb{W},N}(\tau) &\coloneqq \label{eq:Iw} \Big[ \partial_m^2\log \langle \mathbb{W} \rangle_{\mathcal{N}=2^*} (m,\tau)\Big]_{m=0}\,,
\end{align}
which is a much more interesting quantity when compared to \eqref{eq:Cw}-\eqref{eq:Ew} and it will be the main character of the present work.

As argued in \cite{Pufu:2023vwo} and then further clarified in \cite{Billo:2023ncz,Billo:2024kri, Dempsey:2024vkf}, the matrix model quantity defined in \eqref{eq:Iw} corresponds to the connected part of a particular integrated correlator of two superconformal primary operators $\mathcal{O}_2(x,Y)$ in the presence of a fundamental half-BPS Wilson-line defect schematically of the form
\begin{equation}
 \mathcal{I}_{\mathbb{W},N}(\tau) = \int \langle \mathcal{O}_2(x_1,Y_1) \mathcal{O}_2(x_2,Y_2)\, \mathbb{W}(Y_3)\rangle_{\rm c} \,{\rm d} \mu(x_i)\,.
\end{equation}
We refer to \cite{Billo:2024kri, Dempsey:2024vkf} for the precise form of the integration measure ${\rm d} \mu(x_1,x_2)$ over the two insertion points, as well as for the precise dependence from the $R$-symmetry polarisations. Our focus here is to understand the exact expression for $\mathcal{I}_{\mathbb{W},N}(\tau)$, and the non-holomorphic automorphic forms that describe this physical observable. 
As already argued for in the previous section, thanks to the electromagnetic duality transformation \eqref{eq:automGen} we expect that
\begin{align}
\mathcal{I}_{\mathbb{L},N}([\rho];\tau) = \Big[ \partial_m^2 \log\langle \mathbb{L}\rangle_{\mathcal{N}=2^*}(m,\tau)\Big]_{m=0}= \mathcal{I}_{\mathbb{W},N}(\rho\cdot \tau) \,,\label{eq:Iell}
\end{align}
and consequently we must have
\begin{equation}
\mathcal{I}_{\mathbb{L},N}([\rho];\tau)= \int \langle \mathcal{O}_2(x_1,Y_1) \mathcal{O}_2(x_2,Y_2)\, \mathbb{L}(Y_3)\rangle_{\rm c} \,{\rm d} \mu(x_i)\,.
\end{equation}
Since we know that the Wilson-line defect corresponds to the coset element $[\rho]= [\mathbbm{1}]$, while the 't Hooft-line defect case corresponds to $[\rho]=[S]$
where $S = \left(\begin{smallmatrix} 0 & -1 \\ 1 & 0 \end{smallmatrix}\right)$, we denote accordingly:
\begin{equation}
\mathcal{I}_{\mathbb{W},N}(\tau) =  \mathcal{I}_{\mathbb{L},N}([\mathbbm{1}];\tau)\,,\qquad \mathcal{I}_{\mathbb{T},N}(\tau) \coloneqq  \mathcal{I}_{\mathbb{L},N}([S];\tau)\,.
\end{equation}
We stress again that as an immediate consequence of electromagnetic duality the Wilson-line defect integrated correlator must be invariant under a $T$ transformation, that is for all $n\in\mathbb{Z}$ we have
 \begin{equation}
  \mathcal{I}_{\mathbb{W}, N} (   \tau+n)=\mathcal{I}_{\mathbb{L}, N} ([ \mathbbm{1} ];   T^n\cdot \tau) = \mathcal{I}_{\mathbb{L}, N} ([ T^n ];    \tau) = \mathcal{I}_{\mathbb{W}, N} (   \tau)\,.
 \end{equation}
However, under $S$-duality the Wilson-line defect is mapped into the 't Hooft-line defect,
 \begin{equation}
 \mathcal{I}_{\mathbb{W}, N} \Big(-\frac{1}{\tau}\Big)=\mathcal{I}_{\mathbb{L}, N} ([ \mathbbm{1} ];   S\cdot\tau) = \mathcal{I}_{\mathbb{L}, N} ([ S ];   \tau) = \mathcal{I}_{\mathbb{T}, N} (   \tau)\,.
 \end{equation}

 Starting from the matrix model definition \eqref{eq:Iw} for the Wilson-line defect integrated correlator, we will derive general expressions for an arbitrary line defect integrated correlator \eqref{eq:Iell} for which no matrix model formulation is currently available. For the special case of the ’t Hooft-line defect with gauge group $SU(2)$, in principle one may compute the integrated 
correlator directly from \eqref{eq:Tvev}.

The analysis we will carry out in the next sections will show that the Wilson-line defect integrated correlator \eqref{eq:Iw} has an exact expression in terms of the given novel automorphic functions satisfying electromagnetic duality \eqref{eq:autoform}. Hence we expect that a direct matrix model calculation for the 't Hooft-line defect integrated correlator will  reproduce our lattice sum results 
\eqref{eq:LatticeSumRes2} simply specialised to the case $[\rho]=[S]$. In the case of $SU(2)$ gauge group where the matrix model expression is available (see eq.(7.66) of \cite{Gomis:2011pf}), this has indeed been verified numerically. 

In section \ref{sec:LargeN} we show that the large-$N$ expansion at fixed-$\tau$ of \eqref{eq:Iw}, already considered in \cite{Pufu:2023vwo}, can be written order by  order in $1/N$ as a finite rational linear combination of a class of beautiful novel automorphic functions which satisfy the electromagnetic duality condition \eqref{eq:autoform}.
Then in section \ref{sec:FiniteN}, we  show that the same class of automorphic objects is responsible for the lattice sum integral representation \eqref{eq:LatticeSumRes2} of the defect integrated correlator \eqref{eq:Iw} at generic $N$.

\section{An Automorphic Large-$N$ Expansion}
\label{sec:LargeN}

In this section we revisit the large-$N$ fixed-$\tau$ expansion for the Wilson-line defect integrated correlator \eqref{eq:Iw} first analysed in \cite{Pufu:2023vwo}. We show that order by order in $1/N$ the line defect correlator can be written as a finite linear combination with rational coefficients of a novel class of automorphic functions whose transformation under $\SL(2,\mathbb{Z})$ is precisely the one dictated from $\mathcal{N}=4$ SYM electromagnetic duality presented in equation \eqref{eq:autoform}.

\subsection{Defect integrated correlators at large $N$}

From the definition of the Wilson-line defect integrated correlator \eqref{eq:Iw}, it is clear that to understand its modular properties in the large-$N$ limit,  we must consider the large-$N$ fixed-$\tau$ expansion of the underlying hermitian matrix model discussed in section \ref{sec:MM}.
This task has been carried out in \cite{Pufu:2023vwo}, where the authors expressed the perturbative and instantonic sectors of the integrated correlator \eqref{eq:Iw} in terms of matrix model $n$-body resolvents, whose large-$N$ expansion can be computed systematically using topological recursion \cite{Eynard:2004mh}.
We will not discuss the details of this analysis here but refer to \cite{Pufu:2023vwo} for further clarifications. However, in appendix \ref{sec:AppMM} we present an alternative calculation of the instanton sectors for the large-$N$ expansion of the integrated correlator \eqref{eq:Iw} which present some discrepancies with the results of \cite{Pufu:2023vwo}. Although these differences do not alter the beautiful story presented in \cite{Pufu:2023vwo}, they will shortly be of importance for us in understanding the systematics of the large-$N$ fixed-$\tau$ expansion in terms of a family of novel automorphic functions.

From the analysis of \cite{Pufu:2023vwo}, we know that the large-$N$ expansion of the Wilson-line defect integrated correlator \eqref{eq:Iw} takes the form
\begin{equation}
\mathcal{I}_{\mathbb{W},N}(\tau) = \mathcal{I}_{\mathbb{L},N}([\mathbbm{1}];\tau) = \sum_{\ell=-1}^\infty N^{-\ell/2 }  \mathcal{I}^{(\ell)}_{\mathbb{W}}(\tau)\,.\label{eq:WlargeN}
\end{equation}
The first three orders in $1/N$ are rather simple and are purely perturbative polynomials in $\tau_2= {\rm Im}\,\tau$, 
\begin{equation}
\mathcal{I}^{(-1)}_{\mathbb{W}}(\tau) = 2 \sqrt{\pi } \tau_2^{-\frac{1}{2}}\,,\qquad \mathcal{I}^{(0)}_{\mathbb{W}}(\tau) = \frac{1}{2}-\frac{\pi ^2}{3}\,,\qquad \mathcal{I}^{(1)}_{\mathbb{W}}(\tau)=\frac{3 }{16 \sqrt{\pi }} \tau_2^{\frac{1}{2}}-\frac{\pi ^{\frac{3}{2}}}{4 }\tau_2^{-\frac{3}{2}}\,.
\end{equation}
These purely perturbative terms which will keep on appearing at every order in the $1/N$ expansion and are somewhat spurious remnants of the Wilson-line defect expectation value \eqref{eq:WilsonLag}, which appears as the normalisation factor for the integrated Wilson-line defect correlator, thus effectively obfuscating the novel automorphic structures arising in the integrated correlator \eqref{eq:Iw}. 
From an $\SL(2,\mathbb{Z})$ electromagnetic duality point of view, any isolated perturbative term $\tau_2^s$ in a Wilson-line defect correlator, being that the integrated two-point function \eqref{eq:Iw} or the previously discussed expectation value \eqref{eq:WilsonVEV}, one and two-point correlators \eqref{eq:Cw}-\eqref{eq:Ew}, corresponds to the simplest family of automorphic functions \eqref{eq:Feasy}  relevant for line defect correlators in $\mathcal{N}=4$ SYM. We refer to appendix \ref{sec:AppF} for more details.

While the Wilson-line defect observables \eqref{eq:WilsonVEV}-\eqref{eq:Cw}-\eqref{eq:Ew} are completely contained in this easier family of automorphic functions \eqref{eq:Feasy}, to fully describe the integrated correlator \eqref{eq:Iw} this class of functions is not enough. We will have to disentangle such perturbative contributions from a genuinely new and different family of automorphic functions. 
The same phenomenon will be even more manifest when discussing the finite-$N$ exact results for the line defect integrated correlator in section \ref{sec:FiniteN}. However, in that case we will be able to isolate such polluting terms and neatly highlight the novel automorphic sectors contributing to \eqref{eq:Iw}.

Already starting with the next $1/N$ order it becomes obvious that the Wilson-line defect integrated correlator \eqref{eq:Iw} does contain something else besides the perturbative $\tau_2^s$ terms.
In particular building on the results of \cite{Pufu:2023vwo} we find
\begin{align}
\mathcal{I}^{(2)}_{\mathbb{W}}(\tau)&= \frac{3 \tau_2 (4 \zeta(3)+1)}{32 \pi }+\frac{\pi ^3}{180 }\tau_2^{-3}- \sum_{k> 0} \cos(2\pi k \tau_1)
\frac{6  \sigma_{-2}(k) }{\pi  \tau_2} K_2(2 k  \pi  \tau_2)\,,\label{eq:I2} \\
\mathcal{I}^{(3)}_{\mathbb{W}}(\tau) &\label{eq:I3} = \mathcal{I}^{(3){\text{-}}i}_{\mathbb{W}}(\tau)+ \mathcal{I}^{(3){\text{-}}ii}_{\mathbb{W}}(\tau)+\frac{63}{1024\pi^{\frac{3}{2}}}\tau_2^{\frac{3}{2}}+\frac{ 5\pi ^{\frac{5}{2}}}{192 }\tau_2^{-\frac{5}{2}} \,,
\end{align}
with
\begin{align}
\mathcal{I}^{(3){\text{-}}i}_{\mathbb{W}}(\tau) \label{eq:I3i} &= -\frac{9}{32 \pi^{\frac{3}{2}}} \zeta(3)\tau_2^{\frac{3}{2}}-\frac{ \pi ^{\frac{5}{2}}}{240 } \tau_2^{-\frac{5}{2}}+ \sum_{k> 0} \cos(2\pi k \tau_1) \frac{9 \sigma_{-2}(k) }{2 \pi \sqrt{\pi \tau_2}} K_2(2 k\pi  \tau_2)\, \,, \\
\mathcal{I}^{(3){\text{-}}ii}_{\mathbb{W}}(\tau) \label{eq:I3ii} &=\frac{15}{32 \pi^{\frac{3}{2}}} \zeta(3)\tau_2^{\frac{3}{2}}-\frac{\pi ^{\frac{9}{2}}}{3780 }\tau_2^{-\frac{9}{2}} +\sum_{k> 0} \cos(2\pi k \tau_1)\frac{45 \sigma_{-2}(k)}{2  \pi k (\pi \tau_2)^{\frac{3}{2}}} K_3(2 k \pi  \tau_2)\,,
\end{align}
and $K_s(y)$ is the modified Bessel function of the second kind and $\sigma_s(k) \coloneqq \sum_{d\vert k} d^s$ is the divisor sigma function.
From this order onward we have that instanton contributions are present in the integrated correlator \eqref{eq:Iw} and do in fact play a crucial role. In particular, we notice the similarity between the Fourier series \eqref{eq:I2} with respect to the $\theta$-angle, $\theta = 2\pi \tau_1$, and that of the modular invariant non-holomorphic Eisenstein series $E(s;\tau)$ conveniently presented in \eqref{eq:EisenFourier}. We have also computed the next coefficient $\mathcal{I}_{\mathbb{W}}^{(4)}(\tau)$ at order $N^{-2}$ as given in \eqref{eq:I4}, where a similar structure is found. 

As anticipated, we find some discrepancies between the large-$N$ instanton sectors derived in \cite{Pufu:2023vwo} and the results here presented in \eqref{eq:I3i}-\eqref{eq:I3ii} (see also \eqref{eq:I4} for the next coefficient $\mathcal{I}_{\mathbb{W}}^{(4)}$ at order $N^{-2}$), although we want to stress that despite this disagreement the main results of \cite{Pufu:2023vwo} are still perfectly valid.
In \cite{Pufu:2023vwo} it was shown that starting at order $N^{-3/2}$, i.e. starting with $\mathcal{I}_\mathbb{W}^{(3)}$, the instanton sectors contain terms proportional to $\sin(2\pi k \tau_1) = \sin( k \theta)$. Such terms are odd under the upper-half-plane involution $\tau\to -\bar\tau$ which is indeed consistent with the fact that in the presence of a Wilson-line defect time-reversal symmetry is broken\footnote{We thank Yifan Wang for discussions on this issue.}.
However, in appendix \ref{sec:AppLargeNMM} we show how to systematically extract the instanton sectors from the large-$N$ matrix model integral and prove that no term proportional to $\sin( k \theta)$ is present.
Furthermore, from the discussion we present in appendix \ref{sec:AppendixDecomp} for the calculation of contributions coming from Nekrasov partition function, it appears clear that for the finite-$N$ matrix model the instanton sectors do not present any term proportional to $\sin( k \theta)$ either. We believe that it would be rather odd if such terms were not to be present at finite $N$ but did appear in the large-$N$ limit.

At this point given the first few orders of the large-$N$ fixed-$\tau$ expansion for the Wilson-line defect integrated correlator \eqref{eq:I2}-\eqref{eq:I3}-\eqref{eq:I4}, we need to understand whether such terms can be identified as elements (or linear combinations thereof) of some special class of automorphic functions.
As discussed in \cite{Chester:2019jas, Dorigoni:2021guq}, each order in the large-$N$ expansion of a particular integrated correlator of four superconformal primary operators $\mathcal{O}_2$ can be written as a finite linear combination with rational coefficients of non-holomorphic Eisenstein series.
However, this cannot possibly be the case for the present discussion for two main reasons.
Firstly, each $E(s;\tau)$ is modular invariant with respect to the $\gamma\in \SL(2,\mathbb{Z})$ action $\tau\to \gamma\cdot\tau$, while the Wilson-line defect integrated correlator is expected to transform to a different line defect operator following the automorphic property \eqref{eq:autoform}.
Secondly, from the Fourier mode expansion for the non-holomorphic Eisenstein series \eqref{eq:EisenFourier}, it appears clear that in the instanton sector the index of the Bessel function and that of the divisor sigma function are related in a way which is not compatible with the matrix model results listed in \eqref{eq:I2}-\eqref{eq:I3}-\eqref{eq:I4}, even though they look similar. 

We therefore conclude that a novel class of real-analytic functions satisfying the automorphic property \eqref{eq:autoform} is required to express the line defect integrated correlator \eqref{eq:Iw}.

\subsection{Large-$N$ expansion and novel automorphic functions}

Guided by this similarity (and differences) between the large-$N$ expansion coefficients $ \mathcal{I}^{(\ell)}_{\mathbb{W}}(\tau)$, see e.g. \eqref{eq:I2}, and the non-holomorphic Eisenstein series we now introduce a novel class of automorphic functions inspired by the Eisenstein series lattice-sum representation \eqref{eq:EisenFourier}. 
We conjecture that the integrated correlator $\mathcal{I}_{\mathbb{L},N}([\rho];\tau)$ for a line defect $\mathbb{L}$ whose electromagnetic charges $(p,q)$ we parametrise by $[\rho] = \left( \begin{smallmatrix} * & * \\ q & p \end{smallmatrix}\right)\in B(\mathbb{Z}) \backslash \SL(2,\mathbb{Z})$ can be written in terms of the family of real-analytic functions defined via the lattice sum
\begin{align}
F_{s_1,s_2,s_3}([\rho];\tau)
&\label{eq:Fnew}  \coloneqq  \frac{(\tau_2/\pi)^{s_1}}{|q \tau+p|^{2s_1}} \sum_{(n,m)\in \mathbb{Z}^2 \setminus \{\mathbb{Z}(q,p)\}}   \frac{(\tau_2/\pi)^{s_2}}{|n \tau+m|^{2s_2}} [\pi (n p - m q)]^{s_3} \,,
\end{align}
where for convergence we must require ${\mbox{Re}}(s_2-s_3)>1$, although we can analytically continue this class of functions to general $s_1,s_2,s_3\in \mathbb{C}$.

Appendix \ref{sec:AppF} is devoted to presenting a detailed analysis of the class of functions defined in \eqref{eq:Fnew}, here we simply state some of their key properties.
Firstly, the lattice sum \eqref{eq:Fnew} is constructed precisely in such a way that under the action of the electromagnetic duality \eqref{eq:GNOcoset} it transforms with the desired automorphic property \eqref{eq:autoform}, i.e. for all $\gamma \in \SL(2,\mathbb{Z})$ we have
\begin{equation}
F_{s_1,s_2,s_3}([\rho];\gamma\cdot\tau) = F_{s_1,s_2,s_3}([\rho\,\gamma];\tau)\,.
\end{equation}
For the particular coset element $[\rho] = [\mathbbm{1}]$ that is the relevant one when discussing a Wilson-line defect, the function $F_{s_1,s_2,s_3}([\mathbbm{1}];\tau)$ admits the Fourier mode decomposition with respect to $\tau_1$ given by
\begin{align}
{F}_{s_1,s_2,s_3}([\mathbbm{1}];\tau) &\notag =  
\sum_{k\in \mathbb{Z}} e^{2\pi i k \tau_1} {F}^{(k)}_{s_1,s_2,s_3}([\mathbbm{1}];\tau_2)  \\
&= \label{eq:FnewFourier} \frac{2 \pi ^{\frac{1}{2}+s_3-s_1-s_2} \Gamma \left(s_2-\frac{1}{2}\right)  \zeta (2 s_2-s_3-1)}{\Gamma (s_2)}\tau_2^{s_1+1-s_2}\\
&\notag\phantom{=} +\frac{8 \pi^{s_3-s_1}}{\Gamma (s_2)}\sum_{k>0}\cos (2 \pi  k \tau_1)   
k^{s_2-\frac{1}{2}}\sigma _{s_3+1-2 s_2}(k) \tau_2^{s_1+\frac{1}{2}} K_{s_2-\frac{1}{2}}(2 k \pi  \tau_2)\,.
\end{align}
In contrast with the non-holomorphic Eisenstein series \eqref{eq:EisenFourier}, we note that in this Fourier mode expansion  the index of the Bessel function and that of the divisor sigma are not constrained any longer hence lifting any `tension' with respect to \eqref{eq:I2}.

Either using the lattice sum representation \eqref{eq:Fnew} specialised to $[\rho ] = [\mathbbm{1}]$, or the Fourier mode expansion \eqref{eq:FnewFourier}, it is easy to show  that ${F}_{s_1,s_2,s_3}([\mathbbm{1}];\tau)$ satisfies the homogeneous Laplace equation
\begin{equation}
[\Delta_\tau -2s_1 \tau_2\, \partial_{2} + (s_1+s_2)(1+s_1-s_2)\big]{F}_{s_1,s_2,s_3}([\mathbbm{1}];\tau) = 0\,,\label{eq:diff0MB}
\end{equation}
with $\Delta_\tau = \tau_2^2(\partial_1^2 +\partial_2^2)$.
Furthermore, we find in \eqref{eq:appFimproved} that it is possible to add to ${F}_{s_1,s_2,s_3}([\mathbbm{1}];\tau)$ a multiple of the homogeneous solution $\tau_2^{s_1+s_2}$ to \eqref{eq:FnewFourier} such that the resulting function has a softer behaviour in the small-$\tau_2$ limit. 
As a consequence of electromagnetic duality \eqref{eq:autoform}, the small-$\tau_2$ (i.e. the strong coupling regime) expansion of the large-$N$ Wilson-line defect correlator is directly related to the large-$\tau_2$ (i.e. the weak-coupling regime) expansion of the 't Hooft-line defect correlator. Importantly \cite{Pufu:2023vwo} showed that such softer small-$\tau_2$ behaviour for each $1/N$ coefficient $\mathcal{I}^{(\ell)}_{\mathbb{W}}(\tau)$ is directly related to the requirement of having a consistent genus expansion for the 't Hooft-line defect correlator. This `dual' large-$N$ behaviour can be inferred from the fact that an 't Hooft-line defect operator corresponds to a $D1$-brane on the holographic side. Hence, a two-point function of local operators $\mathcal{O}_2$ in the presence of an 't Hooft-line defect can be interpreted as a scattering process of two gravitons from a $D1$-brane in string theory. In this case open-string states propagate as intermediate states, therefore the weak coupling expansion is controlled by the open-string coupling constant (in contrast with the closed-string expansion for the $F$-string dual to the Wilson-line case).\footnote{A similar behaviour can be seen for the integrated four-point function $\langle \mathcal{O}_2\mathcal{O}_2  \mathcal{D} \mathcal{D}\rangle$ studied in \cite{Brown:2024tru}, where $\mathcal{D}$ is the determinant operator with scaling dimension $N$, holographically dual to a $D3$-brane. From the superstring side, this four-point function is interpreted as a scattering process of two gravitons off a $D3$-brane propagating along a geodesic \cite{Jiang:2019xdz, Brown:2024tru}. This interpretation leads to the same large-$N$ behaviour as \eqref{eq:D1-scattering}. }  More precisely, we expect that the 't Hooft expansion for the 't Hooft-line integrated correlator should behave as
\ie \label{eq:D1-scattering}
\mathcal{I}_{\mathbb{T}, N} (\tilde{\tau}_2 ) = \sum_{g=0}^{\infty} N^{1-g}  \, \mathcal{I}^{(g)}_{\mathbb{T}} (\tilde{\lambda} ) \, , 
\fe
where $\tilde{\tau}_2$ is related to ${\tau}_2$ via $S$ duality (for simplicity we set $\tau_1=0$), i.e. $\tilde{\tau}_2=1/\tau_2$ or equivalently $\tilde{\lambda}=4\pi N / \tilde{\tau}_2 = (4\pi  N)^2/\lambda$ with $\lambda \coloneqq g_{_{{\rm  YM}}}^2N$; in the dual 't Hooft limit we consider $N\gg1$ with $\tilde{\lambda}$ fixed. 

Therefore, we are quite naturally led to consider the family of `improved' automorphic functions,
\begin{align}
 &\widehat{F}_{s_1,s_2,s_3}([\mathbbm{1}];\tau) \coloneqq F_{s_1,s_2,s_3}([\mathbbm{1}];\tau) -4\pi^{s_3-s_1-s_2} \zeta(2s_2)\zeta(-s_3) \tau_2^{s_1+s_2}\phantom{\int}\label{eq:Fimproved}\\
 &\notag =
  \frac{2 \pi ^{\frac{1}{2}+s_3-s_1-s_2} \Gamma \left(s_2-\frac{1}{2}\right)  \zeta (2 s_2-s_3-1)}{\Gamma (s_2)}\tau_2^{s_1+1-s_2}-4\pi^{s_3-s_1-s_2} \zeta(2s_2)\zeta(-s_3) \tau_2^{s_1+s_2}\\
&\notag\phantom{=} +\frac{8 \pi^{s_3-s_1}}{\Gamma (s_2)}\sum_{k>0}\cos (2 \pi  k \tau_1) 
k^{s_2-\frac{1}{2}}\sigma _{s_3+1-2 s_2}(k) \tau_2^{s_1+\frac{1}{2}} K_{s_2-\frac{1}{2}}(2 k \pi  \tau_2)\,.
\end{align}
As previously commented on, the additional term $\tau_2^{s_1+s_2}$ is a homogeneous solution to \eqref{eq:diff0MB}, and the coefficient is determined by requiring a  softer small-$\tau_2$ behaviour when $\tau_1=0$. Indeed, in appendix \ref{sec:AppF} we show that the factor proportional to $\tau_2^{s_1+s_2}$ cancels out precisely the most singular term originating from the infinite sum of instanton contributions in the small-$\tau_2$ expansion (with $\tau_1=0$). 

We now proceed to rewrite the large-$N$ expansion coefficients $\mathcal{I}_\mathbb{W}^{(\ell)}(\tau)$ as finite linear combinations of  $\widehat{F}_{s_1,s_2,s_3}([\mathbbm{1}];\tau)$ with rational coefficients.
It had already been noted in \cite{Pufu:2023vwo}, that the building blocks appearing at each given order in $1/N$ seem to be solution to homogeneous Laplace equations of the same form as \eqref{eq:diff0MB}.
In particular for the orders given in \eqref{eq:I2} and \eqref{eq:I3}, we have
\begin{align}
\Big(\Delta_\tau + 3 \tau_2 \partial_2 - 3 \Big)\, \mathcal{I}_\mathbb{W}^{(2)}(\tau) &=0 \label{eq:I2diff}\,,\\ 
 \Big(\Delta_\tau + 2 \tau_2 \partial_2 - \frac{15}{4} \Big)\, \mathcal{I}_\mathbb{W}^{(3){\text -}i}(\tau) &\label{eq:diff3i}=0 \,,\\ 
  \Big(\Delta_\tau + 4 \tau_2 \partial_2 - \frac{27}{4} \Big)\, \mathcal{I}_\mathbb{W}^{(3){\text -}ii}(\tau) &\label{eq:diff3ii}=0 \,.
 \end{align}
Each of these Laplace equations can be combined with \eqref{eq:diff0MB} to construct two possible automorphic functions $\widehat{F}_{s_1,s_2,s_3}([\mathbbm{1}];\tau)$ which solves the same equation.
Finally, we use the Fourier mode decomposition \eqref{eq:Fimproved} to show that all $\mathcal{I}_\mathbb{W}^{(\ell)}(\tau)$ are indeed captured by this novel family of automorphic objects, at least up to the order $1/N^2$ here computed. 

Omitting the argument of $\widehat{F}_{s_1,s_2,s_3}([\mathbbm{1}];\tau)$ since it is understood here that we are focusing on the Wilson-line defect correlator, we can rewrite \eqref{eq:I2}-\eqref{eq:I3i}-\eqref{eq:I3ii} in terms of $\widehat{F}_{s_1,s_2,s_3}$ as
\begin{align}
 \mathcal{I}^{(2)}_{\mathbb{W}}(\tau) & = - 
 \widehat{F}_{-\frac{3}{2},-\frac{3}{2},-2} +\frac{3}{32 \pi } \tau_2\,, \\
 \mathcal{I}^{(3)}_{\mathbb{W}}(\tau) & = \mathcal{I}^{(3){\text{-}}i}_{\mathbb{W}}(\tau)+ \mathcal{I}^{(3){\text{-}}ii}_{\mathbb{W}}(\tau)+\frac{63}{1024\pi^{\frac{3}{2}}}\tau_2^{\frac{3}{2}}+\frac{ 5\pi ^{\frac{5}{2}}}{192 }\tau_2^{-\frac{5}{2}}\,, \end{align}
 with
\begin{align}
  \mathcal{I}^{(3){\text -}i}_{\mathbb{W}}(\tau) & = \frac{3}{4} \widehat{F}_{-1,-\frac{3}{2},-2}\,,\qquad\qquad  \mathcal{I}^{(3){\text -}ii}_{\mathbb{W}}(\tau)  = -\frac{3}{2} \widehat{F}_{-2,-\frac{5}{2},-4}\,.
\end{align}
At order $1/N^2$ the calculation is presented in appendix \ref{sec:AppLargeNMM} and the resulting contribution $\mathcal{I}^{(4)}_{\mathbb{W}}(\tau)$, given in \eqref{eq:I4}, also admits the decomposition in terms of the automorphic functions $\widehat{F}_{s_1,s_2,s_3}$,
\begin{align} \label{eq:IW4}
\mathcal{I}^{(4)}_{\mathbb{W}}(\tau) &\notag  = -\frac{3}{32} \widehat{F}_{-\frac{1}{2},-\frac{3}{2},-2}-\frac{1}{8} \widehat{F}_{-\frac{5}{2},-\frac{3}{2},-2}-\frac{1}{4} \widehat{F}_{-\frac{3}{2},-\frac{5}{2},-2}+3\, \widehat{F}_{-\frac{3}{2},-\frac{5}{2},-4}+\frac{1}{4} \widehat{F}_{-\frac{7}{2},-\frac{5}{2},-4}\\
  &\phantom{=} +\frac{1}{2} \widehat{F}_{-\frac{5}{2},-\frac{7}{2},-4}-\frac{15}{4} \widehat{F}_{-\frac{5}{2},-\frac{7}{2},-6}+\frac{27 \tau_2^2}{512 \pi ^2}-\frac{1}{128}-\frac{13 \pi ^2}{384 \tau_2^2} -\frac{\pi ^4}{864 \tau_2^4}\,.
  \end{align}
We stress that although the expressions just displayed do present some spurious purely perturbative powers of $(\tau_2/\pi)$ with rational coefficients at each order in $1/N$, 
our large-$N$ results are nonetheless very non-trivial.\footnote{These pure perturbative terms are described by the simpler automorphic functions \eqref{eq:Feasy}, which can also be seen as special cases of $\widehat{F}_{s,0,0}([\mathbbm{1}];\tau)$ where an analytic continuation in the parameters $s_2$ and $s_3$ is understood.} We managed to write each coefficient $\mathcal{I}_\mathbb{W}^{(\ell)}(\tau)$ in a manifestly $\SL(2,\mathbb{Z})$ automorphic way, and the class of functions $\widehat{F}_{s_1,s_2,s_3}$ captures the complete instanton sectors as well as all perturbative coefficients $(\tau_2/\pi)^s$ whose coefficients are non-trivial rational multiples of Riemann odd zeta values. 

It is also important to note that in the small-$\tau_2$ limit, the term proportional to $1/\tau_2^4$ in \eqref{eq:IW4} cancels precisely the contribution arising from the infinite sum of instanton with $\tau_1=0$. This is a consistency check of the large-$N$ behaviour of 't Hooft-line integrated correlator as shown in \eqref{eq:D1-scattering}, which can be obtained by an $S$-transformation of \eqref{eq:IW4}. 

 The above large-$N$ results resemble very closely the large-$N$ fixed-$\tau$ expansion of the integrated correlator of four $\mathcal{O}_2$ considered in \cite{Chester:2019jas, Dorigoni:2021bvj, Dorigoni:2021guq}, where the functions $\widehat{F}_{s_1,s_2,s_3}$ are replaced by non-holomorphic Eisenstein series.  
Inspired by these results,  our large-$N$ analysis strongly suggests that at finite $N$, the line defect integrated correlator \eqref{eq:Iw} should be given by a formal series over the same automorphic functions \eqref{eq:Fnew}.
In the next section we show that this is indeed the case and present the conjectural lattice sum integral representation \eqref{eq:LatticeSumRes2} for the line defect integrated correlator $\mathcal{I}_{\mathbb{L},N}([\rho];\tau)$ with arbitrary $N$.

\section{Exact Integrated Line Defect Correlators at Finite $N$}
\label{sec:FiniteN}

In this section we provide general arguments leading to the conjectural lattice sum integral representations \eqref{eq:LatticeSumRes2} 
for the line defect integrated correlator \eqref{eq:Iw} in~$\mathcal{N}=4$ SYM with gauge group~$G=SU(N)$ at finite $N$.
In particular, starting from the matrix model definition of the line defect integrated correlator we show that 
\eqref{eq:LatticeSumRes2} hold when the gauge group is $SU(2)$ and $SU(3)$, and determine explicitly the function $\mathcal{B}_N(t_1, t_2, t_3)$ for these cases.

\subsection{General lattice sum integral representation}

To analyse the line defect integrated correlator $\mathcal{I}_\mathbb{L}([\rho];\tau)$ at finite $N$ we first want to simplify the object under consideration and remove any ``polluting'' effects which may appear in the weak coupling expansion $\tau_2\gg1$.
From the matrix model definition \eqref{eq:Iw} we see that  $\mathcal{I}_\mathbb{W}(\tau)$ can be rewritten as
\begin{equation} \label{eq:matrix2}
\mathcal{I}_\mathbb{W}(\tau) =\Big[ \frac{\partial_m^2 \langle \mathbb{W} \rangle_{\mathcal{N}=2^*} -\langle \mathbb{W} \rangle_{\mathcal{N}=2^*}\,  \partial_m^2 Z_N (m,\tau) }{\langle \mathbb{W} \rangle_{\mathcal{N}=2^*}}\Big]_{m=0}\,,
\end{equation}
where the $\mathcal{N}=2^*$ SYM partition function $Z_N(m,\tau)$ is defined in \eqref{Zdef} and the $\mathcal{N}=2^*$ half-BPS fundamental Wilson loop expectation value $\langle \mathbb{W} \rangle_{\mathcal{N}=2^*}(m,\tau)$ is given in terms of the matrix model integral \eqref{eq:WilsonVEVN2}.

From the expression  \eqref{eq:WilsonLag} for the Wilson-line defect expectation value, we see that the denominator in the above expression besides removing an exponential factor, it introduces in the weak coupling expansion $\tau_2\gg1$ an infinite number of perturbative corrections coming from the expansion of the Laguerre polynomial, thus polluting the weak coupling expansion of $\mathcal{I}_\mathbb{W}(\tau)$.
Furthermore, given that the Wilson-line defect expectation value is only a function of $\tau_2 = {\mbox{Im}}(\tau)$, it can be easily made automorphic with respect to electromagnetic duality using the easier class of functions presented in \eqref{eq:Feasy}, i.e. from \eqref{eq:WilsonLag} we deduce
\begin{align}
\langle \mathbb{L} \rangle(\tau)  &\notag = \frac{1}{N}\exp\Big[ \frac{\pi}{2\tau_2}\frac{N-1}{N} \Big]  L^1_{N-1}\left( -\frac{\pi}{\tau_2}\right) \Big\vert_\rho = \frac{1}{N}\exp\Big[ \frac{\pi}{2 \,{\rm{Im} (\rho \cdot \tau)}}\frac{N-1}{N} \Big]  L^1_{N-1}\Big( -\frac{\pi}{\rm{Im} (\rho \cdot \tau)}\Big) \\
&= \frac{1}{N}\exp\Big[ \frac{\pi |q\tau+p|^2}{2\tau_2}\frac{N-1}{N} \Big]  L^1_{N-1}\left( -\frac{\pi |q\tau+p|^2}{\tau_2}\right)\,,
\end{align}
which we have checked explicitly in \eqref{eq:THooftS} for the 't Hooft-line defect case $(p,q)=(0,1)$.

Rather than carrying along for the ride this unnecessary automorphic factor, we define a cleaned-up version of the line defect integrated correlator and introduce a `reduced' version of the integrated correlator,
\begin{equation}
\widetilde{\mathcal{I}}_{\mathbb{L},N}([\rho];\tau)  \coloneqq   \frac{L^1_{N-1}( -\frac{\pi}{\rm{Im} (\rho \cdot \tau)})}{N} \,\mathcal{I}_{\mathbb{L},N}([\rho];\tau)
\label{eq:LRedM}\, .
\end{equation}
We proceed to justify the lattice sum integral representation for the reduced correlator $\widetilde{\mathcal{I}}_{\mathbb{L},N}([\rho];\tau)$, and then an expression for $\mathcal{I}_{\mathbb{L},N}([\rho];\tau)$ can of course be obtained inverting the above relation. 

From our large-$N$ analysis of section \ref{sec:LargeN}, we have seen that order by order in $1/N$ we can write the line defect integrated correlator as a finite linear combination of automorphic forms \eqref{eq:Fnew}. As commented earlier, in previous works \cite{Dorigoni:2021guq,Dorigoni:2022zcr,Alday:2023pet} on integrated correlators of four superconformal primary operators of the stress tensor it has been observed that in the transition from large-$N$ to finite $N$ the automorphic building blocks do not change in nature, being that non-holomorphic Eisenstein series as in \cite{Dorigoni:2021guq,Dorigoni:2022zcr} or generalised Eisenstein series in \cite{Alday:2023pet}.
However, while at large-$N$ it is possible to isolate order by order in $1/N$ finite rational linear combinations of the relevant automorphic building blocks, at finite $N$ one must include a formal infinite series over such building blocks, which has to be resummed, possibly via a lattice sum integral representation. 

Guided from experience, we assume that a similar phenomenon takes place for the present discussion of the line defect integrated correlator and work under the verifiable hypothesis that at finite-$N$ the correlator $\widetilde{\mathcal{I}}_{\mathbb{L},N}([\rho];\tau)$ admits a representation as a (formal) infinite series involving solely the automorphic functions $F_{s_1,s_2,s_3}([\rho];\tau)$ presented in \eqref{eq:Fnew}.
In appendix \ref{sec:AppF} we discuss various properties for the novel automorphic building blocks, in particular we highlight here their integral representation,
\begin{align}
&\notag {F}_{-s_1-s_3,s_2,-2s_3}([\rho];\tau)= \\
& \sum_{(n,m)\in \mathbb{Z}^2 \setminus\{\mathbb{Z}(q,p)\}}  \int_0^\infty e^{-t_1 \frac{{\rm Im}(\rho\cdot\tau)}{\pi} }  e^{- t_2 \pi \frac{|m+n\tau|^2}{\tau_2}}  e^{-t_3 \pi  {\rm Im}(\rho\cdot\tau) (np-mq)^2} \,\frac{t_1^{s_1-1}}{\Gamma(s_1)}\cdot \frac{t_2^{s_2-1}}{\Gamma(s_2)}\cdot \frac{t_3^{s_3-1}}{\Gamma(s_3)}\, {\rm d}^3 t\,,\label{eq:FIntRepMB}
\end{align}
valid for ${\rm Re}\,s_i > 0$, with $i=1,2,3$, where this particular combination of indices has been chosen for future convenience.
Once again inspired by the results of the integrated four-point correlator \cite{Dorigoni:2021guq}, given \eqref{eq:FIntRepMB}, it is natural to assume that the finite-$N$ expression for the integrated correlator $\widetilde{\mathcal{I}}_{\mathbb{L},N}([\rho];\tau)$ will be a lattice sum expression akin to the equation above without any constraint on the summation variables, i.e. $(n,m) \in \mathbb{Z}^2$, and where the monomial $t_1^{s_1-1}t_2^{s_2-1} t_3^{s_3-1}$ gets replaced by a more general function of $t_1,t_2,t_3$. More precisely, we conjecture
\begin{align}
\widetilde{\mathcal{I}}_{\mathbb{L},N}([\rho];\tau) 
&\label{eq:LatticeSumResMB}=  \sum_{(n,m)\in \mathbb{Z}^2}\int_0^\infty e^{-t_1 \frac{{\rm Im}(\rho\cdot\tau)}{\pi} }  e^{- t_2 \pi \frac{|m+n\tau|^2}{\tau_2}}  e^{-t_3 \pi  {\rm Im}(\rho\cdot\tau) (np-mq)^2} \mathcal{B}_N(t_1, t_2,t_3) \, {\rm d}^3 t\,.
\end{align}
Note that despite the explicit appearance of the representative~$\rho\in {\rm SL}(2,\mathbb{Z})$ of the equivalence class~$[\rho]$, equation \eqref{eq:LatticeSumResMB} is actually well-defined on the coset $B(\mathbb{Z})\backslash {\rm SL}(2,\mathbb{Z})$ parametrising all possible electromagnetic charges of the line defects here considered since we have
\begin{equation}
{\rm Im}(\rho\cdot\tau) = \frac{\tau_2}{|q\tau+p|^2}\qquad \forall \,\rho \in [ \rho ] =\left(\begin{matrix}  * &  *  \\ q & p \end{matrix}\right)\in B(\mathbb{Z})\backslash {\rm SL}(2,\mathbb{Z})\,.\label{eq:ImTau}
\end{equation}
From the general expression \eqref{eq:LatticeSumResMB} combined with \eqref{eq:LRedM} and \eqref{eq:ImTau} we obtain the lattice sum integral representation \eqref{eq:LatticeSumRes2} presented in the introduction, which is straightforward to specialise to the case of the integrated correlator of a half-BPS 't Hooft line defect,
\begin{align}
\mathcal{I}_{\mathbb{T},N}(\tau) &= \mathcal{I}_{\mathbb{L},N}([S];\tau) 
\label{eq:LatticeSumtHooft}  \\
&\notag = \frac{N}{L^1_{N-1}( -\frac{\pi |\tau|^2}{\tau_2})}  \sum_{(n,m)\in \mathbb{Z}^2}  \int_0^\infty e^{-t_1 \frac{\tau_2}{\pi |\tau|^2} }  e^{- t_2 \pi \frac{|n\tau+m|^2}{\tau_2}}  e^{-t_3 m^2\pi\frac{ \tau_2}{|\tau|^2}  } \mathcal{B}_N(t_1, t_2,t_3) \,{\rm d}^3t\,.
\end{align}

Let us analyse more in detail the general lattice sum integral representation \eqref{eq:LatticeSumResMB}.
As we will see, it is straightforward to integrate out $t_1$, which then leads to an alternative representation,
\begin{align}
\widetilde{\mathcal{I}}_{\mathbb{L},N}([\rho];\tau) 
&\label{eq:LatticeSumResMB2}= \sum_{(n,m)\in \mathbb{Z}^2}   \int_0^\infty e^{- t_2 \pi \frac{|m+n\tau|^2}{\tau_2}}  e^{-t_3 \pi {\rm Im}(\rho\cdot\tau) (np-mq)^2} {\widetilde{\mathcal{B}}}_N({\rm Im}(\rho \cdot \tau); t_2,t_3) \,{\rm d}^2 t\, ,
\end{align}
with ${\widetilde{\mathcal{B}}}_N(y; t_2,t_3)$ related to $\mathcal{B}_N(t_1, t_2,t_3)$ via 
\begin{equation}\label{eq:WN}
{\widetilde{\mathcal{B}}}_N(y; t_2,t_3) \coloneqq  \int_0^\infty e^{ - t_1 y} \mathcal{B}_N(t_1, t_2,t_3)\, {\rm d}t_1\, .
\end{equation}
An immediate consequence of the conjectural lattice sum integral representations \eqref{eq:LatticeSumResMB}-\eqref{eq:LatticeSumResMB2} is that by performing a double Poisson summation over $n$ and $m$ and then changing integration variables to 
\ie  \label{eq:inversion}
t_2\to (t_2+t_3)^{-1}\, , \qquad
t_3 \to t_2^{-1} - (t_2+t_3)^{-1} \, , 
\fe 
we see that the functions $\mathcal{B}_N$ and ${\widetilde{\mathcal{B}}}_N$ must satisfy the inversion relations
\begin{align}
\mathcal{B}_N(t_1, t_2,t_3) &\label{eq:inv1}= \frac{1}{[t_2(t_2+t_3)]^{3/2}} \mathcal{B}_N\Big( t_1, \frac{1}{t_2+t_3}, \frac{1}{t_2} - \frac{1}{t_2+t_3}\Big)\,,\\
{\widetilde{\mathcal{B}}}_N(y, t_2,t_3) &\label{eq:inv2}= \frac{1}{[t_2(t_2+t_3)]^{3/2}} {\widetilde{\mathcal{B}}}_N\Big(y, \frac{1}{t_2+t_3}, \frac{1}{t_2} - \frac{1}{t_2+t_3}\Big)\,.
\end{align}

It is useful to specialise \eqref{eq:LatticeSumResMB} to the case $[\rho] = [\mathbbm{1}]$, equivalently $(p,q)=(1,0)$, relevant for the Wilson-line defect case.
For this particular case we write $\widetilde{\mathcal{I}}_{\mathbb{W},N}(\tau) \coloneqq \widetilde{\mathcal{I}}_{\mathbb{L},N}([\mathbbm{1}];\tau) $ and divide the lattice sum \eqref{eq:LatticeSumResMB} as
\begin{align}
\widetilde{\mathcal{I}}_{\mathbb{W},N}(\tau) 
&\notag=  \sum_{\substack{ n=0 \\ m\in \mathbb{Z}  \!\phantom{\vert} } }  \int_0^\infty e^{- t_1 \frac{\tau_2 }{ \pi }}  e^{-   \frac{t_2 \pi m^2 }{ \tau_2} }  \mathcal{B}_N(t_1, t_2,t_3) \, {\rm d}^3 t \\
&\label{eq:LatticeSumResSplit} \phantom{=} + \sum_{\substack{ n\neq 0\\ m\in \mathbb{Z}}}  \int_0^\infty e^{-t_1 \frac{\tau_2 }{ \pi} }  e^{-  t_2 \frac{\pi [ ( n \tau_1+ m)^2+ (n \tau_2)^2]}{\tau_2}} e^{-  t_3\pi n^2 \tau_2}  \mathcal{B}_N(t_1, t_2,t_3) \, {\rm d}^3 t\,.
\end{align}
Furthermore, we can combine the integral representation \eqref{eq:FIntRepMB} with the Taylor expansion
\begin{equation} \label{eq:BN}
\mathcal{B}_N(t_1, t_2,t_3) = \sum_{s_1,s_2,s_3=1}^\infty d^{(N)}_{s_1,s_2,s_3} \frac{t_1^{s_1-1}}{\Gamma(s_1)}\cdot \frac{t_2^{s_2-1}}{\Gamma(s_2)}\cdot \frac{t_3^{s_3-1}}{\Gamma(s_3)}\,,
\end{equation}
and rewrite the second term in \eqref{eq:LatticeSumResSplit} to arrive at 
\begin{align}
\widetilde{\mathcal{I}}_{\mathbb{W},N}(\tau) 
&\label{eq:LatticeSumResSplit2}=  \sum_{\substack{ n=0 \\ m\in \mathbb{Z}  \!\phantom{\vert} } }  \int_0^\infty e^{- t_1 \frac{\tau_2 }{ \pi} }  e^{-   \frac{t_2 \pi m^2}{ \tau_2} }  \mathcal{B}_N(t_1, t_2,t_3) \, {\rm d}^3 t + \sum_{s_1,s_2,s_3=1}^\infty d^{(N)}_{s_1,s_2,s_3} F_{-s_1-s_3,s_2,-2s_3}([\mathbbm{1}];\tau)\,.
\end{align}
To better understand the first term in the above expression, it is useful to consider the large-$\tau_2$ asymptotic expansion of \eqref{eq:LatticeSumResSplit}. This can be done in a standard manner by performing a Poisson resummation with respect to the summation variable $m\to \hat{m}$, which also yields the Fourier mode expansion where the Fourier mode is $k = \hat{m} n $ hence only the second term in \eqref{eq:LatticeSumResSplit} will contribute to the $k\neq 0$ Fourier sectors.
The Fourier expansion of \eqref{eq:LatticeSumResSplit} can be written neatly as 
\begin{align}
&\widetilde{\mathcal{I}}_{\mathbb{W},N}(\tau) = \sum_{k\in \mathbb{Z}} e^{2 \pi i k \tau_1}\, \widetilde{\mathcal{I}}^{(k)}_{\mathbb{W},N}(\tau_2) \,,\label{eq:FourierDecomp}
\end{align}
where the Fourier zero-mode is given by
\begin{align}
\widetilde{\mathcal{I}}^{(0)}_{\mathbb{W},N}(\tau_2) &\label{eq:It0}= \sum_{\substack{ n=0 \\ \hat{m}\in \mathbb{Z}  \!\phantom{\vert} } }\sqrt{\tau_2}  \int_0^\infty e^{- t_1 \frac{\tau_2 }{ \pi} }  e^{- t_2 \pi \hat{m}^2 \tau_2 }  \,t_2^{-\frac{3}{2}}\, \mathcal{B}_N(t_1, t_2^{-1},t_3)  \, {\rm d}^3 t \\
&\notag \phantom{=} +  \sum_{\substack{ n\neq 0\\ \hat{m}=0}}  \sqrt{\tau_2} \int_0^\infty e^{-t_1 \frac{\tau_2 }{\pi} }  e^{-  (t_2+t_3)  \pi n^2 \tau_2} \,t_2^{-\frac{1}{2}}\,  \mathcal{B}_N(t_1, t_2,t_3)  \, {\rm d}^3 t\,.
\end{align}
Similarly, the Fourier non-zero modes with $k=\hat{m}n\neq 0$ are entirely captured by the second term in \eqref{eq:LatticeSumResSplit} and are exponentially suppressed at large-$\tau_2$ as manifest in
\begin{equation}
\widetilde{\mathcal{I}}^{(k)}_{\mathbb{W},N}(\tau_2) = \sum_{\substack{ n,\hat{m} \neq 0 \\ n\hat{m} = k}}\sqrt{\tau_2}  \,e^{ - 2\pi |k|\tau_2} \int_0^\infty e^{-t_1 \tau_2 / \pi }  e^{-\pi \tau_2 \big(|n| \sqrt{t_2} - \frac{|\hat{m}|}{\sqrt{t_2}} \big)^2 }e^{-  t_3\pi n^2 \tau_2} \,t_2^{-\frac{1}{2}}\,  \mathcal{B}_N(t_1, t_2,t_3)  \, {\rm d}^3 t\,.\label{eq:Itk}
\end{equation}

Based on the particular finite-$N$ examples we will analyse promptly, it appears that, in a direct analogy with \cite{Dorigoni:2021guq}, the $(n,\hat{m})=(0,0)$ contribution to the zero mode sector \eqref{eq:It0} vanishes identically, i.e.
\begin{equation}
 \int_0^\infty t_2^{-\frac{1}{2}}\,  \mathcal{B}_N(t_1, t_2,t_3)\, {\rm d} t_2 {\rm d} t_3 =0\,, \qquad \forall\, t_1 \in \mathbb{R}\,.\label{eq:00id}
\end{equation}
Furthermore, this observation combined with the inversion relation \eqref{eq:inv1} leads to the result that the two terms in \eqref{eq:It0} are actually identical, that is the Fourier zero-mode must be given by 
\begin{align}
\widetilde{\mathcal{I}}^{(0)}_{\mathbb{W},N}(\tau_2) 
&\label{eq:It0simp} =4\sum_{\hat{m}=1 }^\infty \sqrt{\tau_2}  \int_0^\infty e^{- t_1 \frac{\tau_2 }{ \pi} }  e^{- t_2 \pi \hat{m}^2 \tau_2 }  \,t_2^{-\frac{3}{2}}\, \mathcal{B}_N(t_1, t_2^{-1},t_3)  \, {\rm d}^3 t\\
&=  4 \sum_{n=1}^\infty  \sqrt{\tau_2} \int_0^\infty e^{-t_1 \frac{\tau_2 }{ \pi} }  e^{-  (t_2+t_3)  \pi n^2 \tau_2} \,t_2^{-\frac{1}{2}}\,  \mathcal{B}_N(t_1, t_2,t_3)  \, {\rm d}^3 t\,.
\end{align}
When combined back with the purely perturbative factor we removed in \eqref{eq:LRedM}, we claim that the zero Fourier mode sector \eqref{eq:It0simp} fully captures the perturbative expansion of the Wilson-line defect integrated correlator, while the non-zero modes \eqref{eq:Itk} reproduce the non-perturbative instanton and anti-instanton contributions.

The strategy to reconstruct the lattice sum integral representation \eqref{eq:LatticeSumResMB} starting from the matrix model formulation discussed in section \ref{sec:EasyCorr} is as follows:
\begin{itemize}
\item[(i)] We start by focusing our attention to the instanton sectors of the Wilson-line defect integrated correlator and uniquely identify the coefficients $d^{(N)}_{s_1,s_2,s_3}$ for the formal expansion \eqref{eq:LatticeSumResSplit2} in terms of automorphic functions $F_{s_1,s_2,s_3}([\mathbbm{1}];\tau)$;
\item[(ii)] Using the integral representation \eqref{eq:FIntRepMB} we resum such a formal infinite series and construct the integrand $\mathcal{B}_N(t_1,t_2,t_3)$ for which we can verify the conjectural identities \eqref{eq:inv1}-\eqref{eq:inv2}-\eqref{eq:00id};
\item[(iii)] Finally we check that the complete perturbative sector is indeed encoded in the zero Fourier mode expression \eqref{eq:It0simp}, thus justifying the unconstrained nature of the complete lattice sum i.e. as a sum over the full lattice $(n,m)\in \mathbb{Z}^2$.
\end{itemize}

To provide evidences that our analysis is indeed correct we now follow the strategy just outlined and present the detailed analysis for the Wilson-line defect correlator \eqref{eq:Iw} computed from the specific matrix model integrals with $N=2$ and $N=3$. It should be stressed that even though only these specific examples of $N$ are here considered explicitly, as we argue below it appears clear that the general strategy should apply to general $N$.

\subsection{Finite-$N$ exact results}

In this subsection, we follow the general strategy outlined in the above subsection and show that the matrix model integral for the integrated two-point correlator in the presence of Wilson-line can be expressed in the form of \eqref{eq:LatticeSumResSplit} for  $N=2$ and $N=3$. We begin by determining the  coefficients $d^{(N)}_{s_1,s_2,s_3}$ in \eqref{eq:LatticeSumResSplit2}. This is done by comparing the expression \eqref{eq:LatticeSumResSplit2} with the explicit results from matrix model integrals, order by order in the large-$\tau_2$ (or weak coupling) expansion. We leave the details of the matrix model computation and their results in appendix \ref{sec:AppMM}.  

We begin by considering the instanton 
contribution to the integrated correlator in the presence of Wilson-line defect, i.e. the Fourier modes $\widetilde{\mathcal{I}}^{(k)}_{\mathbb{W},N}(\tau_2)$ with $k\neq 0$ in \eqref{eq:FourierDecomp}. From \eqref{eq:matrix2} and the matrix model integrals \eqref{eq:ZN2s}-\eqref{eq:WilsonVEVN2}, the $k$-instanton contribution to \eqref{eq:LRedM} can be expressed as
\begin{equation} \label{eq:k-matrix}
\widetilde{\mathcal{I}}^{(k)}_{\mathbb{W},N}(\tau_2) =\frac{L^1_{N-1}( -\frac{\pi}{\tau_2})}{N}\, e^{- 2\pi |k| \tau_2} \sum_{\substack{p,q>0\\ pq = k}}\frac{\exg{ \mathbb{W}(a)I_{p\times q}(a)}-\exg{ \mathbb{W}(a)}\exg{ I_{p\times q}(a)}}{\exg{ \mathbb{W}(a)}}  \,,
\end{equation}
where the overall factor comes from our definition \eqref{eq:LRedM}, and as in \eqref{eq:WilsonVEV} we denote with $\mathbb{W}(a)$ the matrix model integrand for the Wilson loop expectation value in $\mathcal{N}=4$ SYM, i.e.
\begin{equation}
\mathbb{W}(a) \coloneqq  \frac{1}{N} \mbox{Tr} \,e^{2\pi a} = \frac{1}{N} \Big(\sum_{i=1}^N e^{2\pi  a_i}\Big)\,.
\end{equation}
The matrix model integrand $I_{p \times q}(a)$ instead denotes the order $m^2$ term in the small-mass expansion of the $k$-instanton contribution to the Nekrasov partition function \eqref{eq:ZnekIns}, shown in \cite{Chester:2019jas} to be given by 
 \begin{equation}
\begin{aligned} \label{eq:IIpq}
I_{p \times q}(a) &\coloneqq  \prod_{j=1}^N \oint  
\prod^p_{a=1} \prod^q_{b=1}  {\Big[z-a_j + i (a+b-2) \Big]^2\over \Big[z-a_j + i (a+b-2) \Big]^2+1 } \times
\left[
\left({2\over p^2}+{2\over q^2} \right) \right.\\
& \phantom{=}\left. + \sum_{j=1}^N
{\frac{i (p+q)(p-q)^2} {(p\,q)[z-a_j+i \,(p+q -1)] [z-a_j+ i \,(q-1) ] [z-a_j+i \,(p-1)]} } 
\right]{{\rm d}z \over 2\pi}  \,,
\end{aligned}
\end{equation}
where the integration contour is a counter-clockwise contour surrounding
the poles at $z = a_j + i$ with $j=1,\ldots,N$. 
For the purpose of determining the coefficients $d^{(N)}_{s_1,s_2,s_3}$, the  expectation values in \eqref{eq:k-matrix}  can be evaluated perturbatively in a large-$\tau_2$ expansion. 
We find that the $k$-instanton 
contribution can be expressed in general as
\begin{equation}  \label{eq:k-instI}  \widetilde{\mathcal{I}}^{(k)}_{\mathbb{W},N}(\tau_2) =  \sum_{s = 2}^\infty \sigma_{1-2s}(k) P_{N,s}(\tau_2,k) \, e^{-2\pi |k| \tau_2} \,,
\end{equation}
where $P_{N,s}(\tau_2,k)$ is a polynomial in $\tau_2^{-1}$ and $k$. From the analysed cases, i.e. $N=2$ and $N=3$, the degree in $\tau_2^{-1}$ of the polynomial $P_{N,s}$ appears to be $2s+N-2$, in particular it does increase as $s$ increases. More explicitly, for $SU(2)$ and $SU(3)$ the first few orders of \eqref{eq:k-instI} are presented in \eqref{eq:SU2Ipq} and \eqref{eq:SU3Ipq}. 

Given the similarity between the above expression and the non-zero Fourier modes \eqref{eq:FnewFourier} of the automorphic functions  ${F}_{s_1,s_2,s_3}([\mathbbm{1}];\tau)$, we can try and write \eqref{eq:k-instI} as an infinite series 
\begin{align}
 \widetilde{\mathcal{I}}^{(k)}_{\mathbb{W},N}(\tau_2) &\notag \stackrel{?}{=}\sum_{s_1,s_2,s_3}  d^{(N)}_{s_1,s_2,s_3} {F}^{(k)}_{-s_1-s_3,s_2,-2s_3}([\mathbbm{1}];\tau_2)\\
 &\label{eq:k-instF}   =  \sum_{s_1,s_2,s_3} d^{(N)}_{s_1,s_2,s_3} \frac{4 \pi^{s_1-s_3}}{\Gamma (s_2)}   k^{s_2-\frac{1}{2}} \sigma _{1-2s_2-2s_3}(k) \tau_2^{s_1+\frac{1}{2}} K_{s_2-\frac{1}{2}}(2 k \pi  \tau_2) \, .
\end{align}
The coefficients $d^{(N)}_{s_1,s_2,s_3}$ are uniquely determined by  comparing \eqref{eq:k-instI} with \eqref{eq:k-instF}. More concretely, we expand the modified Bessel functions in \eqref{eq:k-instF} at large $\tau_2$ and then impose that the coefficient of each divisor sigma function $\sigma_{1-2s}$ in \eqref{eq:k-instI} must equal that of \eqref{eq:k-instF} as a polynomial in  $k$ and $\tau_2$,~i.e. we impose that \eqref{eq:k-instF} equals \eqref{eq:k-instI} separately for each term $\sigma_{1-2s} \,k^a \tau_2^{b}$. This process yields a system of linear equations for the coefficients $d_{s_1,s_2,s_3}^{(N)}$ which turn out to have a unique solution. It is worth noting that since the degree in $\tau_2^{-1}$ of the polynomials $P_{N,s}(\tau_2,k)$ increases with $s$, this leads to having an infinite set of linear equations so that it is absolutely non-trivial that a solution actually does in fact exist. 

Implementing the above general procedure and using the explicit results for $\widetilde{\mathcal{I}}^{(k)}_{\mathbb{W},N}(\tau)$ for $N=2$ and $N=3$, as presented more in detail in appendix \ref{sec:AppendixDecomp} we find for $SU(2)$
\begin{equation}\label{eq:su2ansatz}
        \widetilde{\mathcal{I}}^{(k)}_{\mathbb{W},2}(\tau_2) = \sum_{s_1, s_2, s_3 =1}^\infty  d^{(2)}_{s_1,s_2,s_3} F^{(k)}_{-s_1-s_3,s_2,-2s_3} (\tau_2)\,,
\end{equation}
with coefficients given by
\begin{align}
\label{eq:su2gnereal} d^{(2)}_{s_1,s_2,s_3} &=(-1)^{s_2+s_3} 4^{-s_3} \left(2 s_2+2 s_3-1\right){}^2 \left[s_3^2+\left(s_1+s_2-1\right)
   s_3+s_1-s_1 s_2+s_2-1\right]\\
&\phantom{=}\notag \times \frac{ \Gamma \left(2 s_3\right) \Gamma \left(s_1+s_2+s_3-1\right)}{\Gamma \left(2
   s_1\right) \Gamma \left(s_3-s_1+2\right) \Gamma \left(s_1+s_3+1\right)} \, . 
\end{align}
Similarly for $SU(3)$, we have 
\begin{equation}\label{eq:ansatzSU3}
    \widetilde{\mathcal{I}}^{(k)}_{\mathbb{W},3}(\tau_2) = \sum_{s_1, s_2, s_3 =1}^\infty  d^{(3)}_{s_1,s_2,s_3} F^{(k)}_{-s_1-s_3,s_2,-2s_3} (\tau_2) \, ,
\end{equation}
with the coefficients given by 
\begin{align}\label{eq:su3gnereal}
&d^{(3)}_{s_1,s_2,s_3} = \frac{(-1)^{s_2+s_3} 4^{-s_3} \left(2 s_2+2 s_3-1\right)^2 \Gamma \left(2 s_3\right) \Gamma \left(s_1+s_2+s_3-2\right)}{3\,  \Gamma \left(2 s_1\right) \Gamma \left(s_3-s_1+3\right) \Gamma \left(s_1+s_3+1\right)}   \\
&\phantom{=}\times \Big\{ s_3^6+s_3^5\big[3 s_1+4 s_2-4\big] + s_3^4\big[2 s_1^2+\left(7 s_2-3\right) s_1+6 s_2^2-9 s_2+6\big]  \cr
&\phantom{=}+s_3^3 \big[-2 s_1^3+2 \left(s_2+9\right) s_1^2+\left(s_2-2\right) \left(3 s_2+14\right) s_1+s_2 \left(s_2 \left(4
   s_2-3\right)-2\right)+13\big] \cr
&\phantom{=}+s_3^2 \big[-3 s_1^4+28 s_1^3-\left(\left(s_2-16\right) s_2+66\right) s_1^2+\left(s_2 \left(\left(22-3 s_2\right) s_2-54\right)+83\right)
   s_1 \big]  \cr
&\phantom{=}+s_3^2\big[+ s_2 \left(s_2 \left(s_2 \left(s_2+5\right)-20\right)+46\right)-48 \big] + s_3 \big[-s_1^5+2 \left(s_2+5\right) s_1^4+\left(\left(s_2-12\right) s_2-20\right) s_1^3 \big] \cr
&\phantom{=} +s_3 \big[\left(9-3 \left(s_2-10\right) s_2\right)
   s_1^2 + \left(s_2-1\right) \left(s_2 \left(s_2 \left(3 s_2-7\right)+18\right)-8\right)-2 s_1 \left(\left(\left(s_2-4\right)
   s_2+8\right) s_2^2+3\right) \big]\cr
&\phantom{=} + \left(s_1-2\right) \left(s_1-1\right) \left(s_2-1\right) \big[s_1^3-\left(s_2+9\right) s_1^2+\left(8-\left(s_2-4\right)
   s_2\right) s_1+\left(s_2-2\right) \left(\left(s_2-1\right) s_2+6\right)\big] \Big\} \, .  \nonumber
\end{align}
Note that from the explicit expressions \eqref{eq:su2gnereal}-\eqref{eq:su3gnereal}, we easily see that for the $s_1$ variable the series in both \eqref{eq:su2ansatz} and \eqref{eq:ansatzSU3} are actually sums over finitely many terms.

At this point we use the integral representation \eqref{eq:FIntRepMB} to rewrite infinite series like \eqref{eq:su2ansatz} and \eqref{eq:ansatzSU3} via the lattice sum integral representation
\begin{align}
   & \sum_{{s_1,s_2,s_3 = 1}}^\infty  d^{(N)}_{s_1,s_2,s_3} F_{-s_1-s_3,s_2,-2s_3}([\mathbbm{1}];\tau) \label{eq:ansatzGenRes} = \!\!\!\sum_{\substack{ (n,m)\in \mathbb{Z}^2 \\ n \neq0 }}  \int_0^\infty e^{ - t_1\frac{\tau_2}{\pi}} e^{ - t_2 \pi \frac{|n\tau+m|^2}{\tau_2}} e^{ - t_3 \pi \tau_2 n^2}  \mathcal{B}_N(t_1,t_2,t_3)\,{\rm d}^3 t\,,
   \end{align}
where we defined $\mathcal{B}_N(t_1,t_2,t_3) $ in \eqref{eq:BN} and we furthermore have included in the above series the Fourier zero-mode of $F_{-s_1-s_3,s_2,-2s_3}$ as well. 

With our conjectural expressions for the coefficients $d^{(2)}_{s_1,s_2,s}$ and $d^{(3)}_{s_1,s_2,s}$ given in \eqref{eq:su2gnereal}-\eqref{eq:su3gnereal}, we simply use the definition \eqref{eq:BN} to obtain the corresponding generating functions, $\mathcal{B}_N(t_1, t_2,t_3)$.  For $SU(2)$ we find, 
\begin{align}
\mathcal{B}_2(t_1, t_2,t_3) &= \mathcal{D}_x \Big[\frac{\, _1F_2\left(\frac{3}{2};2,\frac{5}{2}\vert -\frac{t_1}{4 x}\right)t_1+6 \, _0{F}_1\left(1\vert -\frac{t_1}{4 x}\right)}{24\, x^{3/2}} \Big] \,,\label{eq:B2}
\end{align}
where for convenience of presentation we have defined the differential operator
\begin{equation}
\mathcal{D}_x F(x)= 4\, t_3^{-7/2} \left\lbrace  \left[t_2 \left(t_2+t_3\right)-1\right]{}^2 \partial_x^2+ t_3 \left[t_2 \left(t_2+t_3\right)+1\right] \partial_x \right\rbrace F(x)\,  ,
\end{equation}
and have introduced the auxiliary variable $x$ is given by, 
\ie
x \coloneqq \frac{(1+t_2)(1+t_2+t_3)}{t_3} \, .
\fe
We note that $x$ is actually invariant under the inversion of $t_2,t_3$ defined in \eqref{eq:inversion}. With this property, it is straightforward to verify that $\mathcal{B}_2(t_1, t_2,t_3)$ indeed obeys the transformation rule \eqref{eq:inv1}. 
From \eqref{eq:B2}, we can furthermore derive the integral representation \eqref{eq:LatticeSumResMB2} expressed in terms of the alternative kernel ${\widetilde{\mathcal{B}}}_N$ defined in \eqref{eq:WN}, which for the case $N=2$ is given by
\begin{align}
&\notag {\widetilde{\mathcal{B}}}_2(y;t_2,t_3)  =  \int_0^\infty e^{-t_1 y} \mathcal{B}_2(t_1,t_2,t_3) {\rm d} y =\frac{ \exp\big[-\frac{t_3}{4 y (t_2+1)  (t_2+t_3+1)}\big]}{8 y^3 (t_2+1)^{11/2}  (t_2+t_3+1)^{11/2}} \times\\
&\notag  \Big\{12y^2 (t_2+1)^2 (t_2+t_3+1)^2 \Big[(t_2-3) (3 t_2-1) (t_2+1)^2+2 (3 (t_2-2) t_2-1) (t_2+1) t_3+t_2 (3 t_2-2) t_3^2\Big]\\
&\notag +2 y (t_2+1) (t_2+t_3+1) \Big[ (t_2-3) (3 t_2-1) (t_2+1)^4+(t_2 (3 t_2 (3 t_2-7)-1)-7) (t_2+1)^2 t_3\\
&\label{eq:Btilde2}   +t_2^2 (3 t_2-7) t_3^3+3 (t_2-2) t_2 (3 t_2-1) (t_2+1) t_3^2\Big] -t_3 [t_2 (t_2+t_3)-1]^2 [t_2 (t_2+t_3+2)+1]\Big\}\,.
\end{align}

Although the identity \eqref{eq:su2ansatz} shows that the integrated correlator can be written as an infinite sum over automorphic functions solely at the level of instanton sectors, as previously argued it is enough to remove the restriction $n\neq 0$ in \eqref{eq:ansatzGenRes} to obtain the lattice sum integral representation \eqref{eq:LatticeSumResMB2} for the complete integrated correlator.
In particular, we stress that the purely perturbative sector for the $N=2$ integrated correlator, derived from the $\mathcal{N}=2^*$ SYM matrix model in \eqref{eq:finalEx2}, can be retrieved identically from the integral representation \eqref{eq:It0simp} expressed in terms of $\mathcal{B}_2(t_1, t_2,t_3)$.
The two different integral representations \eqref{eq:finalEx2} and  \eqref{eq:It0simp} for the purely perturbative sector are simply related via the integral transform identity (see e.g. \cite{Dorigoni:2022zcr})
\begin{equation}\label{eq:IntTrId}
\int_0^\infty \frac{\tilde{B}(w)}{\sinh^2(w)} {\rm d} w = \sum_{\hat{m}=1}^\infty \int_0^\infty e^{- \hat{m}^2 t} B(t) {\rm d}t\,,\quad {\rm where}\quad B(t) = \frac{2 }{\sqrt{\pi t^3}}\int_0^\infty e^{-\frac{w^2}{t}} w\, \partial_w \tilde{B}(w) {\rm d}w\,.
\end{equation}

As briefly reviewed in section \ref{sec:EasyCorr}, the expectation value of a half-BPS 't Hooft loop in $\mathcal{N}=2^*$ SYM can be written via supersymmetric localisation \cite{Gomis:2011pf} in terms of a matrix model. However, monopole bubbling effects are only under full control for the case of $SU(2)$ gauge group. It would be extremely interesting to compute the integrated correlator of an 't Hooft line defect in $\mathcal{N}=4$ SYM starting from the $\mathcal{N}=2^*$ SYM supersymmetric localisation results of \cite{Gomis:2011pf} as in \eqref{eq:Iell} and check against the predicted expression \eqref{eq:LatticeSumtHooft} at finite coupling $\tau$. For general $N$, our expression \eqref{eq:LatticeSumtHooft} would provide for extremely stringent constraints against monopole bubbling effects in the $\mathcal{N}=2^*$ SYM theory with gauge group $SU(N)$.

Similar arguments hold for $SU(3)$.
Using the coefficients \eqref{eq:su3gnereal}, we find 
\begin{align}
 \mathcal{B}_3(t_1, t_2,t_3) = \mathcal{D}_x \Big\{&\frac{1}{960\, x^{11/2}} \Big[  t_1^2 \left[-24 (2 r+1) x+32 x^2+15\right] \,
   _1F_2\left(\frac{3}{2};3,\frac{7}{2}\Big\vert-\frac{t_1}{4 x}\right)  \cr 
&   +  2 t_1^2
   \left[-24 (2 r+1) x+32 x^2+15\right] \,
   _1F_2\left(\frac{5}{2};3,\frac{7}{2}\Big\vert-\frac{t_1}{4 x}\right) \cr 
   & +120 x^2
   \left[-24 (2 r+1) x+32 x^2+15\right] \, _0F_1\left(;1\Big\vert-\frac{t_1}{4
   x}\right)  \cr 
   &   +20 x t_1 \left[-48 (2 r+1) x+64 x^2+30\right] \,
   _0F_1\left(;2\Big\vert-\frac{t_1}{4 x}\right)  
   \cr &   +20 x t_1\left[  -16 (9 r+10) x^2+3 (68 r+49) x+96
   x^3-60\right] \Big] \Big\}\, , \label{eq:B3}
\end{align}
where the auxiliary variable $r\coloneqq t_2/t_3$ is also invariant under the inversion \eqref{eq:inversion}. It is then easy to check that $\mathcal{B}_3(t_1, t_2,t_3)$ obeys the transformation rule \eqref{eq:inv1}.
From \eqref{eq:B3} it is possible to derive the alternative integration kernel $\widetilde{\mathcal{B}}_3$ as defined in \eqref{eq:WN}. The structure is very similar to the $N=2$ case presented in \eqref{eq:Btilde2}, only the expression for $\widetilde{\mathcal{B}}_3$ is more cumbersome than its $N=2$ counterpart, and hence we do not present it here.

We stress again that although to derive the identity \eqref{eq:ansatzSU3} we only used the instanton sectors for both integrated correlator and automorphic functions, by simply removing the restriction $n\neq0$ from the lattice sum the expression \eqref{eq:ansatzSU3} gets promoted to the lattice sum integral representation \eqref{eq:LatticeSumResMB2} for the complete integrated correlator.
The purely perturbative sector for the $N=3$ integrated correlator, derived from the $\mathcal{N}=2^*$ SYM matrix model in \eqref{eq:finalEx3}, is reproduced identically from the integral representation \eqref{eq:It0simp} in terms of $\mathcal{B}_3$, thanks to the integral transforms identity \eqref{eq:IntTrId}.

These particular cases, $N=2$ and $N=3$, combined with the large-$N$ fixed-$\tau$ results derived in section \ref{sec:LargeN}, provide very strong evidences for the general conjectured lattice sum integral representation \eqref{eq:LatticeSumResMB2} for the line defect integrated correlator at arbitrary $N$ and $\tau$.

\section{Conclusions and Outlook}
\label{sec:conclusion}

In this paper, we investigated correlation functions of local operators in the presence of a half-BPS line defect for $\mathcal{N}$ = 4 SYM theory with gauge group $SU(N)$. These correlators can be obtained by taking derivatives with respect to either the coupling constant, $\tau$, or the deformation mass parameter, $m$, of the vacuum expectation value for a corresponding line defect in $\mathcal{N}=2^*$ SYM.
We considered two distinct classes of such physical observables. The first class is obtained by considering only derivatives with respect to the coupling constant $\tau$. This procedure yields non-integrated line defect correlators with some specific insertions of local operators, such as the superconformal primary operator $\mathcal{O}_2$ or the chiral Lagrangian $\mathcal{O}_\tau$, and with very particular kinematics. 

The second class of line defect correlators, which is the primary focus of this paper, has been introduced in \cite{Pufu:2023vwo} and it is constructed by taking two derivatives with respect to the mass deformation parameter of the expectation value of a half-BPS fundamental Wilson line defect in $\mathcal{N}=2^*$ SYM and then set the mass to zero. This yields a two point function of the superconformal primary operators $\mathcal{O}_2$ in the presence of a half-BPS Wilson-line defect where the space-time dependence of the local operators is integrated out using a carefully chosen integration measure.
The integration measure is implicitly determined by supersymmetric localisation formula and has been  worked out explicitly in \cite{Billo:2024kri, Dempsey:2024vkf}. In contrast with the integrated four point correlation functions of local operators  introduced in \cite{Binder:2019jwn,Chester:2020dja}, relatively little is known about these integrated line defect correlators. In particular, since line defect operators transform non-trivially under $\mathcal{N}=4$ SYM electromagnetic duality, such integrated correlators are no longer modular invariant functions of the coupling constant $\tau$ and it is unclear which class of automorphic objects, if any, we have to consider to describe these physical observables. The main goal of this paper is precisely addressing this question.

We constructed a novel class of automorphic functions whose properties under ${\rm SL}(2,\mathbb{Z})$ electromagnetic duality are exactly those of fundamental line defects in $\mathcal{N}=4$ SYM and showed that the particular integrated line defect correlators considered can be expressed in terms of elements belonging to this family of functions. 
This novel family of automorphic functions is introduced by considering a non-standard Poincar\'e series approach, where the associated seed functions display an explicit dependence from the electromagnetic charges $(p,q)$ carried by the line defect. The mathematical details of our constructions can be found in appendix \ref{sec:AppF}.

To support the proposal that integrated line defect correlators can be expressed in terms of this class of automorphic functions, we present compelling evidences in both the large-$N$ expansion at fixed $\tau$ and the finite-$N$ results for the fundamental Wilson-line defect integrated correlator:
\begin{itemize}
\item[(i)] In the large-$N$ fixed-$\tau$ limit, following the analysis of \cite{Pufu:2023vwo} we worked out the large-$N$ expansion of the integrated line defect correlators up to order $1/N^2$. We explicitly verified that each order in the $1/N$ expansion can be written as a finite linear combination with rational coefficients of the proposed automorphic functions.

\item[(ii)] For finite $N$, we exploit the instanton corrections to the integrated line defect correlator to conjecture a lattice sum integral representation which can be written as a formal series of exactly the same automorphic functions appearing at large-$N$. To provide evidences in support of our proposal, we present a detailed analysis for the cases with gauge group $SU(2)$ and $SU(3)$. 
Our expressions are very reminiscent of the exact results derived in \cite{Dorigoni:2021bvj,Dorigoni:2021guq} for the integrated correlators of four superconformal primary operators $\mathcal{O}_2$.
\end{itemize}
We stress that both at large $N$ and at finite $N$, our lattice sum integral representation for the integrated line-defect correlators in terms of automorphic objects is not a mere rewriting of matrix model results. The proposed expressions manifest very non-obvious and extremely intriguing modular structures for non-local physical observables due to the presence of a line-defect. 

One of the key consequences of our analysis is that we can predict the integrated correlators of two local $\mathcal{O}_2$ operators in the presence of a line defect with general electromagnetic charges, i.e. not necessarily a Wilson-line defect. This is a highly non-trivial result since a putative localisation formula is not known for general line defect. For the case of an 't Hooft loop with gauge group $SU(2)$ a concrete localisation formula does exist \cite{Gomis:2011pf}. It would be interesting to compute the $SU(2)$ integrated correlator of an 't Hooft line defect in $\mathcal{N}=4$ SYM starting from the $\mathcal{N}=2^*$ SYM supersymmetric localisation results of \cite{Gomis:2011pf} and check against our predicted expression \eqref{eq:LatticeSumtHooft} at finite coupling $\tau$. We have verified this agreement by performing a numerical evaluation of the matrix model integral for the $SU(2)$ 't Hooft line defect integrated correlator.

There are many future directions worth of attention. 
Firstly, the large-$N$ expansion at fixed $\tau$ is almost certainly an asymptotic formally divergent series in $1/N$. In the context of integrated correlators of four superconformal primary operators, it was shown in \cite{Dorigoni:2022cua} 
that a consistent large-$N$ expansion necessitates the presence of additional exponentially suppressed yet modular invariant corrections.
We believe something similar should happen for the large-$N$ expansion of integrated line defect correlators where, based on the results of \cite{Dorigoni:2022cua}, it is tantalising to conjecture that at large-$N$ we should encounter exponentially suppressed, yet automorphic non-perturbative corrections possibly of the form
\begin{equation}
\widetilde{D}_N(s_1,s_2,s_3;[\rho];\tau) = \frac{\tau_2^{s_1}}{|q\tau+p|^{2s_1}} \sum_{(n,m)\neq \mathbb{Z}(q,p)} e^{-\sqrt{\frac{4N |n\tau+m|^2}{\tau_2}}} \frac{\tau_2^{s_2}}{|n\tau+m|^{2s_2}} (np-mq)^{s_3}\,.
\end{equation}
When the systematic of the large-$N$ expansion for the integrated line defect correlator is under control, a suitable modification of the resummation procedure introduced in \cite{Dorigoni:2024dhy} should yield such non-perturbative and automorphic large-$N$ corrections.

On the other hand at finite $N$, the mathematical and physical structures of integrated line defect correlators remain largely unexplored. In contrast, integrated four-point correlators of local operators exhibit a plethora of elegant properties that have been well-studied. For instance, such physical observables have been shown to satisfy intriguing Laplace-difference equations \cite{Dorigoni:2021guq,Dorigoni:2022zcr,Dorigoni:2023ezg,Alday:2023pet}. These equations both link in a non-trivial manner different integrated correlators and provide for a set of recursive relations allowing us to determine some of these integrated correlators for all classical and exceptional gauge groups starting from the $SU(2)$ gauge group case.
We believe that the homogeneous Laplace equation \eqref{eq:DiffEqGen} satisfied by the elements of the presented class of automorphic functions will almost certainly play a key role in understanding possible similar Laplace-difference equations satisfied by the integrated line defect correlators at finite-$N$.
Thanks to the proposed lattice sum integral representation \eqref{eq:LatticeSumRes2}, rather than studying directly the integrated line defect correlator for arbitrary $N$ we can analyse the easier integrand functions $\mathcal{B}_N$, here presented explicitly for $N=2$ and $N=3$. Any putative Laplace-difference equation for the integrated correlator will be translated into a much simpler relation satisfied by the $\mathcal{B}_N$'s. For this reason it is definitely worth extending our analysis to integrated line-defect correlators for gauge groups with higher ranks by exploiting the general methodology outlined in section \ref{sec:FiniteN}.

Furthermore, in this paper we have only considered line defects in the fundamental representation of the gauge group $SU(N)$. It should be possible to generalise our results to line defects belonging to other representations of the gauge group as well as extending our analysis to more general classical (or possibly even exceptional) gauge groups. Importantly as discussed in the introduction, electromagnetic duality for extended operators does in general need to keep track of the global form of the gauge group as well as of possible discrete theta angles. 
However, for our observables to be sensitive to such data we have to consider $\mathcal{N}=4$ SYM on more complicated space-time manifolds, such as $\mathbb{RP}^4$ considered in
\cite{Wang:2020jgh,Caetano:2022mus,Zhou:2024ekb}.
We do not know which class of automorphic objects could possibly describe integrated line defect correlators in either higher representations or in such non-trivial space-time backgrounds.

From a broader perspective, we highlight that the $\mathcal{N}=4$ SYM bootstrap programme has been extended to include correlation functions in the presence of defects operators, see e.g. \cite{Liendo:2016ymz,Liendo:2018ukf,Ferrero:2021bsb, Barrat:2021yvp}. As discussed in \cite{Pufu:2023vwo}, the integrated correlators considered in this paper impose strong and exact (as functions of the coupling $\tau$) constraints on the two-point function line defect correlators. Similar to the case of integrated four-point functions of local operators, integrated line defect correlators furnish extremely valuable inputs for the defect-CFT conformal bootstrap programme both from an analytic and a numerical point of view. In this spirit, it would be of interest to consider integrated correlators involving more general superconformal primary operators with higher scaling dimensions. Unfortunately there is no obvious supersymmetric matrix model that would generalise \eqref{eq:Iw} to  these more general integrated line defect correlators. However, it may be feasible to try and address this problem by the approach similar to \cite{Brown:2023zbr} with inputs coming from high-order perturbative results for these more general integrated line defect correlators. Obtaining results for line defect correlators involving higher-dimension local operators would also lead to pathways for studying potential hidden relations between defect correlators in $\mathcal{N}=4$ SYM, analogous to what has been discovered for four-point correlators of local operators \cite{Caron-Huot:2018kta, Caron-Huot:2021usw}.

\section*{Acknowledgements}
We thank Gus Brown, Michael B.~Green, Abhiram Kidambi, Axel Kleinschmidt, Sungjay Lee, Hong Lü, Francesco Galvagno, Zhengxian Mei, Martin Raum, Victor A.~Rodriguez, Jaroslav Scheinpflug, Yifan Wang, Mitchell Woolley, and Junbao Wu for helpful discussions and to Francesco Galvagno for comments on the draft. We are particularly grateful to the
Galileo Galilei Institute for Theoretical Physics for the hospitality and the INFN for
partial support during the
GGI programme “Resurgence and Modularity in QFT and String Theory”.
DD and CW research was partially supported by the Munich Institute for Astro-, Particle and BioPhysics (MIAPbP) which is funded by the Deutsche Forschungsgemeinschaft (DFG, German Research Foundation) under Germany's Excellence Strategy – EXC-2094 – 390783311.
DD would like to thank the Albert Einstein Institute and in particular Axel Kleinschmidt, Hermann Nicolai and Stefan Theisen
for the hospitality and financial support during the final stages of this project. 
ZD would like to thank Tianjin University for hospitality, where a preliminary version of this work was presented. HX would like to thank New College, Oxford, and the organisers of Bootstrap 2024 in Madrid, for their hospitality during various stages of this work.
ZD is supported by an STFC Consolidated Grant, ST$\backslash$T000686$\backslash$1 ``Amplitudes, strings \& duality". CW is supported by a Royal Society University Research Fellowship,  URF$\backslash$R$\backslash$221015 and partly a STFC Consolidated Grant, ST$\backslash$T000686$\backslash$1 ``Amplitudes, strings \& duality".

\appendix

\section{Matrix Model Details}
\label{sec:AppMM}

In this appendix we summarise some of the more technical details regarding the matrix model calculations for the line defect integrated correlator following the setup introduced in section \ref{sec:EasyCorr}.

The main object under consideration is the $SU(N)$ hermitian matrix model whose partition function is given by
\begin{equation}
Z_N(\tau_2) \coloneqq   \int   e^{- 2\pi \tau_2 \sum_i a_i^2} \,\vdm(a_i)\,{\rm d}^{N-1} a  \,,\label{eq:NormApp}
\end{equation}
where the integral is over $N-1$ real variables parametrising the $SU(N)$ Cartan subalgebra and $\vdm(a_i)$ denotes the square of the standard Vandermonde determinant
\begin{equation}\label{eq:vdmApp}
\vdm(a_i) \coloneqq  \prod_{i < j}|a_i-a_j|^2 \,.
\end{equation}
Expectation values in the this matrix model are denoted using the double parenthesis notation $\llangle \mathcal{O}(a_k)\rrangle$ and are defined as 
 \begin{equation}\label{eq:MMApp}
 \llangle \mathcal{O}(a_k)\rrangle \coloneqq \frac{1}{  Z_{N}( \tau_2)} \int   e^{- 2\pi \tau_2 \sum_i a_i^2} \,\vdm(a_i)\, \mathcal{O}(a_k)\,{\rm d}^{N-1} a \,,
 \end{equation}
where in our normalisation we have $\llangle 1 \rrangle =1$.

From the $\mathcal{N}=2^*$ SYM partition function \eqref{eq:ZN2s} and the definition \eqref{eq:WilsonVEVN2} for the half-BPS Wilson-line defect expectation value, we see that the $\mathcal{N}=4$ SYM integrated correlator \eqref{eq:Iw} in the presence of a half-BPS Wilson-line defect can be decomposed as
\begin{equation} 
    \mathcal{I}_{\bW,N} (\tau) = \mathcal{I}_{\bW, N}^{ \text{pert}} (\tau) + \mathcal{I}_{\bW, N}^{ \text{inst}} (\tau)\,,
\end{equation}
where for the perturbative part we have
\begin{equation} \label{eq:IWpert}
   \mathcal{I}_{\bW, N}^{ \text{pert}} (\tau) = \frac{\llangle \bW(a_{i}) {Z}^{\prime \prime}_{\text{pert}}(0, a_{ij})\rrangle}{\llangle \bW(a_{i})\rrangle} -  \llangle {Z}^{\prime \prime}_{\text{pert}}(0, a_{ij})\rrangle\,,
\end{equation}
while for the instanton part we find
\begin{equation}\label{eq:IWinstGen}
    \mathcal{I}_{\bW, N}^{ \text{inst}} (\tau) = \frac{\llangle \bW(a_{i}) {Z}^{\prime \prime}_{\text{inst}}(0, a_{ij})\rrangle}{\llangle \bW(a_{i})\rrangle} -  \llangle {Z}^{\prime \prime}_{\text{inst}}(0, a_{ij})\rrangle + c.c.\,.
\end{equation}
In the above formulas ${Z}^{\prime \prime}_{\text{pert}}(0, a_{ij})$ and ${Z}^{\prime \prime}_{\text{inst}}(0, a_{ij})$ denote the two mass derivatives of the one-loop perturbative and non-perturbative instanton contributions of the partition functions in the supersymmetric localisation formula as given in \eqref{Zdef}, i.e.
\ie
Z^{\prime \prime}_{\text{pert}}(0, a_{ij}) \coloneqq \partial_m^2 \hat{Z}^{\rm pert}_N(m, a_{ij}) \vert_{m=0} \, , \qquad 
Z^{\prime \prime}_{\text{inst}}(0, a_{ij}) \coloneqq \partial_m^2 \hat{Z}^{\rm inst}_N(m, a_{ij}) \vert_{m=0} \, . \label{eq:Zpp}
\fe
Explicitly, they can be expressed as follows. For the perturbative part defined in \eqref{Zpertdef}, we have \cite{Russo:2013kea}, 
\ie
Z^{\prime \prime}_{\text{pert}}(0, a_{ij}) = K'(a_{ij})\, , \qquad {\rm with} \qquad
K'(z) \coloneqq  \int_0^{\infty} {2w [1-\cos(2wz)] \over \sinh^2(w)} \,{\rm d}w\,. \label{eq:Kprime}
\fe
The instanton part has been worked out in  \cite{Chester:2019jas} and it can be expressed as 
\ie
Z^{\prime \prime}_{\text{inst}}(0, a_{ij})  = \sum_{k=1}^\infty e^{2\pi i k \tau_1} Z^{\prime \prime(k)}_{\text{inst}}(0, a_{ij})  = {\sum_{k=1}^\infty e^{2\pi i k \tau} }\Big(\sum_{\substack{p,q>0\\ pq = k}} I_{p \times q}(a_i) \Big)\, , \label{eq:ZnekIns}
\fe
where $I_{p \times q}(a_i) $ is given as a contour integral in \eqref{eq:IIpq}. With these ingredients, in the next subsections we will compute  $\mathcal{I}_{\bW, N}^{ \text{pert}} (\tau)$ and   $\mathcal{I}_{\bW, N}^{ \text{inst}} (\tau)$ separately.

\subsection{Perturbative expansion at finite $N$}

We begin with the analysis for the perturbative contribution \eqref{eq:IWpert} to the integrated Wilson-line defect correlator. The disconnected term in \eqref{eq:IWpert}, given by $\langle Z^{\prime \prime}_{\text{pert}}(0, a_{ij})\rangle$, is essentially the perturbative contribution to the integrated four-point correlator which has been evaluated in \cite{Chester:2019pvm}, therefore we only focus on the first term in \eqref{eq:IWpert}. 

We start by noting that the ratio $\llangle \bW(a_{i}) Z^{\prime \prime}_{\text{pert}}(0, a_{ij})\rrangle /\llangle \bW(a_{i})\rrangle$ is independent on whether we consider an $SU(N)$ or a $U(N)$ gauge group and, thanks to invariance under permutations of the $a_i$, we can further simplify its evaluation by using the relation
\ie
 \llangle \bW(a_{i}) Z^{\prime \prime}_{\text{pert}}(0, a_{ij})\rrangle  = \sum_{i,j,k} {1\over N} \llangle e^{2\pi a_k} K'(a_{ij}) \rrangle &= \sum_{i,j} \llangle e^{2\pi a_1} K'(a_{ij}) \rrangle \, .
\fe
Using \eqref{eq:Kprime} the above expression becomes 
\ie \label{eq:totK}
\llangle \bW(a_{i}) Z^{\prime \prime}_{\text{pert}}(0, a_{ij})\rrangle  = \int^{\infty}_0\Big( \sum_{i,j} \llangle e^{2\pi a_1} \left[ 1 - \cos(2w a_{ij})\right] \rrangle \Big){2w \,{\rm d}w \over \sinh^2(w)}  \, ,
\fe
which we now proceed to evaluate term by term.

Expanding out the matrix model expectation value appearing at the integrand in \eqref{eq:totK} we can use
\ie
 \sum_{i,j}  \llangle e^{2\pi a_1}  \rrangle =  N^2\llangle e^{2\pi a_1}  \rrangle \, ,
\fe
while the contribution coming from $\cos(2w a_{ij})$ can be expressed in terms of
\ie \label{eq:terms}
& \sum_{i,j} \llangle e^{2\pi a_1} e^{2iw a_{ij}} \rrangle  
=N \llangle e^{2\pi a_1}  \rrangle + \Big[ \sum_{i>j>1} \llangle e^{2\pi a_1} e^{2iw a_{ij}} \rrangle  + \sum_{j>1} \llangle e^{2\pi a_1} e^{2iw a_{j1}} \rrangle+(w \rightarrow -w)\Big]  \cr 
&=N \llangle e^{2\pi a_1}  \rrangle +(N-1) \left[  {(N-2)\over 2} \llangle e^{2\pi a_1} e^{2iw a_{32}} \rrangle  +  \llangle e^{2\pi a_1} e^{2iw a_{21}} \rrangle+(w \rightarrow -w)\right] \, . 
\fe
Each expectation value can be computed using the method of Hermite orthogonal polynomials \cite{Mehta:1981xt}. In general, if the insertion $\mathcal{O}_n(a)$ only depends on a subset of the integration variables $a_i$, we can perform the gaussian integration over the variables that do not appear in $\mathcal{O}_n(a)$ and subsequently use 
\ie
\llangle \mathcal{O}_n(a) \rrangle  = {1\over N!} \sum_{\sigma \in S_N} \sum_{\mu \in S_n} (-1)^{|\mu|} \int \left( \prod_{i=1}^n  d a_i  { p_{\sigma(i)-1}(a_i)p_{\mu(\sigma(i))-1}(a_i) \over h_{\sigma(i)-1} }  e^{-2y \sum_i a_i^2} \right) \mathcal{O}_n(a) \,, 
\fe
where 
\ie
h_n &\coloneqq \sqrt{2\pi} n! (4y)^{-n-1/2}\,, \qquad
p_n(a)  \coloneqq (8y)^{-n/2} H_n(\sqrt{2y} a)\, ,
\fe
where $H_n(x)$ is the Hermite polynomial. Importantly, these functions satisfy the integral identities, 
\ie\label{eq:MMidentities}
\int da\, p_m(a) p_n (a) e^{-2y a^2} = h_n \delta_{mn}\, , \qquad 
\int da\, e^{-a^2 + x a} H_m(a)H_n(a) = e^{x^2\over 4} 2^m \sqrt{\pi} m! x^{n-m} L_m^{n-m}(-x^2/2)\, ,
\fe
where $L_m^{n}(x)$ is the generalised Laguerre polynomial. 

We now consider each term in \eqref{eq:terms} separately. 
Firstly we have 
\ie
\llangle e^{2\pi a_1}  \rrangle &= {1\over N!} \sum_{\sigma \in S_N}  \int     d a_1  { p_{\sigma(1)-1}(a_1)p_{\sigma(1)-1}(a_1) \over h_{\sigma(1)-1} }  e^{-2y a_1^2}   e^{2\pi a_1} \cr
&={1\over N!} \sum_{\sigma \in S_N}  \int     d a_1  {(8y)^{1-\sigma(1)} H_{\sigma(1)-1}(\sqrt{2y}a_1)H_{\sigma(1)-1}(\sqrt{2y}a_1) \over \sqrt{2\pi} (\sigma(1)-1)! (4y)^{-\sigma(1)+1/2}  }  e^{-2y a_1^2}   e^{2\pi a_1} \, , 
\fe
which after having used \eqref{eq:MMidentities} it reduces to
\ie\label{eq:Wvev}
\llangle e^{2\pi a_1}  \rrangle &=e^{\pi \over 2\tau_2} {1\over N!} \sum_{\sigma \in S_N}        L_{\sigma(1)-1} \left(-{\pi \over \tau_2} \right)={1\over N}\, e^{\pi \over 2 \tau_2}        L_{N-1}^1 \left(-{\pi \over \tau_2} \right)\, ,
\fe
a well-known result \cite{Drukker:2000rr} for the expectation value of a circular half-BPS fundamental Wilson-line defect $\langle \mathbb{W} \rangle$ in $\mathcal{N}=4$ SYM $U(N)$. For the case of $SU(N)$ we simply need replacing the exponential factor $e^{\pi^2 \over 2y}$ in \eqref{eq:Wvev} by $e^{(N-1)\pi^2 \over 2Ny}$; this minor difference between $U(N)$ and $SU(N)$ cancels out in the final result of the integrated Wilson-line correlator.

Similar computations can be done for the other terms in \eqref{eq:terms}, and we find that the expectation value $\llangle e^{2\pi a_1} e^{2iw a_{21}}  \rrangle$ can be written as a sum of products of two Laguerre polynomials, whereas $\llangle e^{2\pi a_1} e^{2iw a_{32}} \rrangle$ is given by a similar sum of products of three Laguerre polynomials.  Putting everything together, we obtain the final expression for the perturbative contribution to the integrated two-point function in the presence of a Wilson-line defect, valid for general $N$ and arbitrary coupling constant, 
\ie \label{eq:finalEx}
\mathcal{I}^{\rm pert}_{\mathbb{W}, N} (\tau_2) = \int_0^{\infty} {2wdw \over \sinh^2(w)} \left[  { P_1(w) +P_2(w) +(w \rightarrow -w) \over \langle e^{2\pi a_1}  \rangle  } - Z^{\prime \prime}(w) \right] \, ,
\fe
where $P_1(w)$ encodes the contribution from $\llangle e^{2\pi a_1} e^{2iw a_{32}} \rrangle$, given by
\begin{align}
& P_1(w) \\
&\notag =e^{\frac{\pi ^2-2 w^2}{2 y}}\!\! \!\sum_{i,j,k=1}^N \!\!\! { (-1)^{j+k} \over 2N }\!\!  \left\lbrace L_{i-1}\left(-\frac{\pi ^2}{y}\right) \left[(-1)^{j+k} L_{j-1}\left(\frac{w^2}{y}\right)
   L_{k-1}\left(\frac{w^2}{y}\right)-L_{j-1}^{k-j}\left(\frac{w^2}{y}\right) L_{k-1}^{j-k}\left(\frac{w^2}{y}\right)\right] \right.
   \cr
  & \notag \quad  +\left(\frac{i\, w}{\pi
   }\right)^{i-j}  L_{i-1}^{j-i}\left(-\frac{\pi ^2}{y}\right) \left[L_{k-1}^{i-k}\left(\frac{w^2}{y}\right)
   L_{j-1}^{k-j}\left(\frac{w^2}{y}\right)-(-1)^{i+k} L_{k-1}\left(\frac{w^2}{y}\right)
   L_{j-1}^{i-j}\left(\frac{w^2}{y}\right)\right] 
   \cr
   & \quad \left. +\left(\frac{i\, w}{\pi }\right)^{i-k} \! (-1)^{j+k}  L_{i-1}^{k-i}\left(-\frac{\pi ^2}{y}\right)
   \left[(-1)^{i+j} L_{j-1}^{i-j}\left(\frac{w^2}{y}\right) L_{k-1}^{j-k}\left(\frac{w^2}{y}\right){-}L_{j-1}\left(\frac{w^2}{y}\right)
   L_{k-1}^{i-k}\left(\frac{w^2}{y}\right)\right]\right\rbrace\, , 
\end{align}
while $P_2(w)$ contains the contribution from $\llangle e^{2\pi a_1} e^{2iw a_{21}}  \rrangle$,
\begin{align}
&P_2(w) =\\
&\notag  e^{\frac{\pi^2-2 w^2-2 i \pi  w}{2 y}} \! \sum_{i,j=1}^N \left[ L_{i-1}\!\left(-\frac{(\pi -i w)^2}{y}\right) L_{j-1}\!\left(\frac{w^2}{y}\right){-}\left( {\pi \over i w} -1 \right)^{j-i}  L_{i-1}^{j-i}\!\left(-\frac{(\pi -i w)^2}{y}\right) L_{j-1}^{i-j}\!\left(\frac{w^2}{y}\right)\right] \, . 
\end{align}
Finally, $Z^{\prime \prime}(w) $ is related to the disconnected term $\llangle Z^{\prime \prime}_{\text{pert}}(0, a_{ij})\rrangle$ which is known to be
\ie
Z^{\prime \prime}(w)  = e^{-\frac{w^2}{y}} \left(L_{n-1}^1\left(\frac{w^2}{y}\right){}^2-\sum_{i,j=1}^N (-1)^{i-j} L_{i-1}^{j-i}\left(\frac{w^2}{y}\right)
   L_{j-1}^{i-j}\left(\frac{w^2}{y}\right)\right) \, . 
\fe
For small values of $N$, the perturbative contribution $\mathcal{I}^{\rm pert}_{\mathbb{W}, N} (\tau_2)$ simplifies. For example, the cases $N=2$ and $N=3$ are used explicitly in the main text and are given by  
\begin{align}
\label{eq:finalEx2}
\mathcal{I}^{\rm pert}_{\mathbb{W}, 2} (\tau_2) =& \frac{1}{y \left(2 y+\pi ^2\right)} \int_0^{\infty}   e^{-\frac{w^2}{y}} \Big\{ y \left[2 y+(\pi -2 i w)^2\right]e^{\frac{-i \pi  w}{y}}+  \cr
 & 2 \left(2 y+\pi ^2\right)  \left(2 w^2-y\right) +y
    \left[2 y+(\pi +2 i w)^2\right]e^{\frac{ i \pi  w}{y}}\Big\}  {2w\, {\rm d}w \over \sinh^2(w)}\, ,
\end{align} 
and similarly 
\begin{align}
    \label{eq:finalEx3}
&\notag\mathcal{I}^{\rm pert}_{\mathbb{W}, 3} (\tau_2) = \frac{1}{y^3 \left(6 y^2+6 \pi ^2 y+\pi ^4\right)} \int_0^{\infty}   e^{-\frac{w^2}{y}} \Big\{ 12  \pi y^2  w \left[\pi ^2 \left(w^2-2 y\right)-\left(w^4-3 w^2 y+3 y^2\right)\right] \sin
   \left(\frac{\pi  w}{y}\right)  \\
 &\notag   \qquad \!\!\!   +2y^2\left[-4 w^6+18 w^4 y-36 w^2 y^2-\pi ^4 \left(w^2-2 y\right)+\pi ^2 \left(13 w^4-30 w^2 y+12 y^2\right)+12 y^3\right]
   \cos \left(\frac{\pi  w}{y}\right)\\
 &  \notag\qquad\!\!\!
+4 y^2 \left(2 w^6-9 w^4 y+18 w^2 y^2-6 y^3\right)   +2 \pi ^2 y \left(6 w^6-31 w^4 y+48 w^2 y^2-12 y^3\right)\\
& \qquad\!\!\!+\pi ^4
   \left(2 w^6-9 w^4 y+14 w^2 y^2-4 y^3\right) \Big\}  {2w\,{\rm d}w \over \sinh^2(w)} \, . 
\end{align} 

\subsubsection*{Large-$N$ expansion}

The perturbative contribution \eqref{eq:IWpert} admits a large-$N$ 't Hooft expansion, 
\ie\label{eq:genus}
\mathcal{I}^{\rm pert}_{\mathbb{W}, N}(\lambda)= \sum_{m=0}^{\infty} {1 \over N^{2m}}  \mathcal{I}^{{\rm pert}, m}_{\mathbb{W}}  (\lambda) \, .
\fe
In the main text we are interested in the large-$N$ fixed-$\tau$ expansion of the Wilson-line defect integrated correlator. To compute the zero-instanton sector in this regime, i.e. the perturbative terms in $\tau_2$, we need examining the expansion of $\mathcal{I}^{{\rm pert},m}_{\mathbb{W},m}(\lambda)$ in the limit of strong coupling $\lambda\gg 1$. This analysis has been discussed in detail in  
\cite{Pufu:2023vwo}, and the results are given in equation (2.25) of that reference hence we do not repeat the computation here.  However, we highlight that we have found some disagreement with \cite{Pufu:2023vwo} at order $1/N^6$ , i.e. for ${\cal I}^{\text{pert}, 3}_{\mathbb{W}}$. Our calculations for the large-$\lambda$ expansion of ${\cal I}^{\text{pert}, 3}_{\mathbb{W}}$ yields
\ie \label{eq:Iw3correct}
     {\cal I}^{\text{pert}, 3}_{\mathbb{W}} (\lambda)  = -\frac{\lambda ^{9/2}}{1935360} -\frac{\lambda
   ^4}{430080}-\frac{27 \lambda
   ^{7/2}}{2293760} +\frac{13 \lambda
   ^3}{80640} +  \cdots \, ,
        \fe
and the discrepancy with the aforementioned reference starts at order $\lambda^4$. In \cite{Pufu:2023vwo} the large-$\lambda$ expansion is obtained by rewriting each genus expansion coefficient in \eqref{eq:genus} as an integral involving products of two Bessel functions, in particular see their equation (B.3) for the order $1/N^6$. The product of two Bessel functions is then replaced by its Mellin-Barnes representation and the large-$\lambda$ expansion is obtained by closing the contour of integration and evaluating the residues. We believe the disagreement we have found is due to the fact that there are additional poles which contribute in this case and that are missed in \cite{Pufu:2023vwo}.\footnote{The discrepancy can also be verified by performing a numerical integration of equation (B.3) in \cite{Pufu:2023vwo} for $\lambda$ large and comparing it with the asymptotic expansion \eqref{eq:Iw3correct}.} As we argue in the main text after equation \eqref{eq:IW4}, the coefficient of the $\lambda^4$ term in equation \eqref{eq:Iw3correct} is indeed consistent with the expected behaviour of the large-$N$ large-$\lambda$ expansion of the 't Hooft-line defect correlator, which can be obtained from the $S$ transformation $\tau\to -1/\tau$ of the Wilson-line defect correlator. 

\subsection{Instanton contributions at large-$N$ and fixed $\tau$}
\label{sec:AppLargeNMM}

In this section we discuss how to compute the instanton contributions to the integrated Wilson-line defect correlator, $\mc{I}_{\mathbb{W},\text{inst}}(\tau)$,  in a large-$N$ expansion at fixed $\tau$.

We start by reminding the reader of some important properties of the large-$N$ expansion for matrix model integrals which can be derived from the method of topological recursion introduced in \cite{Eynard:2004mh}, see also \cite{Eynard:2007kz,Eynard:2008we}. 
As it will become clear shortly, to analyse the large-$N$ asymptotic expansion of the $k$-instanton sector of the Wilson-line defect integrated correlator \eqref{eq:IWinstGen}, we need to control all matrix model correlation functions of the form 
\begin{equation}\label{trtrtr}
 \exlc{{\rm Tr}\frac{1}{A_x^{n_1}}{\rm Tr}\frac{1}{A_x^{n_2}}\dots {\rm Tr}\frac{1}{A_x^{n_k}}}\, ,\end{equation}
computed from the hermitian matrix model \eqref{eq:MM} where we defined the diagonal matrix $A_x\coloneqq a- x\cdot\mathbbm{1}$ with $a=\mathrm{diag}(a_1,\dots,a_N)$ and $x\in \mathbb{C}$.
With the subscript ${\rm c}$ we denote the connected contribution to the correlation function. Using the differential identity \begin{equation}\label{trick}\frac{1}{A_x^n}=\frac{1}{(n-1)!}\partial^{n-1}_x\frac{1}{A_x}\, ,\end{equation}
we can rewrite the generic matrix model contribution \eqref{trtrtr} as
\begin{align}
\exlc{{\rm Tr}\frac{1}{A_x^{n_1}}{\rm Tr}\frac{1}{A_x^{n_2}}\dots {\rm Tr}\frac{1}{A_x^{n_k}}} =N^{2-k}  (-1)^k \Big[ \Big(\prod_{i=1}^k \frac{1}{(n_i-1)!} \partial_{x_i}^{n_i-1}\Big) W^k(x_1,\dots,x_k)\Big]_{x_1=\dots=x_k=x} \, , 
\end{align}
where we have introduced the $k$-body matrix model resolvents
\begin{align}
W^k(x_1,\dots,x_k)\coloneqq  N^{k-2} (-1)^k \exlc{ {\rm Tr}\frac{1}{A_{x_1}} \dots {\rm Tr}\frac{1}{A_{x_k}}} = N^{k-2} \exlc{\Big(\sum_{i_1=1}^N\frac{1}{x_1-a_{i_1}}\Big) \cdots \Big(\sum_{i_k=1}^N\frac{1}{x_k-a_{i_k}}\Big)} \label{eq:resolv} \, .
\end{align}
Thanks to topological recursion \cite{Eynard:2004mh,Eynard:2007kz,Eynard:2008we}, it is possible to show that each resolvent admits the large-$N$ genus expansion
\begin{align}
W^k(x_1,\dots,x_k) = \sum_{\ell=0}^{\infty} N^{-2\ell} W_\ell^k(x_1,\dots,x_k)  \, . 
\end{align}
and surprisingly all higher-body, higher-genus resolvents can be constructed from the one-body genus zero resolvent
\begin{equation}
W^1_0(x_1) = 2 \mu (\mu x_1) \Big( 1 - \sqrt{1 - (\mu x_1)^{-2}}\Big)\,,\label{eq:W10}
\end{equation}
where we defined
\begin{equation}
 \mu \coloneqq \frac{2\pi}{\sqrt{\lambda}}  = \sqrt{\frac{\pi \tau_2}{N}}=\sqrt{\frac{y}{N}}\,.
 \end{equation}
The functions $W_\ell^k(x_1,\dots,x_k)$ are homogeneous in the $x_i$ and can be expressed as 
\ie \label{eq:WF-rel}
W_\ell^k(x_1,\dots,x_k) = \mu^k F_\ell^k(\mu x_1,\dots,\mu x_k) \, .
\fe
We then conclude that the generic matrix model term \eqref{trtrtr} at the leading order in the large-$N$ expansion with fixed $\tau$ grows like
\ie 
\exlc{{\rm Tr}\frac{1}{A_x^{n_1}}{\rm Tr}\frac{1}{A_x^{n_2}}\dots {\rm Tr}\frac{1}{A_x^{n_k}}}\sim \mu^{n_1+\dots+n_k}N^{2-k}\sim N^{2-k-(n_1+\dots+n_k)/2}\, ,
\fe
since each contribution $F_\ell^k(t_1,...,t_k)$ in \eqref{eq:WF-rel} is finite in the limit $t_i\to 0 $.

We now move to show that the large-$N$ expansion at fixed $\tau$ of the $k$-instanton sector \eqref{eq:IWinstGen} for the integrated Wilson-line defect can be extracted from the expansion at large $A_x$ of matrix model correlators precisely of the form \eqref{trtrtr}.
We begin by analysing the large-$N$ expansion of the $k$-instanton contribution to \eqref{eq:IWinstGen} coming from the small-mass expansion of Nekrasov partition function \eqref{eq:ZnekIns}, denoted by $I_{p\times q}$.
Following \cite{Pufu:2023vwo}, we introduce a slightly different integral representation for  $I_{p\times q}$ compared to equation \eqref{eq:IIpq}, 
\begin{equation}\label{Ipq} I_{p\times q}(a) = \int^{\infty+i \epsilon}_{-\infty+i \epsilon}\Biggl[ \big(c_{p, q} + \mathcal{B}(z) \big) e^{\mathcal{A}(z)}  - c_{p, q} \Biggr]  \frac{{\rm d}z}{2 \pi}\,,\end{equation}
 where the constant $c_{p, q}$ and the functions $\mathcal{A}$ and $\mathcal{B}$ are defined as
\ie\label{eq:cAB}
c_{p, q} &\coloneqq 2  \left( \frac{1}{p^2} + \frac{1}{q^2} \right) \, , \qquad
\mathcal{A}(z) 
\coloneqq \sum_{j=1}^{N}\log{\frac{[z-a_j+(p-1)i]\,[z-a_j+(q-1)i]}{(z-a_j-i)[z-a_j+(p+q-1)i]}}\, , \\
\mathcal{B}(z) &\coloneqq  \sum_{j=1}^N \frac{ i (p+q)(q-p)^2 }{pq\,[z  -  a_j  +  (p  +  q  -  1)i ]\,[z  -  a_j  +  ( q  -  1)i ] \,[z  -  a_j + (p  -  1)i ] } \, .
\fe
The integration contour is chosen in such a way as to separate the poles located at $z= a_j+i$ from the remaining poles located at $z=a_j-i(p+q-1)$, i.e. we need $-(p+q-1)<\epsilon<1$.
Note that  when compared to the original contour integral appearing in \eqref{eq:k-matrix}, the integration contour for $I_{p\times q}$ has now been changed to a horizontal line in the complex $z$-plane.
This change in the integration contour can be performed only after having subtracted the constant term $c_{p,q}$ in \eqref{Ipq} as to make the integrand decay sufficiently fast at infinity $|z|\to\infty$. This subtraction does not modify the contour integral \eqref{eq:IIpq} and it allows us to `open' up the integration contour to any horizontal line \cite{Chester:2019jas} separating the poles located at $z=a_j+i$ from the remaining ones located at $z=a_j-i(p+q-1)$.

At this point we find it convenient to shift the integral to a more symmetric configuration by changing integration variables to $z=x +(1-\frac{p+q}{2})i$, so that the $x$ integration contour can be chosen to lie on the real $x$-axis and $I_{p\times q}$ becomes
\ie \label{eq:Ipqim}
I_{p\times q}(a) =\int_{-\infty}^{\infty} \Big[ \left(c_{p,q}+\frac{8sd^2}{k}{\rm Tr}\frac{-iA_x+s\,\mathbbm{1}}{(A_x^2+d^2\,\mathbbm{1})(A_x^2+s^2\,\mathbbm{1})}\right)\mathrm{Det}\Big(\frac{A_x^2+d^2\,\mathbbm{1}}{A_x^2+s^2\,\mathbbm{1}}\Big)-c_{p,q}\Big]\frac{{\rm d}x}{2\pi} \, , 
\fe
where we defined $s\coloneqq(p+q)/2$ and $d\coloneqq(p-q)/2$. It is important to note that the imaginary part of the integrand is proportional to $iA_x$ and cannot possibly produce a non-vanishing contribution after integration over $x$ since the original expression for the Nekrasov partition function is real. Although we do not have a direct proof for arbitrary $N$, it is straightforward to verify this fact by working at finite $N$ and evaluate explicitly the contribution originating from the factor proportional to $iA_x$ which vanishes identically after integrating over $x$. Therefore, in what follows we simply discard the term proportional to $i A_x$, and work with the expression
\ie \label{eq:IpqNew}
I_{p\times q}(a)=\int_{-\infty}^{\infty} \Big[ \left(c_{p,q}+ \frac{8 s^2 d^2}{k} {\rm Tr}\frac{1}{(A_x^2+d^2\,\mathbbm{1})(A_x^2+s^2\,\mathbbm{1})}\right)\mathrm{Det}\Big(\frac{A_x^2+d^2\,\mathbbm{1}}{A_x^2+s^2\,\mathbbm{1}}\Big)-c_{p,q} \Big]\frac{{\rm d}x}{2\pi}\, . 
\fe
We find this form for the $k$-instanton contribution coming from Nekrasov partition function to be the most convenient for deriving the large-$N$ expansion at fixed $\tau$.

As previously discussed, in the large-$N$ limit we have that the variable $x$ appears in the matrix model resolvents \eqref{eq:WF-rel} always in the combination $\mu x$ with $\mu = \sqrt{y/N}$. 
We then introduce the rescaled variable $t= \mu x$ and consider the expansion for large $A_x = a_i - t/\mu $ in the complex $t$ plane.
As a function of the complex variable $t$, the integrand of \eqref{eq:IpqNew} has poles located at $t=\mu (a_j \pm i\,s)$, so the large $A_x$ expansion converges only when
\ie 
|t-\mu a_j|>\mu s\qquad \forall\, a_j\, . 
\fe
This means that in the $N\to \infty$ limit the excluded region shrinks to the segment $t\in [-1,1]$ on the real axis. One can furthermore show that the contribution to the integral coming from this interval can be ignored in the $N \to \infty$ limit since it is exponentially suppressed in $N$. 

From our expression \eqref{eq:IpqNew}, we immediately see that the functions $\mathcal{A}$ and $\mathcal{B}$ (with a slight abuse of notation we use the same symbol $\mathcal{B}$ even if we have discarded the term $-i A_x$ which vanishes upon integration) defined in \eqref{eq:cAB} can be rewritten conveniently as
\begin{align} \label{eq:newAB}
\mathcal{A}(x)={\rm Tr}\log{\frac{A_x^2+d^2\mathbbm{1}}{A_x^2+s^2\mathbbm{1}}}\, , \qquad \mathcal{B}(x)=\frac{8s^2d^2}{k^2} {\rm Tr}\frac{1}{(A_x^2+s^2\mathbbm{1})(A_x^2+d^2\mathbbm{1})} \, ,
\end{align}
so that it becomes a straightforward task to expand either of them as an infinite sums of powers of traces ${\rm Tr}\,{A_x^{-2n}}$. 

We can now consider the full $k$-instanton sector of the Wilson-line defect integrated correlator \eqref{eq:k-matrix} where we replace $I_{p\times q}$ with \eqref{eq:IpqNew}.
To compute the large-$N$ expansion of \eqref{eq:k-matrix} we furthermore need the identities  
\begin{align}\label{Wtrtr}
\exg{\mathbb{W}e^\mathcal{A}}-\exg{ \mathbb{W}} \exg{e^\mathcal{A}} &=\Big[\exg{\mathbb{W}e^\mathcal{A}}_c-\exg{ \mathbb{W}}\Big]e^{ \langle\hspace{-0.19em}\langle e^\mathcal{A}\rangle\hspace{-0.19em}\rangle_c-1}\, ,\\
\exg{ \mathbb{W}e^\mathcal{A}\mathcal{B}}-\exg{ \mathbb{W}}  \exg{ e^\mathcal{A}\mathcal{B}} &= \Big[ \exg{ \mathbb{W}e^\mathcal{A}\mathcal{B}}_c+\Big( \exg{ \mathbb{W}e^\mathcal{A}}_c  - \exg{ \mathbb{W}} \Big) \exg{ e^\mathcal{A}\mathcal{B}}_c \Big]e^{ \langle\hspace{-0.19em}\langle e^\mathcal{A}\rangle\hspace{-0.19em}\rangle_c-1}\, . \label{Wtrtr2}
\end{align}
We then conclude that equation \eqref{eq:k-matrix} can be rewritten as
\begin{align} 
& \frac{\exg{ \mathbb{W}I_{p\times q}}-\exg{ \mathbb{W}}\exg{ I_{p\times q}}}{\exg{ \mathbb{W}}} \cr 
=&\, \frac{1}{\mu \exg{ \mathbb{W}}}\!\int\limits_{\,\,\,\mathbb{R}\setminus (-1,1)} \Big[\left( \exg{ \mathbb{W}e^\mathcal{A}}_c  -\exg{ \mathbb{W}} \right) \left( c_{p,q}+\exg{ e^\mathcal{A}\mathcal{B}}_c \right)+\exg{ \mathbb{W}e^\mathcal{A}\mathcal{B}}_c \Big] e^{ \langle\hspace{-0.19em}\langle e^\mathcal{A}\rangle\hspace{-0.19em}\rangle_c-1}\,\frac{{\rm d}t}{2\pi}\, . \label{eq:WIpq}
\end{align}
To extract the large-$N$ fixed-$\tau$ expansion of \eqref{eq:WIpq} we need to analyse each term separately. 
Firstly, as already noted in \cite{Chester:2019jas}, at large $N$ and fixed $\tau$ we find  
\ie 
e^{ \langle\hspace{-0.19em}\langle e^\mathcal{A}\rangle\hspace{-0.19em}\rangle_c-1} = \exp \left[ 2 k y\left( 1-\frac{|t|}{\sqrt{t^2-1}} \right) \right]+O\(\frac{1}{N}\)  \, ,\label{eq:expAasy}
\fe
using the notation $y= \pi \tau_2 = 4\pi^2 / g_{_{{\rm YM}}}^2$.
The presence of $|t|$ is due to the requirement that all the resolvents, at all orders in the genus expansion, must be analytic functions on the complex plane except for a cut on the segment $[-1,1]$.

Secondly, in \eqref{eq:WIpq} we immediately see the presence of Wilson-line defect insertions which do not immediately seem to fall in the same category as the correlators \eqref{trtrtr} previously discussed. We now show that these terms can be reconstructed as well from the resolvents and at a given order in $1/N$ only a finite number of terms need to be kept in the expansion. 
We keep the present discussion rather schematic in nature without worrying about the precise details, while in the next subsections we will show concretely how to extract the large-$N$ expansion of \eqref{eq:WIpq} up to order $1/N^2$ 

Since we have already argued that both $\mathcal{A}$ and $\mathcal{B}$ can be expanded in terms of mult-trace correlators \eqref{trtrtr}, the typical contribution to \eqref{eq:WIpq} that we still need to analyse takes the form, 
\begin{align}
&\label{eq:GenCont}C \coloneqq \exlc{\mathbb{W}\, {\rm Tr}\frac{1}{A_x^{n_1}}\dots {\rm Tr}\frac{1}{A_x^{n_k}}}=
 (-1)^k \Big[ \Big(\prod_{i=1}^k \frac{1}{(n_i-1)!} \partial_{x_i}^{n_i}\Big) \exlc{\mathbb{W}(a)\,{\rm Tr}\frac{1}{A_{x_1}}\dots {\rm Tr}\frac{1}{A_{x_k}}} \Big]_{x_1=\dots=x_k=x}\,.
\end{align}
 We now proceed to outline a procedure for evaluating the large-$N$ expansion of the above expression. Since the expressions will become rather cumbersome whenever we use the symbol $\sim$ rather than an equal sign it means that we have dropped numerical factors which do not affect the large-$N$ scaling.
 Firstly, from the definition of the resolvents \eqref{eq:resolv} we immediately see that the matrix model expectation value with an additional insertion of $\mathbb{W}(a)=N^{-1}{\rm Tr}\,e^{2\pi a}$ can be rewritten as an inverse Laplace transform of an augmented resolvent, i.e. we can write 
\begin{equation}
 \exlc{\mathbb{W}\,{\rm Tr}\frac{1}{A_{x_1}}\dots {\rm Tr}\frac{1}{A_{x_k}}} = (-N)^{-k} \int_{{\rm Re}\,w>1}  W^{k+1}(w,x_1,\dots,x_k)e^{2\pi w} \, \frac{{\rm d}w}{2\pi i}\,.\label{eq:InvLap}
\end{equation}
Hence the generic contribution \eqref{eq:GenCont} can be written as
\ie  
C \sim N^{-k}\Big[\partial_{x_1}^{n_1-1}\dots \partial_{x_k}^{n_k-1}\int_{{\rm Re}\,w>1}  W^{k+1}(w,x_1,\dots,x_k)e^{2\pi w} \, \frac{{\rm d}w}{2\pi i }\Big]_{x_1=...=x_k=x}\,, 
\fe
where the integral is performed along any vertical line such that ${\rm Re}\, w >1$ . Using \eqref{eq:WF-rel} and making a change of integration variable $w \to w/\mu$ combined with $t_i = \mu x_i$,  we obtain
\ie 
C \sim \sum_{\ell=0}^\infty N^{-k-2\ell}\mu^{n_1+\dots +n_k} \Big[ \partial_{t_1}^{n_1-1}\dots \partial_{t_k}^{n_k-1}\int_{{\rm Re}\,w>1}    F^{k+1}_\ell (w,t_1,\dots,t_k) \, e^{\frac{2\pi w}{\mu} } \,{\rm d}w \Big]_{t_1=...=t_k=t}\, .
\fe
In the complex $w$-plane, $F^{k+1}_\ell (w,t_1,...,t_k)$ has a branch cut along the interval $[-1,1]$ with a leading singularity of the form
\begin{equation}
\Big[\partial_{t_1}^{n_1-1}\dots \partial_{t_k}^{n_k-1} F^{k+1}_\ell (w,t_1,...,t_k)\Big]_{t_1=...=t_k=t} = \frac{\hat{F}^{k+1}_{\ell}(w,t)}{(w^2-1)^{k+3\ell-\half}}\,,
\end{equation}
with $\hat{F}^{k+1}_{\ell}(w,t)$ a polynomial in $w$ such that $\hat{F}^{k+1}_{\ell}(1,t)\neq 0$. 
We then arrive at
\ie 
C & \sim \sum_{\ell=0}^\infty N^{-k-2\ell}\mu^{n_1+\dots+ n_k}\int_{{\rm Re}\,w>1}  \frac{\hat{F}^{k+1}_{\ell}(w,t)}{(w^2-1)^{k+3\ell-\half}}e^{\frac{2\pi w}{\mu}} \cr
& \sim \sum_{\ell=0}^\infty N^{-k-2\ell}\mu^{n_1+\dots +n_k}\int_{{\rm Re}\,w>1}  \hat{F}^{k+1}_{\ell}(w,t)\partial_w^{(k+3\ell-1)}\[ \frac{p_{k+3\ell}(w)}{(w^2-1)^{\half}}\]\, e^{\frac{2\pi w}{\mu}} {\rm d}y \, ,
\fe
where we have rewritten  the singular term $(w^2-1)^{-k-3\ell+\half}$ in terms of a differential operator acting on $ p_{k+3\ell}(w)\,(w^2-1)^{-\half}$, with $p_{k}(w)$ a polynomial of degree $k-1$ given by
\begin{equation}
p_k(w) =\frac{\sqrt{\pi }  }{2^k \Gamma \left(k-\frac{1}{2}\right)}\sum_{\ell=0}^k \binom{k}{\ell} (-w)^{k-\ell} \left\lbrace[ (-1)^{\ell}-1](w^2-1)^{\frac{\ell+1}{2}}-[(-1)^{\ell}+1] w (w^2-1)^{\frac{\ell}{2}}\right\rbrace\,.
\end{equation}
We integrate by parts $k+3\ell-1$ times and then close the contour of integration to the left half-plane arriving at
\ie 
C \sim \sum_{\ell=0}^\infty N^{-k-2\ell}\mu^{n_1+\dots +n_k}\int_{\gamma} \partial_w^{(k+3\ell-1)}\big[ \hat{F}^{k+1}_{\ell}(w,t)e^{\frac{2\pi w}{\mu}}\big]\frac{p_{k+3\ell}(w)}{(w^2-1)^{\half}}\, {\rm d} w\, ,
\fe
where the contour of integration $\gamma$ is a path circling counterclockwise the segment $ [-1,1]$ in the complex-$w$ plane. The integral has thus been reduced to the evaluation of a simple discontinuity across the branch cut of the function $(w^2-1)^{-\half}$. 
The derivatives acting on $\hat{F}^{k+1}_{l}e^{2\pi w/\mu}$ produce in general many terms which can be organised in powers of $1/N$, i.e.
\ie 
C \sim \sum_{\ell=0}^\infty N^{-k-2\ell}\mu^{n_1+\dots +n_k}\int_\gamma \mu^{1-3\ell-k}\Big(\sum_{r=0}^{k+3\ell-1} \mu^r \partial_w^{r}\hat{F}^{k+1}_\ell(w,t) \Big)\frac{e^{\frac{2\pi w}{\mu}}p_{k+3\ell}(w)}{(w^2-1)^{\half}} {\rm d} w \, .
\fe

Remembering that $\mu = \sqrt{y/N}$ we see that higher powers of $\mu$ correspond to more suppressed terms at large-$N$.
Keeping here for simplicity only the leading contribution, i.e. only the $r=0$ term, we are left with computing the final $w$-integral.
To evaluate the $w$-integral we first change variable to $w\to 1-w$ and, since we are interested only in an asymptotic expansion of the integral\footnote{We have already omitted exponentially suppressed terms when we discarded the integration domain $-\mu^{-1}<x<\mu^{-1}$} as a function of $\mu$, we can extend the integration domain to $w\in[0,\infty]$ given that the difference is exponentially suppressed at large-$N$. Lastly, we simply expand the integrand $\partial_w^{r}\hat{F}^{k+1}(w,t) $ as a power series in $w$ near $w=0$ and integrate term by term to arrive at
\ie 
C &\sim  \sum_{\ell=0}^\infty N^{-k-2\ell}\mu^{n_1+\dots +n_k}\mu^{1-3\ell-k}\Big[ \int_{0}^{\infty}\Big(\sum_{m=0}^\infty c_m(t)w^{m-\half} \Big)e^{-\frac{2\pi w}{\mu}}e^{\frac{2\pi}{\mu}}{\rm d}w +\sum_{r=1}^\infty \mu^r \big(...\big)\Big]\cr
&\sim \sum_{\ell=0}^\infty N^{-\frac{k+\ell+n_1+...+n_k}{2}}\mu^{\frac{3}{2}}\Big( c_0(t)+\frac{c_1(t)\mu}{4\pi}+\dots\Big)e^{\frac{2\pi}{\mu}} + ... \,. \fe 
Finally, the $t$-integral has to be performed as in \eqref{eq:WIpq} and will produce finite expressions thanks to the asymptotic behaviour \eqref{eq:expAasy}.
From this schematic argument, it becomes clear that to obtain any given $1/N^\ell$ order in the large-$N$ fixed-$\tau$ expansion of the instanton sector \eqref{eq:WIpq} for the Wilson-line defect integrated correlator we only need to keep finitely many terms.

In the next section we will work out the numerical details omitted in the present rough arguments, and use this procedure to compute exactly the first few orders in the large-$N$ expansion of \eqref{eq:WIpq}.

\subsubsection*{Order \texorpdfstring{$1/N$}{1/N}}
We now follow the procedure outlined in the previous section and derive the exact coefficients for the first few $1/N$ orders in the large-$N$ expansion of the instanton sector for the Wilson-line defect integrated correlator as defined in \eqref{eq:WIpq}. 
The leading large-$N$ asymptotics of each term appearing in \eqref{eq:WIpq} are given by
\ie 
\exg{\mathbb{W}e^{\mathcal{A}}}_c-\exg{ \mathbb{W}}\sim e^{\frac{2\pi}{\mu}}\mu^{\tfrac{3}{2}}N^{-\tfrac{5}{2}}\, , \quad  \exg{ e^\mathcal{A} \mathcal{B}}_c\sim N^{-1}\, ,   \quad \exg{\mathbb{W}e^\mathcal{A}\mathcal{B}}_c\sim  e^{\frac{2\pi}{\mu}}\mu^{\tfrac{3}{2}}N^{-\tfrac{7}{2}} \, ,
\fe 
so at leading order we have
\ie  \label{eq:WIpq-lead}
\frac{\exg{ \mathbb{W}I_{p\times q}}-\exg{ \mathbb{W}}\exg{ I_{p\times q}}}{\exg{ \mathbb{W}}}\Big\vert_{N^{-1}}=\frac{c_{p,q} }{\mu\exg{ \mathbb{W}}}\!\int\limits_{\,\,\,\mathbb{R}\setminus (-1,1)}  \exg{\mathbb{W}\mathcal{A}}_c \,  e^{\llangle \mathcal{A}\rrangle}\, \frac{{\rm d}t}{2\pi}\,, 
\fe
where $\vert_{N^{-1}}$ denotes the fact that we only keep terms contributing at most to order $N^{-1}$.
The integrand can furthermore be expanded at leading order to produce 
\ie 
\exg{\mathbb{W}\mathcal{A}}_c &=(d^2-s^2) \exlc{\mathbb{W}\,{\rm Tr}\frac{1}{A_x^2}}  = -{ k \over N} \partial_x\exg{{\rm Tr}\,e^{2\pi a}\,{\rm Tr}\frac{1}{A_x }} \, ,
\fe
which, as anticipated in \eqref{eq:InvLap}, we rewrite as an inverse Laplace transform of the resolvent,
\ie
\exg{\mathbb{W}\mathcal{A}}_c  & =\frac{k\mu}{N}\partial_t\int_{{\rm Re}w>1}   W_0^2(w,t/\mu)\,e^{2\pi w} \,\frac{{\rm d}w}{2\pi i}\, .
\fe
Using topological recursion \cite{Eynard:2004mh,Eynard:2008we}, we compute the two-body genus zero resolvent $W_0^2(x_1,x_2)$ from \eqref{eq:W10} eventually arriving at
\ie 
\int_{{\rm Re}w>1} W_0^2(w,t/\mu)e^{2\pi w}\,\frac{{\rm d}w}{2\pi i}=e^{\frac{2\pi}{\mu}}\mu^{\tfrac{3}{2}}\frac{\mathrm{sign}(t)}{\sqrt{t^2-1}}\[\frac{1}{4 \pi  (t-1)}\] + \ldots \, . 
\fe
We substitute this expression back in \eqref{eq:WIpq-lead} and write the leading order instanton correction as 
\ie
\frac{\exg{ \mathbb{W}I_{p\times q}}-\exg{ \mathbb{W}}\exg{ I_{p\times q}}}{\exg{ \mathbb{W}}}\Big\vert_{N^{-1}} = -6
e^{2 k y}{c_{p,q}}\frac{k}{N^2}\frac{e^{\frac{2\pi}{\mu}} \mu^{\tfrac{3}{2}}}{8\pi^2}\int_1^{\infty} \frac{t}{(t^2-1)^{\tfrac{5}{2}}}\exp \left[ -2  k y \frac{|t|}{\sqrt{t^2-1}} \right]\,{\rm d}t \, , 
\fe
where first we split the $t$-integral into the two domains $t\in(-\infty,-1)$ and $t\in(1,\infty)$, and then combined both of them into a single integral. The integral can be performed explicitly to obtain, 
\ie
\frac{\exg{ \mathbb{W}I_{p\times q}}-\exg{ \mathbb{W}}\exg{ I_{p\times q}}}{\exg{ \mathbb{W}}}\Big\vert_{N^{-1}} =-  \frac{3 (p^2+q^2)}{2 y k^2}e^{2  k y}K_2(2  k y) \frac{1}{N}  \, , 
\fe
where $K_2(y)$ denotes the modified Bessel function of the second kind. Remembering that the sum over $p$ and $q$ is constrained by the fact that $ pq = k$, we finally find the instanton contributions at the leading order in the large-$N$ expansion \eqref{eq:WlargeN},
\begin{equation}\label{ordine1}
\mc{I}^{(2)\text{-inst}}_{\mathbb{W}}(\tau)=- \sum_{k> 0} \cos(2\pi k \tau_1)
\frac{6  \sigma_{-2}(k) }{\pi  \tau_2} K_2(2 k  \pi  \tau_2)\,,
\end{equation}
where $\sigma_n(k)=\sum_{d|k}d^n$ is a sum over the divisors of $k$ and we have expressed the result in terms of $\tau=\tau_1+i \tau_2$. 

\subsubsection*{Higher-order terms}

Instanton corrections at higher-order in the large-$N$ expansion can be obtained in a similar fashion. We now present all the terms which contribute to the instanton sectors up to order $1/N^2$.
Firstly, working up to order $1/N^2$ we can simplify the integrand of \eqref{eq:WIpq} using
\ie
\exg{\mathbb{W}e^\mc{A}}_c-\exg{\mathbb{W}} &=\exg{\mathbb{W}\mc{A}}_c =-(s^2-d^2)\exg{\mathbb{W}\,{\rm Tr}\frac{1}{A_x^2}}_c+\frac{(s^4-d^2)}{2}\exg{\mathbb{W}\,{\rm Tr}\frac{1}{A_x^4}}_c\, , \cr
\exg{e^\mc{A}\mc{B}}_c &=\exg{\mathcal{B}}=\frac{8s^2d^2}{k^2}(s^2-d^2)\exg{{\rm Tr}\frac{1}{A_x^4}}\,  ,\cr
\exg{\mathbb{W}e^\mc{A}\mc{B}}_c &=\exg{\mathbb{W}\mc{B}}_c=\frac{8s^2d^2}{k^2}(s^2-d^2)\exg{\mathbb{W}\,{\rm Tr}\frac{1}{A_x^4}}_c \, ,
\fe
where all of these identities hold to the order $1/N^2$ we are working at.
We then use the inverse Laplace transform \eqref{eq:InvLap} combined with the resolvents genus expansion up to order $1/N^2$ to write
\begin{align} \exg{\mathbb{W}\,{\rm Tr}\frac{1}{A_x^2}}_c &=-\frac{\mu}{N}\partial_t\int_{{\rm Re}w>1} \[ W_0^2(w,t/\mu)+\frac{1}{N^2}W_1^2(w,t/\mu)+\frac{1}{N^4}W_2^2(w,t/\mu)\]e^{2\pi w} \frac{{\rm d}w}{2\pi i}\, , \cr 
\exg{\mathbb{W}\,{\rm Tr}\frac{1}{A_x^4}}_c &=-\frac{\mu^3}{6N}\partial_t^3\int_{{\rm Re}w>1}  W_0^2(w,t/\mu) e^{2\pi w}\,\frac{{\rm d}y}{2\pi i}\, , \\
\exg{{\rm Tr}\frac{1}{A_x^4}} &=-\frac{N\mu^3}{6}\partial_t^3W_0^1(t/\mu)=\frac{N\mu^4|t|}{(t^2-1)^{5/2}} \, . \nonumber
\end{align}
Once again, the resolvents can be computed using topological recursion yielding
\begin{align} 
\int_{{\rm Re}w>1}W_0^2(w,t/
\mu)e^{2\pi w}\frac{{\rm d}w}{2\pi i} &=e^{\frac{2\pi}{\mu}} \mu^{\tfrac{3}{2}}\frac{\mathrm{sign}(t)}{\sqrt{t^2-1}}\[\frac{1}{4 \pi  (t-1)}-\frac{3 \mu (t+3)}{64 \pi ^2 (t-1)^2}-\frac{15 \mu^2 [(t-10) t-23]}{2048 \pi ^3 (t-1)^3}\] \, , \notag \\
\int_{{\rm Re}w>1} W_1^2(w,t/\mu) e^{2\pi w} \frac{{\rm d}w}{2\pi i}&=e^{\frac{2\pi}{\mu}} \mu^{-\tfrac{3}{2}}\frac{\mathrm{sign}(t)}{\sqrt{t^2-1}}\[\frac{\pi ^2}{48 (t-1)}-\frac{\pi  \mu (5 t-9)}{256 (t-1)^2}\] \, ,  \\
\int_{{\rm Re}w>1} W_2^2(w,t/\mu) e^{2\pi y} \frac{{\rm d}w}{2\pi i}& =e^{\frac{2\pi}{\mu}} \mu^{-\tfrac{9}{2}}\frac{\mathrm{sign}(t)}{\sqrt{t^2-1}}\[\frac{\pi ^5}{1152 (t-1)}\]\, . \notag
\end{align}
We then substitute this expression back in \eqref{eq:WIpq} expanded up to order $1/N^2$ and perform the $t$-integral as previously discussed.
In this way we finally find the instanton contributions up to order $1/N^2$ in the large-$N$ expansion \eqref{eq:WlargeN}, i.e.
\begin{equation}
\mathcal{I}^{\text{inst}}_{\mathbb{W},N}(\tau) = N^{-1}\,\mathcal{I}^{(2)\text{-inst}}_{\mathbb{W}}(\tau)+N^{-\frac{3}{2}}\,\mathcal{I}^{(3)\text{-inst}}_{\mathbb{W}}(\tau)+N^{-2}\,\mathcal{I}^{(4)\text{-inst}}_{\mathbb{W}}(\tau) +O(N^{-\tfrac{5}{2}})\,,
\end{equation}
where we defined
\begin{align}
&\mathcal{I}^{(2)\text{-inst}}_{\mathbb{W}}(\tau)=- \sum_{k> 0} \cos(2\pi k \tau_1)
\frac{6  \sigma_{-2}(k) }{\pi  \tau_2} K_2(2 k  \pi  \tau_2)\,, \\
&\mathcal{I}^{(3)\text{-inst}}_{\mathbb{W}}(\tau) = \sum_{k> 0} \cos(2\pi k \tau_1)\Big[ \frac{9 \sigma_{-2}(k) }{2 \pi \sqrt{\pi \tau_2}} K_2(2 k\pi  \tau_2)+\frac{45 \sigma_{-2}(k)}{2  \pi k (\pi \tau_2)^{\frac{3}{2}}} K_3(2 k \pi  \tau_2)\Big]\,,
\\
&\mathcal{I}^{(4)\text{-inst}}_{\mathbb{W}}(\tau)   =\\
&\sum_{k=1}^{\infty} \cos (2 \pi  k \tau_1 ) \left[  -\frac{\left(3\tau_2^2+4 \pi ^2\right) 
   \sigma _2(k)}{16 \pi ^2 k^2\tau_2^2} K_2(2 \pi  k\tau_2)+ \frac{5 \left(\pi ^2\tau_2^2 \sigma
   _4(k)-\left(12\tau_2^2+\pi ^2\right)
   \sigma _2(k)\right)}{4 \pi ^3 k^3\tau_2^3} K_3(2 \pi  k\tau_2) \right. \cr
  &\qquad\qquad\qquad\qquad +   \left. \frac{35 \left(2 \pi ^2 \sigma _4(k)-15 \sigma
   _2(k)\right)}{8 \pi ^4 k^4\tau_2^2}  K_4(2 \pi  k\tau_2) \right] \,, \nonumber
\end{align}
As stressed in the main text, the result of our large-$N$ expansion for the instanton sectors differs from that of \cite{Pufu:2023vwo} in that we do not find contributions proportional to $\sin(k\theta)$.
The instanton contributions at order $1/N^2$ were not computed in \cite{Pufu:2023vwo} and are here presented as a further check of our proposed large-$N$ expansion in terms of a novel class of automorphic functions.

To obtain the complete $1/N^2$-order term we must combine the instanton sectors just computed with the perturbative part contained in 
\eqref{eq:Iw3correct} 
and equation (2.25) of \cite{Pufu:2023vwo} for the lower order terms, this yield 
\begin{align} 
&\mathcal{I}^{(4)}_{\mathbb{W}}(\tau)  \label{eq:I4}= \frac{\pi ^6}{56700 \tau _2^6} -\frac{\pi ^4}{1680 \tau _2^4}-\frac{\pi ^2}{30 \tau _2^2} - \frac{(4 \zeta (3) + 1)}{128}  + \frac{9 \tau _2^2 (7 \zeta (3)+10 \zeta (5)+3)}{512 \pi ^2}\\
    &-3 \sum_{k=1}^{\infty} \cos (2 \pi  k \tau_1 ) \left[  \frac{\left(3\tau_2^2+4 \pi ^2\right) 
   \sigma _2(k)}{16 \pi ^2 k^2\tau_2^2} K_2(2 \pi  k\tau_2)- \frac{5 \left(\pi ^2\tau_2^2 \sigma
   _4(k)-\left(12\tau_2^2+\pi ^2\right)
   \sigma _2(k)\right)}{4 \pi ^3 k^3\tau_2^3} K_3(2 \pi  k\tau_2) \right. \cr
   &\qquad\qquad\qquad\qquad\quad - \left. \frac{35 \left(2 \pi ^2 \sigma _4(k)-15 \sigma
   _2(k)\right)}{8 \pi ^4 k^4\tau_2^2}  K_4(2 \pi  k\tau_2) \right] \,. \nonumber
\end{align}
We remark that in the above expression only the first term, proportional to $1/\tau _2^6$, has not been computed directly from the matrix model. This term has been determined by requiring a softer behaviour in the small-$\tau_2$ limit (with $\tau_1=0$) dictated by the 't Hooft expansion of the 't Hooft-line defect integrated correlator (see the discussion around \eqref{eq:D1-scattering}). 
Following an analysis similar to \cite{Pufu:2023vwo}, or alternatively using the asymptotic expansion \eqref{eq:tHooftExp}, it is straightforward to verify that in the limit $\tau_2\to 0$ the $1/\tau _2^6$ term and the $1/\tau _2^4$ term cancel against equal and opposite terms originating from the infinite sum of instantons contributions, which stop being exponentially suppressed when $\tau_2\to 0$.  It would be interesting, although rather laborious, to compute the term $1/\tau _2^6$ directly from the matrix model by a large-$N$ large-$\lambda$ expansion, which would correspond to the term $\lambda^6/N^8$ in the 't Hooft genus expansion at large-$\lambda$.

\section{A Class of Automorphic Functions for Line Defects}
\label{sec:AppF}

In this appendix we present more details on the novel class of automorphic functions, $F_{s_1,s_2,s_3}$, which appear in the analysis of the integrated two-point defect correlator \eqref{eq:Iw}.

More in general, we want to define a family of real-analytic functions, $F$, defined on the upper-half plane parametrised by $\tau\in \mathbb{H}$ and carrying as input an additional label associated with a coset element $[\rho] = \left(\begin{smallmatrix} * & * \\ q & p \end{smallmatrix}\right)\in B(\mathbb{Z})\backslash {\rm PSL}(2,\mathbb{Z}) $, corresponding to the electromagnetic charges $(p,q)$ of the defect.
Crucially, we require that under an action of $\SL(2,\mathbb{Z})$ on the modular parameter $\tau$ these class of functions $F$ transforms with the automorphic property \eqref{eq:autoform} which can be interpreted as electromagnetic duality, i.e. we want to construct real-analytic functions $F$ such that 
\begin{align}
&F: B(\mathbb{Z})\backslash {\rm PSL}(2,\mathbb{Z}) \times \mathbb{H} \rightarrow \mathbb{R}\,,\\
&F([\rho];\gamma\cdot \tau) = F([\rho \,\gamma];\tau) \,, \qquad \forall \gamma \in \SL(2,\mathbb{Z})\,.\label{eq:Fautom}
\end{align}

Firstly, we can immediately construct an easy class of functions satisfying the automorphic property \eqref{eq:Fautom}, namely
\begin{equation}
F^{{\rm easy}}_s( [\rho]; \tau) \coloneq (\tau_2/\pi)^s \big\vert_{\rho} = \pi^{-s} \Big(\mbox{Im}\, \rho\cdot \tau \Big)^s = \frac{(\tau_2/\pi)^s}{|q\tau+p|^{2s}}\,,\label{eq:Feasy}
\end{equation}
were $s\in \mathbb{C}$ and we used the standard $\vert_\rho$ action to denote $f(\tau)\vert_\rho \coloneqq f(\rho\cdot\tau)$.
Note that \eqref{eq:Feasy} is really defined on the equivalence class $[\rho]$ rather than a particular representative thereof and one can easily check that under an ${\rm SL}(2,\mathbb{Z})$ action on $\tau$ the function $F^{{\rm easy}}_s$ transforms as \eqref{eq:Fautom}.

If we specialise $F^{{\rm easy}}_s$ to the case $[\rho] = [\mathbbm{1}]$ corresponding to the Wilson-line defect, we understand the reason why we denoted this class of functions with the superscript `{\rm easy}' since $F^{{\rm easy}}_s( [\mathbbm{1}]; \tau) = (\tau_2/\pi)^s$ is simply a term in the perturbative expansion.
Although not particularly exciting, we see that the Wilson-line defect expectation value \eqref{eq:WilsonLag}, the one- and two-point defect correlators \eqref{eq:Cw}-\eqref{eq:Ew} can all be expressed solely in terms of \eqref{eq:Feasy}.

We now define now a more interesting class of automorphic functions relevant for the integrated defect correlator \eqref{eq:Iw} satisfying \eqref{eq:Fautom} and based on a Poincar\'e series approach. 
Poincar\'e series are a fundamental tool in constructing modular invariant functions and in their simplest instance they take the form
\begin{equation}
F(\tau) = \sum_{\gamma'\in B(\mathbb{Z})\backslash {\rm PSL}(2,\mathbb{Z})} \varphi(\gamma'\cdot \tau)\,.\label{eq:PoinSer}
\end{equation}
However, for the Poincar\'e series to be well-defined over the coset $B(\mathbb{Z})\backslash {\rm PSL}(2,\mathbb{Z})$ we must require that the function $\varphi(\tau)$, usually called Poincar\'e seed, to be invariant under the Borel stabiliser $B(\mathbb{Z})$, i.e. it must be periodic with respect to $\tau_1 = {\mbox{Re}}\,\tau$ so that $\varphi(\tau+n) = \varphi(\tau)$.
A simple Poincar\'e seed one can consider is $\varphi(\tau) = \tau_2^s$ and the associated Poincar\'e series \eqref{eq:PoinSer} is referred to as the non-holomorphic (or real-analytic) Eisenstein series.

Since we are interested in automorphic forms satisfying \eqref{eq:Fautom}, we have to modify the Poincar\'e series definition \eqref{eq:PoinSer} which would otherwise produce modular invariant functions.
For the electromagnetic charges coset element $[\rho]$ to `talk' with the Poincar\'e series, we have to somehow intertwine the two coset elements $[\rho]$ and $[\gamma']$.
Crucially, we notice that given two elements $[\rho] = \left(\begin{smallmatrix} * & * \\ q & p \end{smallmatrix}\right)\in B(\mathbb{Z})\backslash {\rm PSL}(2,\mathbb{Z}) $ and $[\gamma'] = \left(\begin{smallmatrix} * & * \\ c & d \end{smallmatrix}\right)\in B(\mathbb{Z})\backslash {\rm PSL}(2,\mathbb{Z}) $, we can construct the combination
\begin{equation}
\langle [\gamma'],[\rho]\rangle \coloneqq (\rho\, \gamma'^{-1})_{21} =  q d - p c \,,\label{eq:Pairing}
\end{equation}
which is manifestly invariant under a right ${\rm SL}(2, \mathbb{Z})$ action, i.e. 
\begin{equation}
\langle [\gamma'],[\rho]\rangle = \langle [\gamma' \gamma ],[\rho \,\gamma]\rangle \qquad \forall\,\gamma\in {\rm SL}(2,\mathbb{Z})\,.\label{eq:sl2inv}
\end{equation}
In \eqref{eq:Pairing} we denote with $(\rho\, \gamma'^{-1})_{21}$ the bottom left entry of the $2\times 2$ matrix $\rho\, \gamma'^{-1}$, and note that this is the only element in that product of matrices which is well-defined on the coset.
As will be important later on, we note that 
\begin{equation}
\langle [\gamma'],[\rho]\rangle  = 0 \qquad \Longleftrightarrow\qquad [\gamma']=[\rho]\,,
\end{equation}
i.e. $q=c$ and $p=d$. 

We can now define a family of real-analytic functions 
via the Poincar\'e series representation:
\begin{equation}
F([\rho];\tau) = (\tau_2/\pi)^{s_1}\big\vert_{\rho} \sum_{\gamma'\in B(\mathbb{Z})\backslash {\rm PSL}(2,\mathbb{Z})} (\tau_2/\pi)^{s_2}\big\vert_{\gamma'} \, f(\langle [\gamma'],[\rho]\rangle)\,,\label{eq:PoinGen}
\end{equation}
where $s_1,s_2$ are complex numbers and $f:\mathbb{Z}\to \mathbb{R}$ is for the moment an arbitrary real-valued function over the set of all integers.
Presently we do not worry about convergence properties of such a Poincar\'e series but shortly we will consider special cases for which \eqref{eq:PoinGen} is indeed convergent.
It is an easy exercise to show that \eqref{eq:PoinGen} obeys the requested automorphic property \eqref{eq:Fautom}.
To this end we consider
\begin{align}
F([\rho];\gamma\cdot\tau) &\notag =F([\rho];\tau)\big\vert_\gamma  = (\tau_2/\pi)^{s_1}\big\vert_{\rho\,\gamma} \sum_{\gamma'\in B(\mathbb{Z})\backslash {\rm PSL}(2,\mathbb{Z})} (\tau_2/\pi)^{s_2}\big\vert_{\gamma'\gamma} \, f(\langle [\gamma'],[\rho]\rangle)\\
 &\notag=(\tau_2/\pi)^{s_1}\big\vert_{\rho\,\gamma} \sum_{\gamma''\in B(\mathbb{Z})\backslash {\rm PSL}(2,\mathbb{Z})} (\tau_2/\pi)^{s_2}\big\vert_{\gamma''} \, f(\langle [\gamma'' \gamma^{-1}],[\rho]\rangle)\\
 &=(\tau_2/\pi)^{s_1}\big\vert_{\rho\,\gamma} \sum_{\gamma''\in B(\mathbb{Z})\backslash {\rm PSL}(2,\mathbb{Z})} (\tau_2/\pi)^{s_2}\big\vert_{\gamma''} \, f(\langle [\gamma''],[\rho\,\gamma]\rangle) = F([\rho\,\gamma];\tau)\,,\label{eq:FgenAuto}
\end{align}
where in the first line we used the fact that
\begin{equation}
(\tau_2/\pi)^s \big\vert_\rho \big\vert_\gamma = \Big(\, \frac{(\tau_2/\pi)^s}{|q\tau+p|^{2s}}\Big)\Big\vert_\gamma = \frac{(\tau_2/\pi)^s}{|(q a + p c)\tau + (q b+p d)|^{2s}}=(\tau_2/\pi)^s \big\vert_{\rho\,\gamma}\,,
\end{equation}
with $\gamma = \left(\begin{smallmatrix} a & b \\ c & d \end{smallmatrix}\right)$, then we changed summation variables $\gamma'\to \gamma'' \coloneqq \gamma' \gamma$ and finally in the last line we made use of the property \eqref{eq:sl2inv}.
We see that the transformation property \eqref{eq:FgenAuto} of the Poincar\'e series \eqref{eq:PoinGen} coincides precisely with the requested \eqref{eq:Fautom}.

Note furthermore that if we isolate in the Poincar\'e sum \eqref{eq:PoinGen} the single term $[\gamma']=[\rho]$, which is the unique point for which the argument of $f$ vanishes, this reduces to a multiple of the easier functions previously constructed in \eqref{eq:Feasy}, i.e.
\begin{equation}
(\tau_2/\pi)^{s_1}\big\vert_{\rho} \sum_{\gamma'\in B(\mathbb{Z})\backslash \SL(2,\mathbb{Z})} \delta([\gamma']-[\rho])\, (\tau_2/\pi)^{s_2}\big\vert_{\gamma'} \, f(\langle [\gamma'],[\rho]\rangle) = f(0)\,(\tau_2/\pi)^{s_1+s_2} \big\vert_{\rho}  = f(0)\, F_{s_1+s_2}^{\rm easy}([\rho];\tau)\,.
\end{equation}
This observation leads us to consider a more refined version of the Poincar\'e sum \eqref{eq:PoinGen} for which we remove from the sum the single element $[\gamma']=[\rho]$ . Furthermore, we specialise the arbitrary function $f:\mathbb{Z} \to \mathbb{R}$ to the particular case $f(n) = (\pi n)^{s_3}$ (the factors of $\pi$ are chosen as to have a nicer normalisation) to arrive at our novel automorphic building blocks
\begin{align}
\widetilde{F}_{s_1,s_2,s_3}([\rho];\tau) &\coloneqq (\tau_2/\pi)^{s_1} \big\vert_\rho \sum_{\substack{\gamma'\in B(\mathbb{Z})\backslash {\rm PSL}(2,\mathbb{Z})\\ \gamma'\neq \rho}} (\tau_2/\pi)^{s_2}\big\vert_{\gamma'} \, \big(\pi \langle [\gamma'],[\rho]\rangle\big)^{s_3}\\
&= \frac{(\tau_2/\pi)^{s_1}}{|q\tau+p|^{2s_1} }\sum_{\substack{
 (c,d)=1,c\geq 0 \\ (c,d)\neq (q,p)}} \frac{(\tau_2/\pi)^{s_2}}{| c \tau+d|^{2s_2}} [\pi( qd - pc)]^{s_3}\,.\label{eq:FtildePS}
\end{align}
  Convergence of the Poincar\'e series requires ${\mbox{Re}}(s_2-s_3)>1$. However we will shortly analytically continue this class of functions to arbitrary $s_1,s_2,s_3\in \mathbb{C}$.

Rather than working with the above Poincar\'e series we find it more convenient to obtain a lattice sum representation.
To transform the Poincar\'e series \eqref{eq:FtildePS} into a lattice sum we multiply it by $2\zeta(2s_2-s_3)$ which we then rewrite in terms of its Dirichlet series $2\zeta(2s_2-s_3) = 2\sum_{N>0} N^{-2s_2+s_3}$. We then change summation variables $(c,d)\to (n,m) = (Nc,Nd)$ so that, under the assumption\footnote{The condition $s_3\in 2 \mathbb{Z}$ appears to be the physically relevant choice for the present paper. However, it is possible to write similar lattice sums for the more general case when $s_3$ is not an even integer.} $s_3\in 2\mathbb{Z}$, we can rearrange the multiple sums as
\begin{equation}
2\sum_{N>0} \sum_{\substack{
 (c,d)=1,c\geq 0 \\ (c,d)\neq (q,p)}} = \sum_{\substack{ (n,m)\in \mathbb{Z}^2 \\ (n,m)\neq \ell (q,p)\,, \forall \ell \in \mathbb{Z}}}\,.
\end{equation}
We finally arrive at the automorphic building blocks considered in the main body of this work and given by the lattice sum representation
\begin{align}
F_{s_1,s_2,s_3}([\rho];\tau)
&\coloneqq2\zeta(2s_2-s_3) \widetilde{F}_{s_1,s_2,s_3}([\rho];\tau)\label{eq:latticepmNew} \\
&\notag
=  \frac{(\tau_2/\pi)^{s_1}}{|q \tau+p|^{2s_1}} \sum_{(n,m)\in \mathbb{Z}^2 \setminus \{\mathbb{Z}(q,p)\}}   \frac{(\tau_2/\pi)^{s_2}}{|n \tau+m|^{2s_2}} [\pi (n p - m q)]^{s_3} \,.
\end{align}
After having introduced this infinite class \eqref{eq:latticepmNew} of automorphic functions satisfying \eqref{eq:Fautom}, we now discuss some of their key properties.

Firstly, we notice that under the involution $\tau \to - \bar{\tau}$ the automorphic functions  $F_{s_1,s_2,s_3}([\rho];\tau)$ are even, i.e.
\begin{equation}
F_{s_1,s_2,s_3}([\rho];-\bar\tau) = + F_{s_1,s_2,s_3}([\rho];\tau)\,,
\end{equation}
It is possible to modify the starting Poincar\'e series \eqref{eq:PoinGen} as to produce a different class of automorphic functions odd under the involution $\tau \to - \bar{\tau}$, thus leading to cuspidal objects in the sense that their asymptotic expansion at the cusp $\tau_2\gg1$ must necessarily be $O(q)$ with $q=\exp(2\pi i \tau)$.

Secondly, we can check by direct calculation that for general $[\rho] \in B(\mathbb{Z})\backslash \SL(2,\mathbb{Z})$ the functions $F_{s_1,s_2,s_3}([\rho];\tau)$ defined in \eqref{eq:latticepmNew}, satisfy the differential equation:
\begin{equation}
\Big[\Delta_\tau -2i s_1\Big( \frac{q\tau+p}{q\bar\tau+p} \tau_2\,\partial_\tau -\frac{q\bar\tau+p}{q\tau+p} \tau_2\,\partial_{\bar{\tau}}\Big) + (s_1+s_2)(1+s_1-s_2) \Big]{F}_{s_1,s_2,s_3}([\rho];\tau) = 0\,.\label{eq:DiffEqGen}
\end{equation}
In particular, when  $[\rho]=[\mathbbm{1}]$ (the case relevant when discussing Wilson-line defect correlators) the above equation reduces to the simpler differential equation
\begin{equation}
[\Delta_\tau -2s_1 \tau_2\, \partial_{2} + (s_1+s_2)(1+s_1-s_2)\big]{F}_{s_1,s_2,s_3}([\mathbbm{1}];\tau) = 0\,,\label{eq:diff0}
\end{equation}
where $\partial_2 = \partial/\partial \tau_2$.
Note that equation \eqref{eq:DiffEqGen} can in fact be derived by acting with $\vert_{\rho}$ on \eqref{eq:diff0} and noting that $2\tau_2\, \partial_2 = -i \tau_2(\partial_\tau-\partial_{\bar\tau})$ where the Maa{\ss} operators $\tau_2 \partial_\tau$ and $\tau_2 \partial_{\bar\tau}$ change the modular weights respectively by $(+1,-1)$ and $(-1,+1)$, i.e. 
\begin{equation}
\big[\tau_2 \partial_\tau F(\tau)\big] \vert_\rho = \frac{q\tau+p}{q\bar\tau +p} \tau_2 \partial_\tau F(\rho\cdot \tau)\,,\qquad \big[\tau_2 \partial_{\bar\tau} F(\tau)\big] \vert_\rho = \frac{q\bar\tau+p}{q\tau +p} \tau_2 \partial_{\bar\tau} F(\rho\cdot \tau)\,,
\end{equation}thus justifying the automorphy factors appearing in \eqref{eq:DiffEqGen}.

Finally we want to derive the Fourier mode expansion of \eqref{eq:latticepmNew} in the special case when $[\rho] = [\mathbbm{1}]$.
We first note that only for the case $[\rho] = [\mathbbm{1}]$, relevant for the Wilson-line defect correlators, we can perform a sensible Fourier series decomposition of \eqref{eq:latticepmNew} with respect to ${\mbox Re}(\tau) = \tau_1$. The reason is simple, if we consider a translation $T^n\cdot\tau = \tau+n$ of a generic function ${F}_{s_1,s_2,s_3}([\rho];\tau)$ this will in general change the coset element, i.e.
\begin{equation}
{F}_{s_1,s_2,s_3}([\rho];\tau+n) = {F}_{s_1,s_2,s_3}([\rho];T^n\cdot\tau) = {F}_{s_1,s_2,s_3}([\rho\, T^n];\tau)\,,
\end{equation}
and $[\rho\,T^n]\neq [\rho]$ unless $[\rho] = [\mathbbm{1}]$. This can be understood in physics terms as a consequence of the Witten effect where for a dyon with electromagnetic charges $(p,q)$ with $q\neq0$, a translation of the $\theta$ angle will induce a modification of the electric charge.

Let us then consider the Fourier mode expansion for \eqref{eq:latticepmNew} when $[\rho] =[\mathbbm{1}]$ and use the integral representation
\begin{equation}
{F}_{s_1,s_2,s_3}([\rho];\tau) = (\tau_2/\pi)^{s_1} \big\vert_\rho \sum_{(n,m)\in \mathbb{Z}^2 \setminus\{\mathbb{Z}(q,p)\}}  \int_0^\infty e^{-t_2 \pi \frac{|n\tau+m|^2}{\tau_2} -t_3 \pi (np-mq)} \frac{t_2^{s_2-1}}{\Gamma(s_2)}\frac{t_3^{-s_3-1}}{\Gamma(-s_3)} \,{\rm d}^2 t\,,\label{eq:FIntRep}
\end{equation}
which we then specialise to $(p,q)=(1,0)$
\begin{equation}
{F}_{s_1,s_2,s_3}([\mathbbm{1}];\tau) = (\tau_2/\pi)^{s_1}  \sum_{\substack{ (n,m)\in \mathbb{Z}^2 \\ n \neq0 }}  \int_0^\infty e^{-t_2 \pi \frac{|n\tau+m|^2}{\tau_2} -t_3 \pi n} \frac{t_2^{s_2-1}}{\Gamma(s_2)}\frac{t_3^{-s_3-1}}{\Gamma(-s_3)} \,{\rm d}^2 t\,.\label{eq:FWIntRep}
\end{equation}
We perform a Poisson resummation for the variable $m\to \hat{m}$ thus arriving at
\begin{equation}
{F}_{s_1,s_2,s_3}([\mathbbm{1}];\tau) = (\tau_2/\pi)^{s_1}  \sum_{\substack{ (n,\hat{m})\in \mathbb{Z}^2 \\ n \neq0 }} e^{2\pi i \tau_1 n\hat{m}} \sqrt{\tau_2}  \int_0^\infty e^{-t_2 \pi n^2  \tau_2 - t_2 \frac{\pi \hat{m}^2 }{\tau_2} - t_3 \pi n} \frac{t_2^{s_2-\frac{3}{2}}}{\Gamma(s_2)}\frac{t_3^{-s_3-1}}{\Gamma(-s_3)} \,{\rm d}^2 t\,.\label{eq:FWIntRepInst}
\end{equation}
The term $\hat{m}=0$ produces the single perturbative term in the Fourier zero-mode sector, while for $\hat{m}\neq0$ we obtain the non-zero Fourier modes, both of which can be readily evaluated from the integral representation above and take the form
\begin{align}
{F}_{s_1,s_2,s_3}([\mathbbm{1}];\tau) &\notag =  
\sum_{k\in \mathbb{Z}} e^{2\pi i k \tau_1} {F}^{(k)}_{s_1,s_2,s_3}([\mathbbm{1}];\tau_2)  \\
&= \label{eq:F0cusp} \frac{2 \pi ^{\frac{1}{2}+s_3-s_1-s_2} \Gamma \left(s_2-\frac{1}{2}\right)  \zeta (2 s_2-s_3-1)}{\Gamma (s_2)}\tau_2^{s_1+1-s_2}\\
&\notag\phantom{=} +\frac{8 \pi^{s_3-s_1}}{\Gamma (s_2)}\sum_{k>0}\cos (2 \pi  k \tau_1)   
k^{s_2-\frac{1}{2}}\sigma _{s_3+1-2 s_2}(k) \tau_2^{s_1+\frac{1}{2}} K_{s_2-\frac{1}{2}}(2 k \pi  \tau_2)\,.
\end{align}
Here $K_s(y)$ is a modified Bessel function of the second kind, while $\sigma_s(k) \coloneqq \sum_{d\vert k} d^s$ denotes the divisor sigma function.  

From the Fourier mode decomposition \eqref{eq:F0cusp}, we can furthermore extract the asymptotic expansion near the origin $\tau_2\to 0$ which, thanks to the identity
${F}_{s_1,s_2,s_3}([\mathbbm{1}];-1/\tau) = {F}_{s_1,s_2,s_3}([S];\tau)$, we can relate to the large $\tau_2\gg1$ expansion for the 't Hooft-line defect, $[\rho] = [S]$, in the particular case $\tau_1=0$.
To proceed we run an argument similar to the analysis provided in \cite{Pufu:2023vwo}.
Firstly, we use the Mellin-Barnes integral representation for the Bessel function
\begin{equation}
K_t(2z) = \int_{{\rm Re} \,s = \gamma} \Gamma\Big(\frac{s+t}{2}\Big) \Gamma\Big(\frac{s-t}{2}\Big) z^{-s} \frac{{\rm d}s}{8\pi i }\,,
\end{equation}
with $\gamma >  {\rm Re}\,t$.
Substituting this expression in \eqref{eq:F0cusp}, we exchange the integral over $s$ with the sum over Fourier modes, $k>0$, arriving at the Dirichlet series
\begin{equation}
\sum_{k>0}\cos(2\pi k \tau_1) \frac{ \sigma_{s_3+1-2s_2}(k) } { k^{s +\frac{1}{2}-s_2}}\,.\label{eq:Dirictau1}
\end{equation}
Although no closed-form expression is known for generic $-1/2<\tau_1<1/2$, it should be possible to adapt the analysis of \cite{Dorigoni:2020oon} to obtain an analytic continuation of \eqref{eq:Dirictau1} for $\tau_1 \in [-1/2,1/2] \cap \mathbb{Q}$.
Here we simply consider the case $\tau_1=0$ for which \eqref{eq:Dirictau1} reduces to the well-known Dirichlet series
\begin{equation}
\sum_{k>0} \frac{ \sigma_{s_3+1-2s_2}(k) } { k^{s +\frac{1}{2}-s_2}} = \zeta \big(s-s_2+\frac{1}{2}\big) \zeta \big(s+s_2-s_3-\frac{1}{2}\big)\,.
\end{equation}
However, note that by having exchanged integration over $s$ with the sum over $k$, we have now introduced additional singularities in the right half-plane ${\rm Re}\,s \geq \gamma$, thus we have to push the $s$-contour of integration past all poles as to obtain an integral representation which is indeed exponentially suppressed as $\tau_2 \gg 1$.

Proceeding as just discussed and performing the change of integration variables $s\to 2s+s_2 +1/2$ we arrive at the alternative rewriting of \eqref{eq:F0cusp} specialised to $\tau_1=0$,
\begin{align}
{F}_{s_1,s_2,s_3}([\mathbbm{1}];i\tau_2) & = \label{eq:F0cusp1} \frac{2 \pi ^{\frac{1}{2}+s_3-s_1-s_2} \Gamma \left(s_2-\frac{1}{2}\right)  \zeta (2 s_2-s_3-1)}{\Gamma (s_2)}\tau_2^{s_1+1-s_2}\\
&\notag\phantom{=} +4\frac{\pi ^{\frac{1}{2} (s_3-2s_1)}\tau_2^{s_1-s_2} }{\Gamma(s_2)} \int_{{\rm Re}\,s = \gamma} \frac{\Gamma (s+s_2)  }{ \Gamma \left(s+s_2-\frac{s_3}{2}\right)}\xi(2s)\, \xi (2s+2s_2-s_3) \tau_2^{-2s} \frac{{\rm d}s}{2\pi i }\,,
\end{align}
where $\xi(s) \coloneqq \pi^{-s/2}\Gamma(s/2)\zeta(s) = \xi(1-s)$ denotes the completed Riemann zeta function and $\gamma$ must be chosen such that the integrand is analytic for $\mbox{Re}\,s \geq \gamma$.
The small-$\tau_2$ asymptotic expansion can now be computed by closing the contour of integration to the left half-plane and collecting residues.
Interestingly, for the case of physical interest we have $s_3\in 2\mathbb{Z}$ and the integrand displays only a finite number of poles which are located at $s=0,\frac{1}{2},-s_2+s_3/2, -s_2+s_3/2+1/2$, coming from the argument of $\xi(s)$ being either $0$ or $1$, as well as at $s= -s_2-m$ with $m\in\{0,...,1-s_3/2\}$ coming from the gamma function at numerator. 
When $\tau_1=0$, we then find the terminating asymptotic expansion near $\tau_2 \to 0$ of \eqref{eq:latticepmNew},
\begin{align}
& \label{eq:tHooftExp} {F}_{s_1,s_2,s_3}([\mathbbm{1}];i \tau_2)= {F}_{s_1,s_2,s_3}([S];\frac{i}{ \tau_2})\phantom{\int_\vert}\\
&\notag = \frac{2 \pi^{s_2-s_1-\frac{3}{2}} \Gamma\big(\frac{s_3+2-2s_2}{2}\big) \Gamma\big(\frac{s_3+1}{2} \big)}{\Gamma(s_2)}\zeta(s_3+2-2s_2) \tau_2^{s_1+s_2-s_3-1} +2 \pi^{s_3-s_1-s_2}  \zeta(2s_2-s_3)\tau_2^{s_1-s_2}\\
&\notag \phantom{=}\,\,\, + \sum_{m=0}^{{\rm max}(0,-\frac{s_3}{2})} 4 \frac{(-1)^m \pi^{s_3-s_1-s_2} \Gamma(s_2+m)}{\Gamma(s_2)\Gamma(m+1)} \zeta(2s_2+2m) \zeta(-s_3-2m) \tau_2^{s_1+s_2+2m} +O(e^{- \frac{2\pi}{\tau_2}})\,.
\end{align}

We note that the single Fourier zero-mode term in \eqref{eq:F0cusp} proportional to $\tau_2^{s_1+1-s_2}$ cancels out in the small-$\tau_2$ expansion. Furthermore, in the small-$\tau_2$ expansion the $m=0$ term, which reduces to $4\pi^{s_3-s_1-s_2} \zeta(2s_2)\zeta(-s_3) \tau_2^{s_1+s_2}$ corresponds to the second homogeneous solution to the differential equation \eqref{eq:diff0}. 
From the analysis of section \ref{sec:LargeN}, it appears clear that for the range of parameters $s_1,s_2$ and $s_3$ relevant for the large-$N$ expansion of the defect integrated correlator, the power $\tau_2^{s_1+s_2}$ is the most singular term as $\tau_2\to 0$.
It is tempting to consider as `fundamental' the shifted building block 
\begin{align}
 &\widehat{F}_{s_1,s_2,s_3}([\mathbbm{1}];\tau) \coloneqq F_{s_1,s_2,s_3}([\mathbbm{1}];\tau) -4\pi^{s_3-s_1-s_2} \zeta(2s_2)\zeta(-s_3) \tau_2^{s_1+s_2}\phantom{\int}\label{eq:appFimproved}\\
 &\notag =
-4\pi^{s_3-s_1-s_2} \zeta(2s_2)\zeta(-s_3) \tau_2^{s_1+s_2}+  \frac{2 \pi ^{\frac{1}{2}+s_3-s_1-s_2} \Gamma \left(s_2-\frac{1}{2}\right)  \zeta (2 s_2-s_3-1)}{\Gamma (s_2)}\tau_2^{s_1+1-s_2}\\
&\notag\phantom{=} +\frac{8 \pi^{s_3-s_1}}{\Gamma (s_2)}\sum_{k>0}\cos (2 \pi  k \tau_1) 
k^{s_2-\frac{1}{2}}\sigma _{s_3+1-2 s_2}(k) \tau_2^{s_1+\frac{1}{2}} K_{s_2-\frac{1}{2}}(2 k \pi  \tau_2)\,,
\end{align}
which still solves the differential equation \eqref{eq:diff0} and now has a milder behaviour as $\tau_2\to0$. Although there seems to be a connection  \cite{Pufu:2023vwo} between such singular terms and the requirement of having a consistent genus expansion of the 't Hooft-line defect correlator (see the discussion around \eqref{eq:D1-scattering}), we do not fully understand the mathematical reasons behind it.  

To conclude this appendix, given the similarity with the present discussion it is worth reminding the reader about the Fourier mode decomposition for the modular invariant non-holomorphic Eisenstein series $E(s;\tau)$,
\begin{align}
&\label{eq:EisenFourier}E(s;\tau) \coloneqq 2\zeta(2s) \sum_{\gamma' \in B(\mathbb{Z}) \backslash {\rm PSL}(2,\mathbb{Z})} (\tau_2/\pi)^s \,\big\vert_{\gamma'} = \sum_{(m,n) \neq\,(0,0)}\frac{(\tau_2/\pi)^s} {|n\tau+m|^{2s}}  \\
&\notag =2\zeta(2s)\, \big(\frac{\tau_2}{\pi}\big)^s {+}  \frac{2\sqrt \pi \,\Gamma(s-\frac{1}{2}) \zeta(2s-1)}{\pi^s\,\Gamma(s)}\, \tau_2^{1-s} {+} \frac{8}{\Gamma(s)}\,\sum_{k>0}\cos (2 \pi  k \tau_1)  
k^{s-\frac{1}{2} } \, \sigma_{1-2s}(k)
\tau_2^{\frac{1}{2}}\,K_{s-\frac{1}{2}}(  2\pi k \tau_2)\,,
\end{align}
which is the solution to the homogeneous Laplace eigenvalue equation,
\begin{equation}
\big[ \Delta_\tau-s(s-1)\big] E(s;\tau)=0\,.
\end{equation}
These modular invariant functions play a key role \cite{Dorigoni:2021guq} in the study of integrated four-point correlators of local operators in $\mathcal{N}=4$ SYM, they do however differ quite significantly from \eqref{eq:latticepmNew} not just from their automorphic properties.
In particular, when comparing the Fourier modes decomposition for the modular invariant Eisenstein series \eqref{eq:EisenFourier} with that of the automorphic functions \eqref{eq:F0cusp} we notice two key differences:
\begin{itemize}
\item[(i)] the Fourier zero-mode of ${F}_{s_1,s_2,s_3}([\mathbbm{1}];\tau)$ does \textit{not} include the analogue of the first term in \eqref{eq:EisenFourier}, i.e. the monomial $\tau_2^{s_1+s_2}$ is absent from \eqref{eq:F0cusp}; however such term is present in \eqref{eq:appFimproved} for the `improved' class of functions $\widehat{F}_{s_1,s_2,s_3}([\mathbbm{1}];\tau)$;
\item[(ii)] while for the non-holomorphic Eisenstein series the index of the divisor sigma function is (minus) twice the index of the Bessel function, the same is in general not true for ${F}_{s_1,s_2,s_3}([\mathbbm{1}];\tau)$.
\end{itemize}

\section{Finite-$N$ Decomposition in Automorphic Functions}
\label{sec:AppendixDecomp}

In this appendix we provide more technical details to show how to decompose the line defect integrated correlator as an infinite series of automorphic functions, elements of the novel class proposed in appendix \ref{sec:AppF}. As outlined in section \ref{sec:FiniteN}, our method relies first in computing the $k$-instanton contribution to the Wilson-line defect integrated correlator, $\mathcal{I}^{(k)}_{\mathbb{W}, N}(\tau)$. Secondly, we impose for arbitrary instanton number $k$ that $\mathcal{I}^{(k)}_{\mathbb{W}, N}(\tau)$ can be written as a linear combination of the non-zero Fourier modes of automorphic functions  $F_{s_1,s_2,s_3}([\mathbbm{1}];\tau)$, given in \eqref{eq:F0cusp}. A careful analysis of the instanton sector produces a finite system of linear equations for the coefficients of this linear combination which yields a unique solution for the cases $SU(2)$ and $SU(3)$ considered.

As discussed in the main text, we consider the 'reduced' version of the integrated correlator, as defined in \eqref{eq:LRedM},~i.e. we want to study
\ie 
\widetilde{\mathcal{I}}^{(k)}_{\mathbb{W}, N}(\tau_2) = \frac{L^1_{N-1}(-\frac{\pi}{\tau_2})} {N} \mathcal{I}^{(k)}_{\mathbb{W}, N}(\tau_2) \, .
\fe
The goal is to express  $\widetilde{\mathcal{I}}^{(k)}_{\mathbb{W}, N}(\tau_2)$ as an infinite sum over the 
Fourier coefficients $F^{(k)}_{s_1,s_2,s_3}([\mathbbm{1}];\tau_2)$ given in \eqref{eq:F0cusp}.
In what follows we determine the precise form of this series for the explicit examples $N=2$ and $N=3$, we stress however that our method can be utilised for higher values of $N$, although at the present time we have yet to understand systematically the $N$ dependence.

Let us start with discussing the line defect integrated correlator with $SU(2)$ gauge group. In this case, the matrix model originating from supersymmetric localisation is a one-dimensional integral. The $k^{{\rm th}}$ Fourier mode $\mathcal{I}_{\bW,2}^{(k) } (\tau_2) $ for $k> 0$ is given by (using the variable $y=\pi\tau_2$)
\begin{equation}\label{app:SU2}
\mathcal{I}_{\bW,2}^{(k) } (\tau_2) = \sum_{n = 0}^{\infty} \llangle Z^{\prime \prime(k)}_{\text{inst}}(0, a_{ij}) \rrangle_n  \left(\frac{2y\, {}_1 F_1(-n-1; \frac{1}{2}| -\frac{\pi^2}{4y}) }{\pi^2+2y} - 1\right) y^{-n}\,,
\end{equation}
where  $ \llangle Z^{\prime \prime(k)}_{\text{inst}}(0, a_{ij}) \rrangle_n$ denotes the coefficient of $1/y^n$ in the large $y$ expansion of $ \llangle Z^{\prime \prime(k)}_{\text{inst}}(0, a_{ij}) \rrangle$.
We derive \eqref{app:SU2} by expressing $Z^{\prime \prime(k)}_{\text{inst}}(0, a_{ij})$ as a power series in $a_1$ (using the fact $a_2=-a_1$ for $SU(2)$)
\ie
Z^{\prime \prime(k)}_{\text{inst}}(0, a_{ij}) =\sum_{n}   C_n \, a_1^{2n} \, .
\fe
The explicit form of the coefficient $C_n$ is not important for the following discussion. 
Then $\mathcal{I}_{\bW,2}^{(k) } (\tau_2)$ can be computed from \eqref{eq:IWinstGen} and takes the form
\ie \label{eq:IWinst}
\mathcal{I}_{\bW,2}^{(k) } (\tau_2) = e^{-2\pi |k|\tau_2}\sum_{n=0}^\infty  C_n \left( {\llangle e^{2\pi a_1} a_1^{2n} \rrangle \over \llangle e^{2\pi a_1} \rrangle} - \llangle  a_1^{2n} \rrangle \right) \, . 
\fe
The above expression then leads to \eqref{app:SU2}  by using the matrix model expectation values
\ie
{\llangle e^{2\pi a_1} a_1^{2n} \rrangle \over \llangle e^{2\pi a_1} \rrangle} = \frac{  \Gamma
   \left(n+\frac{3}{2}\right) \,
   _1F_1\left(-n-1;\frac{1}{2}|-\frac{1}{4 y}\right)}{4^{n-1}\, y^{n-1} \pi ^{2 n+\frac{1}{2}} (2
   y+1)}\, , \qquad 
\llangle a_1^{2n} \rrangle = 2^{1-2 n} \pi ^{-2 n-\frac{1}{2}} y^{-n} \Gamma
   \left(n+\frac{3}{2}\right) \, .
   \fe

The relation \eqref{app:SU2} is extremely useful since $\llangle Z^{\prime \prime(k)}_{\text{inst}}(0, a_{ij}) \rrangle$ 
has been evaluated for any instanton number $k$ in \cite{Dorigoni:2021guq},
\begin{align} 
\llangle Z^{\prime \prime(k)}_{\text{inst}}(0, a_{ij}) \rrangle  = &\notag \, 2 e^{-2\pi |k| \tau_2} \sum_{\substack{p,q>0\\ pq = |k|}}\Big[-\frac{p+q}{p\,q} + 2(p+q)y(1+2(p-q)^2 y) - 4 y^{3/2}\sqrt{\pi} e^{(p+q)^2y} \big((p^2+q^2)\\
&\label{eq:SU2-k} + (p^2-q^2)^2 y\big) \text{erfc}((p+q)\sqrt{y})\Big]\,,   
\end{align}
where $\text{erfc}(z) \coloneqq \frac{2}{\sqrt{\pi}} \int_z^\infty e^{-t^2} dt$ is the complementary error function. With \eqref{eq:SU2-k} at hand, it is then straightforward to obtain the $k$-instanton   contribution $\mathcal{I}_{\bW,2}^{(k) } (\tau_2)$, or equivalently $\widetilde{\mathcal{I}}^{(k)}_{\mathbb{W}, 2}(\tau_2)$, in the large-$y$ expansion. We give the first few terms in the large $y$ expansion
\ie
&\widetilde{\mathcal{I}}^{(k)}_{\mathbb{W}, 2}(\tau_2) {=} \\
&\sum_{\substack{p,q>0\\ pq = k}} \!\!\Big[\frac{3 \pi ^2 \left(3 p^2-10 p q+3 q^2\right)}{2 y^2 (p+q)^5} {+} \frac{\pi ^2 \left(-150 p^2+420 p q-150 q^2\right){+}\pi ^4 \left(3 p^4-4 p^3 q-14 p^2 q^2-4 p q^3+3 q^4\right)}{4 y^3 (p+q)^7} \cr 
&\qquad \quad -\frac{15 \pi ^2 \left(-147 p^2+378 p q-147 q^2\right)+15 \pi ^4 \left(5 p^4-4 p^3 q-18 p^2 q^2-4 p q^3+5 q^4\right)}{8 y^4(p+q)^9} +... \Big]\,.
\fe

Next, we replace $q$ by $k/p$ such that we have a single sum over all positive divisors $p$ of $k$. We then consider a large $k$ expansion and perform the sum over $p$ yielding divisor sigma functions $\sigma_{s}(k)$ with different indices. We conclude by grouping all terms according to the index of the divisor sigma function which accompanies them, i.e. we arrive at
\begin{align}
    \label{eq:SU2Ipq}
\widetilde{\mathcal{I}}^{(k)}_{\mathbb{W}, 2} = &\notag \sigma_{-3}(k)\Big(\frac{3 \pi ^4}{4 y^3}+\frac{9 \pi ^2}{2 y^2} \Big) - \sigma_{-5}(k) \Big[\frac{5 \pi ^6}{16 y^5}+\frac{75 \pi ^4}{8 y^4} + \frac{25 \pi ^2 \left(\pi ^2 k+6\right)}{4 y^3} + \frac{75 \pi ^2 k}{2 y^2}\Big] \\
&+\sigma_{-7}(k)\Big[\frac{7 \pi ^8}{64 y^7}+\frac{49 \pi ^6}{8
   y^6}+\frac{7 \left(28 \pi ^6 k+840
   \pi ^4\right)}{64 y^5}+\frac{7 \left(840 \pi ^4 k+2520 \pi
   ^2\right)}{64 y^4}\\
   &\notag \qquad\qquad \,\,\,+ \frac{7 \left(224 \pi ^4 k^2+3360 \pi ^2 k\right)}{64
   y^3}+\frac{147 \pi ^2 k^2}{y^2} \Big]+ ...\, . \nonumber
\end{align}
We observe that only odd negative indices of the divisor sigma functions appear and their coefficients are given by polynomials in $1/y$ and $k$. At this point we compare the above expression with the Fourier expansion of $F_{s_1,s_2,s_3}(\tau)$ given in \eqref{eq:F0cusp} and are led to the following ansatz, 
\begin{align}\label{app:su2ansatz}
    \widetilde{\mathcal{I}}_{\bW,2}^{(k)} &=  \sum_{s = 2}^\infty \sum_{s_1,s_2} \tilde{d}^{(2)}_{s_1,s_2,s} F^{(k)}_{-s_1,s_2,2s_2-2s}\\
    & \notag=  \sum_{s = 2}^\infty \sum_{s_1,s_2} \tilde{d}^{(2)}_{s_1,s_2,s}  \frac{4   \pi ^{-2 s+2 s_1+2 s_2-\frac{1}{2}} }{\Gamma (s_2)}k^{s_2-\frac{1}{2}} \sigma_{1-2 s}(k) y^{\frac{1}{2}-s_1} K_{s_2-\frac{1}{2}}(2 k y)\,,
\end{align}
with $s_1$ and $s_2$ taking values in a finite set for a given $s$.
In order to solve for the unknown coefficients $\tilde{d}^{(2)}_{s_1,s_2,s}$ we match the index of the divisor sums on both sides, as well as the powers of $y$ and $k$. Working at fixed index $s$ for the divisor sigma function, this yields a finite system of linear equations in the coefficients $\tilde{d}^{(2)}_{s_1,s_2,s}$ which has surprisingly a unique solution for which we list all the non-trivial coefficients for $2\leq s \leq 4$, 
\begin{equation}
\begin{aligned}
        {\tilde{d}}^{(2)}_{2,1,2} &= \frac{9}{4},\ {\tilde{d}}^{(2)}_{3,1,2} = \frac{3}{8}, \\
        {\tilde{d}}^{(2)}_{2,2,3} &= -\frac{75}{4},\ {\tilde{d}}^{(2)}_{3,1,3} =-\frac{75}{8},\ {\tilde{d}}^{(2)}_{3,2,3} = -\frac{25}{8},\ {\tilde{d}}^{(2)}_{4,1,3} =-\frac{25}{8},\ {\tilde{d}}^{(2)}_{5,1,3} = -\frac{5}{32},\\
        {\tilde{d}}^{(2)}_{2,3,4} &= 147,\ {\tilde{d}}^{(2)}_{3,2,4} = \frac{147}{2},\ {\tilde{d}}^{(2)}_{3,3,4} = \frac{49}{2},\ {\tilde{d}}^{(2)}_{4,1,4} = \frac{735}{16},\\
        {\tilde{d}}^{(2)}_{4,2,4} &= \frac{441}{16},\ {\tilde{d}}^{(2)}_{5,1,4} = \frac{735}{32},\
        {\tilde{d}}^{(2)}_{5,2,4} = \frac{49}{32},\ {\tilde{d}}^{(2)}_{6,1,4} = \frac{147}{64},\ {\tilde{d}}^{(2)}_{7,1,4} = \frac{7}{128} \, .
\end{aligned}
\end{equation}
We note that the summation indices $s_1, s_2$ in \eqref{app:su2ansatz} only appear in certain ranges, namely 
\ie
1 \leq s_2\leq s-1 \, , \quad 2 \leq s_1 \leq 2s-2s_2 + 1 \, , \quad s_1 + s_2 \geq s +1 \, .\label{eq:range}
\fe
By studying the coefficients ${\tilde{d}}^{(2)}_{s_1,s_2,s}$ for all values of $2\leq s \leq 25$ and $s_1,s_2$ as above, we were able to find the conjectural expression
\ie \label{eq:su2genOld}
{\tilde{d}}^{(2)}_{s_1,s_2,s} = \frac{(-1)^{-s} (2 s-1)^2  \left[ s (s_1+2 s_2-2)-(2
  s_2-1) (s_1+s_2-1) \right] \Gamma (2 s-2 s_2)
  \Gamma (s_1+s_2-1)}{ 2^{2 s-2 s_2}   \Gamma (s_1+1) \Gamma (2
  s-s_1-2 s_2+2) \Gamma (2
  (s_1+s_2-s))} \, ,
\fe
and the ansatz \eqref{app:su2ansatz} combined with the summation ranges \eqref{eq:range} becomes
\begin{equation}\label{app:su2ansatz2}   \widetilde{\mathcal{I}}^{(k)}_{\mathbb{W},2}(\tau_2) =  \sum_{s = 2}^\infty   \sum_{\substack{1 \leq s_2\leq s-1 \\ 2 \leq s_1 \leq 2s-2s_2 + 1 \\ s_1 + s_2 \geq s +1 } } {\tilde{d}}^{(2)}_{s_1,s_2,s} F^{(k)}_{-s_1,s_2,2s_2-2s} (\tau_2)\,.
\end{equation}
We find it more convenient to change summation variables $(s_1,s_2,s) \to (s_1+s_3,s_2,s_2+s_3)$ so that \eqref{app:su2ansatz2}  becomes
\begin{equation}\label{app:su2ansatzApp}
        \widetilde{\mathcal{I}}^{(k)}_{\mathbb{W},2}(\tau_2) =   \sum_{ s_1, s_2, s_3 =1}^\infty  d^{(2)}_{s_1,s_2,s_3} F^{(k)}_{-s_1-s_3,s_2,-2s_3} (\tau_2)\,,
\end{equation}
where the new coefficients are simply given by $d^{(2)}_{s_1,s_2,s_3} = {\tilde d}^{(2)}_{s_1+s_3,s_2,s_2+s_3} $, 
\begin{align}
\label{eq:su2App} d^{(2)}_{s_1,s_2,s_3} =&(-1)^{s_2+s_3} 4^{-s_3} \left(2 s_2+2 s_3-1\right){}^2 \left[s_3^2+\left(s_1+s_2-1\right)
   s_3+s_1-s_1 s_2+s_2-1\right]\\
&\notag \times \frac{ \Gamma \left(2 s_3\right) \Gamma \left(s_1+s_2+s_3-1\right)}{\Gamma \left(2
   s_1\right) \Gamma \left(s_3-s_1+2\right) \Gamma \left(s_1+s_3+1\right)} \, . 
\end{align}
Note that due to the factor $\Gamma \left(s_3-s_1+2\right)$ at denominator in \eqref{app:su2ansatzApp}, the sum over $s_1$ ranges only between $1\leq s_1 \leq s_3+1$. We present the expression \eqref{app:su2ansatzApp} in the main text, see \eqref{eq:su2ansatz}, since we find it the most convenient to construct a lattice sum integral representation for the full integrated line defect correlator.


The same method works for the case $N=3$ as well, so we will be brief. The main difference between  $N=3$ and $N=2$ is that we no longer have a compact formula like \eqref{app:SU2} for the $k^{\rm th}$ Fourier mode sector, hence we must go back to the matrix integral and consider
\begin{equation}
    \mathcal{I}_{\bW,3}^{(k)} (\tau_2)= \frac{\llangle (\sum_{i=1}^3 e^{2\pi a_i})\, Z^{\prime \prime(k)}_{\text{inst}}(0, a_{ij}) \rrangle}{\llangle \sum_{i=1}^3 e^{2\pi a_i}\rrangle} -  \llangle Z^{\prime \prime(k)}_{\text{inst}}(0, a_{ij}) \rrangle\, , 
\end{equation}
and we evaluate this expression as a perturbative series for $y\gg 1$. 
We then run the same procedure and group every term according to the index of the divisor sigma function it is accompanied by, thus obtaining the following expression for $\widetilde{\mathcal{I}}_{\bW,3}^{(k)}$:  
 \begin{align} \label{eq:SU3Ipq}
\widetilde{\mathcal{I}}_{\bW,3}^{(k)} = & \, \sigma_{-3}(k)\Big(\frac{\pi ^6}{2 y^4}+\frac{6 \pi ^4}{y^3}+\frac{12 \pi ^2}{y^2} \Big) \\
    &- \sigma_{-5}(k) \Big( \frac{5 \pi ^8}{24 y^6}+\frac{10 \pi ^6}{y^5} + \frac{5 \left(30 \pi ^6 k+420 \pi ^4\right)}{24 y^4}+\frac{5\left(360 \pi ^4 k+720 \pi ^2\right)}{24 y^3}+\frac{150 \pi ^2 k}{y^2} \Big) + ...\,. \nonumber
   \end{align} 
Again we observe that that only odd negative indices appear for the divisor sigma functions and their coefficients are polynomials in $1/y$ and $k$. We make an ansatz similar to \eqref{app:su2ansatz},
\begin{equation}
\widetilde{\mathcal{I}}_{\bW,3}^{(k)} = \sum_{s = 2}^\infty \sum_{s_1,s_2} {\tilde d}^{(3)}_{s_1,s_2,s} F^{(k)}_{-s_1,s_2,2s_2-2s} \,.\label{eq:su3ansApp}
\end{equation}
We impose that the polynomial in $1/y$ and $k$ multiplying a particular divisor sigma function with a given fixed index $s$ in \eqref{eq:SU3Ipq} equals that of \eqref{eq:su3ansApp}. This yields a finite linear system of equations for the coefficients of the corresponding polynomials in $1/y$ and $k$ which produces a unique solution for the coefficients ${\tilde{d}}^{(3)}_{s_1,s_2,s}$. We list here all the non-trivial coefficients for $2\leq s \leq 4$, 
\ie
       \tilde{d}^{(3)}_{2,1,2} &= 6,\ \tilde{d}^{(3)}_{3,1,2} = 3,\ \tilde{d}^{(3)}_{4,1,2} = \frac{1}{4}, \\
        \tilde{d}^{(3)}_{2,2,3} &= -75,\ \tilde{d}^{(3)}_{3,1,3} =-\frac{75}{2},\ \tilde{d}^{(3)}_{3,2,3} = -\frac{75}{2},\ \tilde{d}^{(3)}_{4,1,3} =-25,\ \tilde{d}^{(3)}_{5,1,3} = -\frac{55}{16},\\
        \tilde{d}^{(3)}_{2,3,4} &= 882,\ \tilde{d}^{(3)}_{3,2,4} = 441,\ \tilde{d}^{(3)}_{3,3,4} = 441,\ \tilde{d}^{(3)}_{4,1,4} = \frac{2205}{8},\\
        \tilde{d}^{(3)}_{4,2,4} &= \frac{2499}{8},\ \tilde{d}^{(3)}_{5,1,4} = \frac{3675}{16},\
        \tilde{d}^{(3)}_{5,2,4} = \frac{735}{16},\ \tilde{d}^{(3)}_{6,1,4} = \frac{343}{8},\ \tilde{d}^{(3)}_{7,1,4} = \frac{77}{32} \, . 
\fe

Compared to \eqref{eq:range}, for the case of $SU(3)$ the ranges for the summation variables $s_1, s_2$ are given by
\ie
1 \leq s_2\leq s-1 \, , \quad 2 \leq s_1 \leq 2s-2s_2 + 2 \, , \quad s_1 + s_2 \geq s +1 \, . 
\fe
Again, after having analysed all coefficients $d^{(3)}_{s_1,s_2,s}$ for $2\leq s \leq 22$ and $s_1,s_2$ in the above ranges, we conjecture the general expression 
\begin{align}\label{eq:su3gnerealApp}
&{\tilde{d}}^{(3)}_{s_1,s_2,s} = \frac{(-1)^s (2 s-1)^2  2^{2 s_2-2 s} \Gamma (2s-2
  s_2) \Gamma
  (s_1+s_2-2)}{3\, \Gamma
  (s_1+1) \Gamma (2 s-s_1-2 s_2+3) \Gamma (2
  (s_1+s_2-s))}   \\
& \times  \left[ 4 s^3 (s_1+2 s_2-3)
  (s_2^2 + (s_1-3) s_2+s_1+2) +2
  s^2 \left( s_1^4+s_1^3 (s_2-11)+s_1^2
  (-11 s_2^2-13 s_2+33) \right. \right. \cr
& \qquad -s_1(21 s_2^3-50 s_2^2+6 s_2+23) -10
  s_2^4+52 s_2^3-83 s_2^2+35
  s_2+6 \left.\right)\cr
&\qquad+s \left(-s_1^5+s_1^4 (15-7
  s_2)+s_1^3 (s_2^2+76
  s_2-68) +s_1^2 (35 s_2^3+13
  s_2^2-192 s_2+120) \right. \cr
& \qquad\left. +\,2 s_1 (22
  s_2^4-71 s_2^3+16 s_2^2+94
  s_2-57) +2 (8 s_2^5-47 s_2^4+78
  s_2^3-10 s_2^2-59 s_2+30) \right)\cr
&\qquad \left. +(2
  s_2-1) (s_1+s_2-2) (s_1+s_2-1)
  (s_1^3-9 s_1^2+s_1 (-3
  s_2^2-3 s_2+8)-2(s_2^3-3
  s_2^2-4 s_2+6)) \right] \, .  \nonumber
\end{align}

Just as we did for $SU(2)$, we change summation variables $(s_1,s_2,s) \to (s_1+s_3,s_2,s_2+s_3)$ so that \eqref{eq:su3ansApp} can be written as
\begin{equation}\label{app:su3ansatzApp}
        \widetilde{\mathcal{I}}^{(k)}_{\mathbb{W},3}(\tau_2) =   \sum_{s_1, s_2, s_3 =1}^\infty  d^{(3)}_{s_1,s_2,s_3} F^{(k)}_{-s_1-s_3,s_2,-2s_3} (\tau_2)\,,
\end{equation}
where the new coefficients are simply given by $d^{(3)}_{s_1,s_2,s_3} = {\tilde d}^{(3)}_{s_1+s_3,s_2,s_2+s_3} $, 
\begin{align}\label{eq:su3gnerealApp2}
&d^{(3)}_{s_1,s_2,s_3} = \frac{(-1)^{s_2+s_3} 4^{-s_3} \left(2 s_2+2 s_3-1\right)^2 \Gamma \left(2 s_3\right) \Gamma \left(s_1+s_2+s_3-2\right)}{3\,  \Gamma \left(2 s_1\right) \Gamma \left(s_3-s_1+3\right) \Gamma \left(s_1+s_3+1\right)}   \\
&\phantom{=}\times \Big\{ s_3^6+s_3^5\big[3 s_1+4 s_2-4\big] + s_3^4\big[2 s_1^2+\left(7 s_2-3\right) s_1+6 s_2^2-9 s_2+6\big]  \cr
&\phantom{=}+s_3^3 \big[-2 s_1^3+2 \left(s_2+9\right) s_1^2+\left(s_2-2\right) \left(3 s_2+14\right) s_1+s_2 \left(s_2 \left(4
   s_2-3\right)-2\right)+13\big] \cr
&\phantom{=}+s_3^2 \big[-3 s_1^4+28 s_1^3-\left(\left(s_2-16\right) s_2+66\right) s_1^2+\left(s_2 \left(\left(22-3 s_2\right) s_2-54\right)+83\right)
   s_1 \big]  \cr
&\phantom{=}+s_3^2\big[+ s_2 \left(s_2 \left(s_2 \left(s_2+5\right)-20\right)+46\right)-48 \big] + s_3 \big[-s_1^5+2 \left(s_2+5\right) s_1^4+\left(\left(s_2-12\right) s_2-20\right) s_1^3 \big] \cr
&\phantom{=} +s_3 \big[\left(9-3 \left(s_2-10\right) s_2\right)
   s_1^2 + \left(s_2-1\right) \left(s_2 \left(s_2 \left(3 s_2-7\right)+18\right)-8\right)-2 s_1 \left(\left(\left(s_2-4\right)
   s_2+8\right) s_2^2+3\right) \big]\cr
&\phantom{=} + \left(s_1-2\right) \left(s_1-1\right) \left(s_2-1\right) \big[s_1^3-\left(s_2+9\right) s_1^2+\left(8-\left(s_2-4\right)
   s_2\right) s_1+\left(s_2-2\right) \left(\left(s_2-1\right) s_2+6\right)\big] \Big\} \, .  \nonumber
\end{align}
Note that due to the factor $\Gamma \left(s_3-s_1+3\right)$ at denominator in the coefficients $d^{(3)}_{s_1,s_2,s_3}$,  in \eqref{app:su3ansatzApp} the sum over $s_1$ ranges only between $1\leq s_1 \leq s_3+2$.


\bibliographystyle{JHEP}
\bibliography{main}
\end{document}